\documentclass[11pt]{article}

\usepackage{amsthm}
\usepackage{graphicx} 
\usepackage{array} 
\usepackage{amsmath, amssymb, amsfonts, verbatim}
\usepackage{hyphenat,epsfig,subfigure,multirow}
\usepackage{nicefrac}
\usepackage{paralist}
\usepackage{enumitem}
\usepackage{nicematrix}
\usepackage{xcolor}

\usepackage[ruled,vlined,linesnumbered]{algorithm2e}
\usepackage{thmtools}
\usepackage{thm-restate}

\DeclareFontFamily{U}{mathx}{\hyphenchar\font45}
\DeclareFontShape{U}{mathx}{m}{n}{
      <5> <6> <7> <8> <9> <10>
      <10.95> <12> <14.4> <17.28> <20.74> <24.88>
      mathx10
      }{}
\DeclareSymbolFont{mathx}{U}{mathx}{m}{n}
\DeclareMathSymbol{\bigtimes}{1}{mathx}{"91}

\usepackage{tcolorbox}
\tcbuselibrary{skins,breakable}
\tcbset{enhanced jigsaw}

\usepackage[normalem]{ulem}
\usepackage[compact]{titlesec}

\definecolor{DarkRed}{rgb}{0.5,0.1,0.1}
\definecolor{DarkBlue}{rgb}{0.1,0.1,0.5}
\usepackage{nameref}
\definecolor{ForestGreen}{rgb}{0.1333,0.5451,0.1333}
\definecolor{Red}{rgb}{0.9,0,0}
\usepackage[linktocpage=true,	pagebackref=true,colorlinks,
linkcolor=DarkRed,citecolor=ForestGreen,bookmarks,bookmarksopen,bookmarksnumbered]{hyperref}
\usepackage[noabbrev,nameinlink]{cleveref}
\crefname{property}{property}{Property}
\creflabelformat{property}{(#1)#2#3}
\crefname{equation}{eq}{Eq}
\creflabelformat{equation}{(#1)#2#3}

\usepackage{bm}
\usepackage{url}
\usepackage{xspace}
\usepackage[mathscr]{euscript}

\usepackage{tikz}
\usetikzlibrary{arrows}
\usetikzlibrary{arrows.meta}
\usetikzlibrary{shapes}
\usetikzlibrary{backgrounds}
\usetikzlibrary{positioning}
\usetikzlibrary{decorations.markings}
\usetikzlibrary{patterns}
\usetikzlibrary{calc}
\usetikzlibrary{fit}
\usetikzlibrary{snakes}

\usepackage{mdframed}
\makeatletter
\def\BState{\State\hskip-\ALG@thistlm}
\makeatother
\usepackage{natbib}
\usepackage{enumitem}
\usepackage{algorithmic}

\usepackage[margin=1in]{geometry}

\renewcommand{\qed}{\nobreak \ifvmode \relax \else
      \ifdim\lastskip<1.5em \hskip-\lastskip
      \hskip1.5em plus0em minus0.5em \fi \nobreak
      \vrule height0.75em width0.5em depth0.25em\fi}

\newcommand{\eps}{\ensuremath{\varepsilon}}
\newcommand{\Kr}{\ensuremath{\mathcal{K}}}

\newcommand{\poly}{\mbox{\rm poly}}

\DeclareMathOperator*{\Prob}{\ensuremath{\textnormal{Pr}}}
\renewcommand{\Pr}{\Prob}

\newenvironment{tbox}{\begin{tcolorbox}[
		enlarge top by=5pt,
		enlarge bottom by=5pt,
		 breakable,
		 boxsep=0pt,
                  left=4pt,
                  right=4pt,
                  top=10pt,
                  arc=0pt,
                  boxrule=1pt,toprule=1pt,
                  colback=white
                  ]
	}
{\end{tcolorbox}}

\newcommand{\II}{\ensuremath{\mathbb{I}}}

\newcommand{\mireal}[1][]{
  \ifx\relax#1\relax%
    \II(\mione \,; \mitwo)%
  \else%
    \II(\mione \,; \mitwo\mid #1)%
  \fi
}

\newcommand{\MM}{\ensuremath{{M}}}

\newcommand{\RR}{\ensuremath{\mathbf{R}}}

\renewcommand{\SS}{\ensuremath{\mathcal{S}}}

\newtheorem{theorem}{Theorem}
\newtheorem{lemma}{Lemma}[section]

\newtheorem{corollary}[theorem]{Corollary}
\newtheorem{claim}[lemma]{Claim}
\newtheorem{fact}[lemma]{Fact}

\newtheorem{definition}[lemma]{Definition}

\newtheorem{problem}{Problem}

\newtheorem*{theorem*}{Theorem}
\newtheorem*{claim*}{Claim}
\newtheorem*{proposition*}{Proposition}
\newtheorem*{lemma*}{Lemma}
\newtheorem*{problem*}{Problem}

\crefname{lemma}{Lemma}{Lemmas}
\crefname{claim}{Claim}{Claims}

\newtheorem{mdresult}{Result}

\newtheoremstyle{restate}{}{}{\itshape}{}{\bfseries}{~(restated).}{.5em}{\thmnote{#3}}
\theoremstyle{restate}

\allowdisplaybreaks

\newcommand{\nnz}{\mathrm{nnz}}

\newcommand{\Z}{\mathbb{Z}}
\newcommand{\F}{\mathbb{F}}

\newcommand{\tp}{\top}

\renewcommand{\AA}{\mathbf{A}}
\newcommand{\BB}{\mathbf{B}}

\newcommand{\DD}{\mathbf{D}}

\renewcommand{\II}{\mathbf{I}}

\renewcommand{\MM}{\mathbf{M}}

\newcommand{\OO}{\mathbf{O}}

\renewcommand{\SS}{\mathbf{S}}

\newcommand{\UU}{\mathbf{U}}
\newcommand{\VV}{\mathbf{V}}
\newcommand{\WW}{\mathbf{W}}
\newcommand{\XX}{\mathbf{X}}

\newcommand{\bb}{\mathbf{b}}
\newcommand{\cc}{\mathbf{c}}

\newcommand{\ee}{\mathbf{e}}
\newcommand{\ff}{\mathbf{f}}

\newcommand{\rr}{\mathbf{r}}
\renewcommand{\ss}{\mathbf{s}}

\newcommand{\uu}{\mathbf{u}}
\newcommand{\vv}{\mathbf{v}}
\newcommand{\ww}{\mathbf{w}}
\newcommand{\xx}{\mathbf{x}}
\newcommand{\yy}{\mathbf{y}}
\newcommand{\zz}{\mathbf{z}}

\newcommand{\diag}{\mathrm{diag}}

\newcommand{\Mbegin}{\left[\begin{matrix}}
\newcommand{\Mend}{\end{matrix}\right]}

\newcommand{\Q}{\mathbb{Q}}

\newcommand{\sgn}{\mathrm{sgn}}
\newcommand{\R}{\mathbb{R}}

\newcommand{\xxhat}{\widehat{\xx}}
\newcommand{\yyhat}{\widehat{\yy}}

\newcommand{\xxtil}{\widetilde{\xx}}
\newcommand{\yytil}{\widetilde{\yy}}

\newcommand{\bbtil}{\widetilde{\bb}}
\newcommand{\rrtil}{\widetilde{\rr}}

\newcommand{\xxbar}{\overline{\xx}}
\newcommand{\yybar}{\overline{\yy}}

\newcommand{\Otil}{\widetilde{O}}
\newcommand{\lam}{\lambda}
\newcommand{\lamtil}{\widetilde{\lambda}}

\newcommand{\approxbar}{\overset{\underline{\hspace{0.6em}}}{\approx}}

\newcommand{\callalg}[1]{\hyperref[alg:#1]{\textsc{#1}}}
\newcommand{\labelalg}[1]{\label{alg:#1}}

\newcommand{\callthm}[1]{\hyperref[#1]{Theorem} \ref{#1}}  

\newcommand{\TimeMatVec}[1]{ \mathrm{T}_{\mathrm{MatVec}(#1)} }
\newcommand{\rep}[1]{ {#1}_{\mathrm{rep}} }

\newcommand{\merr}[1]{ \mathcal{E}({#1}) }
\newcommand{\fl}{\mathrm{fl}}

\newcommand{\calI}{\mathcal{I}}
\newcommand{\calL}{\mathcal{L}}

\newcommand{\uuhat}{\widehat{\uu}}

\newcommand{\ssbar}{\overline{\ss}}

\newcommand{\la}{\langle}
\newcommand{\ra}{\rangle}

\usepackage{bbm}

\usepackage[hang,flushmargin]{footmisc}
\makeatletter
\newcommand{\algorithmfootnote}[2][\footnotesize]{%
  \let\old@algocf@finish\@algocf@finish
  \def\@algocf@finish{\old@algocf@finish
    \leavevmode\rlap{\begin{minipage}{\linewidth}
    #1#2
    \end{minipage}}%
  }%
}
\makeatother

\newcommand{\SSigma}{\mathbf{\Sigma}}

\usepackage{mathdots}

\title{Numerical Linear Algebra in Linear Space}
\author{Yiping Liu\footnote{(liu-yp23@mails.tsinghua.edu.cn) Tsinghua University. 
 \smallskip} 
\\ {\small Tsinghua University} 
\and
Hoai-An Nguyen\footnote{(hnnguyen@andrew.cmu.edu) Computer Science Department, Carnegie Mellon University.
Supported in part by an NSF GRFP fellowship grant number DGE2140739 and NSF CAREER Award CCF-2330255. \smallskip} \\ {\small Carnegie Mellon University}  \and
Junzhao Yang\footnote{(junzhaoy@andrew.cmu.edu) Computer Science Department, Carnegie Mellon University.
 \smallskip} \\ {\small Carnegie Mellon University} 
}

\date{}

\begin{document}
\maketitle

\pagenumbering{arabic}

\begin{abstract}
We present a randomized linear-space solver for general linear systems $\mathbf{A} \mathbf{x} = \mathbf{b}$ with $\mathbf{A} \in \mathbb{Z}^{n \times n}$ and $\mathbf{b} \in \mathbb{Z}^n$, without any assumption on the condition number of $\mathbf{A}$. 
For matrices whose entries are bounded by $\poly(n)$, the solver returns a $(1+\epsilon)$-multiplicative entry-wise approximation to vector $\mathbf{x} \in \mathbb{Q}^{n}$ using $\widetilde{O}(n^2 \cdot \mathrm{nnz}(\mathbf{A}))$ bit operations and $O(n \log n)$ bits of working space (i.e., linear in the size of a vector), where $\mathrm{nnz}$ denotes the number of nonzero entries. Our solver works for right-hand vector $\bb$ with entries up to $n^{O(n)}$. 
To our knowledge, this is the first linear-space linear system solver over the rationals that runs in $\widetilde{O}(n^2 \cdot \mathrm{nnz}(\mathbf{A}))$ time.

We also present several applications of our solver to numerical linear algebra problems, for which we provide algorithms with efficient polynomial running time and near-linear space. In particular, we present results for linear regression, linear programming, eigenvalues and eigenvectors, and singular value decomposition.

\end{abstract}

\newpage

\section{Introduction}
The memory usage of numerical linear algebra is of fundamental importance to both theory and practice. In practical settings, one of the primary goals of linear algebra packages is to accelerate computation by exploiting the sparse structure of real-world matrices, while keeping memory usage well below quadratic \cite{George73nested, DEGLL99supernodal, Liu92multifrontal, TSW16survey}.
Theoretically, the space complexity has been widely studied in various models. 
Csanky~\cite{C1976fast} and Berkowitz~\cite{B1984computing} showed that $\mathsf{det} \in \mathsf{NC}^2$, yielding a linear system solver requiring only $O(\log^2 n)$ space but using  superpolynomial time---e.g., $n^{O(\log n)}$. 
Recently, Raz~\cite{R2019fast} proved a lower bound for the streaming setting: even for learning parity (i.e., solving a linear system over $\F_2$), any one-pass learner must either use at least $\Omega(n^2)$ bits of memory or use exponentially many samples. Sharan, Sidford and Valiant \cite{SSV19memory} extend this memory-sample tradeoff to the continuous domain, proving a lower bound for linear regression in the streaming setting.

In this paper, we study fast space-efficient algorithms for numerical linear algebra problems. We measure space complexity in terms of \emph{working space}\footnote{In this model, algorithms cannot modify the inputs in place or reread the outputs.}---excluding the input and output space. 
We first focus on the problem of solving linear systems, $\AA \xx = \bb$.
Typical textbook approaches include Gaussian elimination and iterative methods.
The primary advantage of iterative methods is their substantially lower working space compared to direct approaches such as Gaussian elimination; however, their running time scales at least proportional to the square root of the condition number, which can itself be polynomial or even exponential. This begs the question: can we solve linear systems using small working space without any running time dependence on the condition number?

There is a setting in which this has been shown to be possible. Over finite fields, Wiedemann~\cite{W1986solving} designed a linear system solver that is both efficient in time and space. The input consists of a matrix $\AA \in \F_p^{n \times n}$ with $\nnz$ nonzero entries, and a vector $\bb \in \F_p^n$, where $\F_p$ denotes the finite field of integers $\{0,1,\ldots, p-1\}$ for a prime $p$. Wiedemann's algorithm outputs $\xx \in \F_p^n$ to the system $\AA \xx \equiv \bb \pmod p$ in $O(n \cdot \nnz )$ field operations and stores $O(n)$ field elements \cite{W1986solving}.

In the more general setting, however, the answer is more unclear. Here, the input consists of matrix $\AA$ and vector $\bb$, both of which have integer entries, and the task is to approximate the solution vector $\AA^{-1} \bb$ whose entries are rational.
There is a long line of work on optimizing the running time of solving linear systems over the rationals \cite{D1982exact, S2005shifted, EGGSV2006solving, BLS2019deterministic, PV2024solving}. However, all of these algorithms require $\Omega(n^{1.5})$ space---
\cite{EGGSV2006solving} requires $\Omega(n^{1.5})$ and the others require $\Omega(n^2)$ space.
Although not explicitly stated in the literature, it is possible to achieve linear working space by invoking \cite{W1986solving} with the Chinese Remainder Theorem (as discussed in \Cref{sec:lin-solve-overview}). However, this approach incurs a running time of $O(n^3 \cdot \nnz)$. Given these limitations---and the fundamental importance of solving linear systems---it is natural to study the following question: 
\begin{center}
    \textit{How fast can linear systems over the rationals be solved using only linear working space?}
\end{center}

In response to this question, we present a randomized solver for general linear systems over the rationals, using $O(n \log n)$ bits of working space---or equivalently, $O(n)$ words\footnote{We assume each input entry is bounded by $\poly(n)$ and is stored in one word.} which is linear in the representation size of a vector---and only running in $\Otil(n^2 \cdot \nnz)$ bit operations. In particular, for sparse matrices, the running time is $\Otil(n^3)$. To the best of our knowledge, no previous algorithm achieves this time complexity under comparable space constraints.
We further observe that significantly improving upon the $\Otil(n^2 \cdot \nnz)$ time bound appears challenging. Even for sparse matrices with a polynomial condition number bound, achieving a solver that runs in $\Otil(n^2)$ time is a major open problem. 
This suggests that matching the efficiency of Wiedemann’s finite field solver which runs in $\Otil(n \cdot \nnz)$ time may be out of reach in the rational setting.  
Furthermore, approaches that rely on fast matrix multiplication typically require storing dense matrix objects, making subquadratic space usage particularly challenging.

Our solver provides strong guarantees: it makes no assumptions on the condition number of $\AA$, and outputs an \emph{entry-wise} approximation of the solution $\AA^{-1} \bb$. This removes the standard assumption of a polynomially-bounded condition number and provides a stronger guarantee than the commonly used norm-wise approximation.
We remark that we do not pursue sublinear space due to the following barrier: obtaining a polynomial-time linear system solver in $O(n^{0.99})$ space with multiplicative guarantees would imply an algorithm for directed connectivity that runs in both polynomial-time and $O(n^{0.99})$ space, which remains open. 

In addition to presenting our solver, we demonstrate its utility by applying it to several other core problems in numerical linear algebra. For each, we develop algorithms that use near-linear working space. We outline these applications below.

\subsection{Our Contributions}
Here we formally describe the results of our paper.
In the following, we take special care to track the space complexity in the number of \emph{bits}, and time complexity in the number of \emph{bit operations}.
In our computational model, we measure only the \emph{working space}, which may be smaller than the size of the input or output. The algorithm cannot modify the inputs in place or re-read the outputs. 

To serve both the community and our own needs, we give a self-contained presentation of Wiedemann's algorithm \cite{W1986solving} for linear systems solving over finite fields and its applications. In particular, we present the algorithm and analysis while tracking the space and time complexity carefully. We also reformulate the algorithm so that we can adapt it for our application of computing determinants, following the approaches of \cite{KS1991wiedemann} and \cite{CEKSTV2002efficient}.
Formal statements are presented in \Cref{sec:ffla}, with proofs deferred to \Cref{sec:wiedemann}.
We use finite-field solvers for linear systems and determinants as subroutines of our main algorithm.

We now present our main result, the linear-space linear system solver.
\begin{theorem}[Linear systems solving, \Cref{thm:LinSolve}, Informal]
    There exists a randomized algorithm that, given $\eps \in (0,1)$, an invertible matrix $\AA \in \Z^{n \times n}$ with entries in $[-U, U]$, and a vector $\bb \in \Z^n$ with entries in $[-U^n, U^n]$, with high probability outputs a $(1+\eps)$-multiplicative entry-wise approximation of $\AA^{-1} \bb$ in $\Otil(n^2 \log(1/\eps) \cdot \nnz(\AA) )$\footnote{We use $\nnz$ to denote the number of nonzero entries in $\AA$. $\Otil$ hides logarithmic factors, $\log^{O(1)}( nU \log (1/\eps))$. \label{note-Otil}} bit operations using $O(n \log (nU))$ bits of space. 
\end{theorem}

We highlight some properties of our solver.
First, our solver gives \emph{multiplicative} and \emph{entry-wise} approximations. This is stronger than the usual notion of norm-wise approximations. In the extremely low error regime $\eps = 2^{- \Theta(n \log(nU))}$, our solver produces exact solutions in $O(n \log (nU))$ bits of working space.
Second, we represent the output using $O(\log(nU/\eps))$-bit floating-point numbers, a format that is more space-efficient for large numbers---e.g., entries of $\AA^{-1} \bb$ can be as large as $2^{\Theta(n)}$. We note that the working space we achieve is smaller than the output size, which requires $O(n \log(nU/\eps))$ bits. 
Third, we require no bound on the condition number of $\AA$---we therefore say our solver works for general linear systems.
Moreover, our solver can accommodate right-hand vectors $\bb$ whose entries are as large as $U^n$.

We then apply our linear-space solver to other numerical linear algebra primitives, including linear regression, linear programming, eigenvalues/eigenvectors, and singular value decomposition. We obtain $\Otil(n)$-space polynomial-time algorithms for all of these problems. 

\begin{corollary}[Linear regression, \Cref{cor:linear-regression}, Informal]
     There exists a randomized algorithm that, given $\eps>0$, $n \ge d$, a full-rank matrix $\AA \in \Z^{n \times d}$ 
    and a vector $\bb \in \Z^{d}$ both with integer entries in $[-U, U]$, with high probability computes an $(1+\eps)$-multiplicative entry-wise approximation of $\arg \min_{\xx \in \R^d}  \|\AA \xx - \bb\|_2$, and runs in $\Otil(d^3 \log(1/\eps) \cdot \nnz(\AA) )$\textsuperscript{\ref{note-Otil}} bit operations using $O(d \log(nU))$ bits of space.
\end{corollary}

By applying our solver to the Interior Point Method (IPM), we obtain the following near-linear space algorithm. Our results build on the analysis of an inverse free path following IPM by Ghadiri, Peng and Vempala \cite{GPV2023bit}.

\begin{theorem}[Linear programming, \Cref{thm:IPM}, Informal]
     There exists a randomized algorithm that, given $\eps \in (0,1)$, full-rank and $\poly(n)$-conditioned $\AA \in \Z^{n \times d}$, $\bb \in \Z^d$, and $\cc \in \Z^n$ all with entries in $[-U, U]$, with high probability outputs $\xxhat \in \R^n$ such that
    \begin{align*}
        \cc^\tp \xxhat \le \min_{\AA^\tp \xx = \bb, \xx \ge 0} \cc^\tp \xx + \eps, \quad \text{ and } \quad
        \| \AA^\tp \xxhat - \bb \|_2 \le \eps,
    \end{align*}
    in $\Otil( d^3n^{0.5} \cdot \nnz(\AA))$ bit operations
     using $O(n \log(\frac{n  U R}{\eps r}))$\footnote{ $r$ and $R$ denote the inner radius and outer radius of the linear program, respectively, defined in \Cref{def:inner-radius-outer-radius}.}
     bits of space.  
\end{theorem}

We apply our solver to eigenvalue and eigenvector computations, as well as singular value decomposition (SVD), obtaining linear-space algorithms for each task in the regime where $\eps > 1 / \poly(n,U)$. Our algorithms build on spectrum perturbation and divide-and-conquer strategies, which we believe are of independent interest, and we provide a full finite-precision error analysis.

\begin{theorem}[\Cref{thm:Spectrum}, \Cref{thm:Eigendecompose}, and \Cref{cor:SVD}, Informal]
    There exist randomized algorithms that, given a matrix $\AA \in \Z^{n \times n}$ with integer entries in $[-U, U]$, in $\Otil(n^3 \cdot \nnz(\AA))$\footnote{Here, $\Otil$ hides logarithm factors, $\log^{O(1)}( nU/\eps)$. We focus on $\eps \ge 1/\poly(n, U)$ for these eigenproblems.} bit operations using $O(n \log(nU/\eps))$ bits of space, with high probability solve the following tasks:
    \begin{itemize}
        \item For symmetric $\AA$, compute all eigenvalues up to $\eps$ additive error. (\Cref{thm:Spectrum})
        \item For symmetric $\AA$, compute $\eps$-approximate corresponding eigenvectors. (\Cref{thm:Eigendecompose})
        \item For general $\AA$, compute $\eps$-approximate SVD, $\UU \SSigma \VV^\tp$. (\Cref{cor:SVD})
    \end{itemize}
\end{theorem}

\subsection{Related Work} \label{sec:RL}
We now give an overview of other related works.

\paragraph{Linear system solvers.} 

We outline the relevant previous work on solving linear systems---namely, approximately computing $\AA^{-1} \bb$ for an integer matrix $\AA$. For simplicity, we omit logarithmic factors for both space and time bounds. 
Dixon~\cite{D1982exact} proposed an algorithm that uses $O(n^\omega + n \cdot \nnz)$ time but $O(n^2)$ space where $\omega$ is the matrix multiplication exponent. A combination of \cite{D1982exact} with \cite{W1986solving} gives an $O(n^2 \cdot \nnz)$ time algorithm but again $O(n^2)$ space. Storjohann~\cite{S2005shifted} and Birmpilis, Labahn, and Storjohann~\cite{BLS2019deterministic} give algorithms with $O(n^\omega)$ time and $O(n^2)$ space. Peng and Vempala~\cite{PV2024solving} presented an algorithm for sparse \emph{polynomially-conditioned} matrices in time asymptotically faster than matrix multiplication for any $\omega > 2$; based on the value of $\omega$ at the time, they achieved $O(n^{2.331645})$ time but $O(n^2)$ space. Eberly, Giesbrecht, Giorgi, Storjohann, and Villard~\cite{EGGSV2006solving} developed an $O(n^{2.5})$ time and $O(n^{1.5})$ space algorithm for polynomially-conditioned, sparse matrices. 

There are also ways of achieving $O(n)$ space and $O(n^4)$ time (for sparse matrices), although they have not been formally written down. For example, one could combine \cite{D1982exact} with \cite{W1986solving} to obtain an algorithm with time increased to $O(n^3 \cdot \nnz)$ but with $O(n)$ space. One could also compute $O(n)$ determinants by \cite{W1986solving} and leverage Cramer's rule to again achieve a running time of $O(n^4)$ for sparse matrices. 

\paragraph{Bit complexity.} Recall that our input matrix $\AA$ has integer entries from $[-U, U]$. Unlike much of the literature, we do not make the assumption of an upper bound on $U$. Instead we carefully track the bit complexity and bits of space used by our algorithms. Bit complexity has been the focus in numerous topics in the literature including diagonalization of a matrix \cite{S2023complexity, DKRS2023bit, BGKS2023pseudospectral} and continuous optimization \cite{GPV2023bit}.

\paragraph{Specialized linear system solvers.} 
There has been a large body of work on solvers for special classes of matrices including 
circulant matrices \cite{G2005topeplitz},  Hankel and Toeplitz matrices \cite{KKM1979displacement, XXG2012superfast, XXCB2014superfast}, $d$-sparse matrices \cite{S2010algorithms}, symmetric diagonally dominant matrices \cite{ST2014nearly, CKMPPRX2014solving, AKP2019solving}, and Laplacians \cite{V2012laplacian}. There has also been work on low-rank linear systems \cite{CLW2018quantum, CGLLTW2020quantum} and on quantum and quantum-inspired algorithms for solving linear systems \cite{ WZP2018quantum, CGLLTW2020quantum, CLW2018quantum,HHL2009quantum, KP2020quantum}. 

\paragraph{Linear regression.}
There has been work in the streaming model for linear regression. Clarkson and Woodruff~\cite{CW2009numerical} give a linear sketch for finding $\xx$ such that 
\[
\|\AA \xx - \bb \|_2 \leq (1 + \eps) \min_{\xx' \in \mathbb{R}^d} \|\AA \xx' -\bb\|_2
\]
where $\AA$ is an $n \times d$ matrix with integer entries that are specified by $O(\log(nd))$ bits. 
The sketch uses $O(d^2 \eps^{-1} \log(1/\delta) \log(nd))$ bits of space where $\delta$ is the probability of failure and directly translates to a one-pass turnstile streaming algorithm. In the turnstile streaming model, insertions and deletions to the matrix $\AA$ are given one-by-one in a stream, and an algorithm cannot store all the updates. Instead, it must process the updates in low space and output an answer at the end of the stream. They also provide a lower bound for one-pass turnstile streaming algorithms for $n \geq d$ and sufficiently large $d$ of $\Omega(d^2 \eps^{-1}\log(nd))$. Dagan, Kur, and Shamir~\cite{DKS2019space} give a similar lower bound for streaming algorithms where the updates are instead rows of $\AA$. We note that in these settings the assumption is that $n$ is much larger than $d$, and it is desirable to avoid large dependencies on $n$ in the space. 

We note that our upper bound of $\Otil(d)$ space for linear regression does not contradict this lower bound. This is because the lower bound applies to one-pass streaming algorithms. In particular, the streaming algorithm can only access each update one time (unless it stores the update), whereas we assume read access to the input and can use additional queries to reduce our space.
There has also been work, primarily through sketching, to reduce the dimension of $\AA$ to speed up the time of linear regression (see \cite{W2014sketching} for more). 

There are gradient based methods that achieve $\Otil(d)$ space but using $O(\poly(\kappa))$ iterations (which translates to $\Otil(\poly(\kappa) \cdot n^2)$ time) such as that of Musco, Musco, and Sidford~\cite{MMS2018stability}. Given that we do not assume a bound on $\kappa$, we offer a significant advantage when we do not have $\kappa = \poly(n)$ (and even for $\kappa = \poly(n)$ when the exponent is large). 

\paragraph{Learning lower bounds.} 
There is a long line of work \cite{R2019fast, R2017time, MM2017mixing, BGY2018time, MM2018entropy, KRT2017time, GRT2018extractor, SSV19memory} on memory-sample trade-off lower bounds for learning problems. Raz~\cite{R2019fast} first proved a memory-sample lower bound for learning parity in the one-pass setting. Learning parity is just solving linear system $\AA \xx = \bb$ over the field $F = \mathbb{F}_2$ with a random $\AA \in F^{m \times n}$ and $\bb \in F^{n}$ where $m \ge n$. The learner is given the rows of $[\AA; \bb]$ sequentially, with each row as a sample.
They showed that if the learner has $0.01n^2$ bits of memory and reads the input with only one pass, then it must see $2^{\Omega(n)}$ samples to solve for $\xx$ with probability at least $2^{-O(n)}$.

Garg, Raz and Tal~\cite{GRT2019time} extended this result to two-passes, giving an $\Omega(n^{1.5})$ memory versus $2^{\Omega(\sqrt{n})}$ samples trade-off lower bound. \cite{LTWY2023tight} improved this result and showed a tight memory-sample lower bound for constant-pass learning, strengthening the result of \cite{R2019fast}.
They also showed an algorithm using $O(n^2/k)$ memory and $O(2^k)$  passes for a parameter $k$ to complement the lower bound.

This upper bound can be improved to $O(n)$ memory and $O(n)$ passes using Wiedemann's algorithm \cite{W1986solving}. In this paper, we can apply our solver to the learning problem over $\Q$. As a corollary (by \Cref{cor:LinSolve-black-box}), our algorithm gives a learner with $O(n \log (nU))$ bits of memory and $O(n^2 \log(1/\eps))$ passes to learn an entry-wise $\eps$-multiplicative approximation for the solution $\AA^{-1} \bb$.

\subsection{Paper Organization}
The organization of the paper goes as follows. In \Cref{sec:TO} we provide a technical overview of our contributions. In \Cref{sec:prelims} we give our preliminaries. In \Cref{sec:ffla} we present the finite field algorithms. In \Cref{sec:rationals} we present an algorithm for computing the determinant and then give our main linear systems solver. In \Cref{sec:app}, we give our applications for linear regression and linear programming. In \Cref{sec:eigen}, we discuss our applications for computing eigenvalues, eigenvectors and SVD. 

\section{Technical Overview} \label{sec:TO}

We provide an overview of our techniques for all of our contributions. We assume that all matrices and vectors have maximum dimension $n$, and that the entries are bounded by $U$.
An algorithm with space usage $O(n \log (nU))$ bits we say uses \textit{linear-space} since under the standard assumption that $U = \poly(n)$, the space is linear to that of a vector in $\R^n$ with entries bounded by $\poly(n,U)$ from an information-theoretic perspective. This is also the space requirement for storing a determinant of an $n \times n$ $01$-matrix.

\subsection{Determinant in Linear Space}
Our linear-space linear systems solver requires computing the determinant. Therefore, computing the determinant must also only use linear space. \cite{W1986solving} and \cite{KS1991wiedemann} showed that determinant over finite fields can be obtained from the minimal polynomial with some preconditioning. 
We present the algorithm, \callalg{DeterminantZp}, that computes the determinant by Wiedemann's algorithm with a simplified preconditioning from \cite{CEKSTV2002efficient} in \Cref{sec:DeterminantZp}. The algorithm runs in $O(n \cdot \nnz)$ time.

For integer matrices, the determinant can be exponentially large and requires $\Omega(n \log (nU))$ bits to store. We can still compute the determinant exactly in linear space using Chinese Remainder Theorem (CRT).  
Specifically, we sample $n$ primes, $p_1,\ldots, p_n$, and compute $\det(\AA) \bmod p_i$ for each using \callalg{DeterminantZp} (\Cref{thm:DeterminantZp}). We then reconstruct the integer determinant via CRT. 
The correctness follows from Hadamard's bound, ensuring that the determinant's magnitude is below the product of the primes, allowing exact recovery. The algorithm runs in $\Otil(n^2 \cdot \nnz)$ bit operations and requires only $O(n \log (nU))$ bits of space. 

\subsection{{Linear-Space General Linear Systems Solver}} \label{sec:lin-solve-overview}
We detail our solver for general linear systems over the rationals which proves \Cref{thm:LinSolve}. 
Given an input matrix $\AA \in \Z^{n \times n}$ and a vector $\bb \in \Z^n$ with integer entries bounded by $U$,
our goal is to efficiently compute an $e^\eps$-multiplicative entry-wise approximate solution.
Just in this overview, for simplicity, we assume that $U \le \poly(n)$ and $\eps \ge 1/\poly(n)$.

Our algorithm is designed to meet several objectives. First, it operates without assumptions on the condition number of $\AA$, other than that it is finite (which follows from $\AA$ being invertible). Second, it ensures a good entry-wise approximation of the solution vector $\xx$. Most critically, it uses only $O(n \log n)$ bits of space and achieves bit complexity $\Otil(n^2 \cdot \nnz)$, where $\nnz$ is the number of nonzero entries in $\AA$. 

The main challenge is optimizing the time while being constrained to $O(n \log n)$ space. Our algorithm builds on finite field methods. However, finite field methods recover $\AA^{-1} \bb \in \Q^n$ from $(\AA^{-1} \bb \bmod P)$ for some big prime $P$, whose entries can be exponential in $n$. Therefore storing this vector would require $\Omega(n^2)$ space. We must therefore make careful modifications to achieve our desired guarantees. We now outline our initial attempts at solving the problem, in the hopes of improving the clarity of the final solution.

\paragraph{The first attempt: directly invoking Wiedemann's algorithm. }
Wiedemann's algorithm solves linear systems over finite fields in linear space and $O(n \cdot \nnz)$ field operations \cite{W1986solving}. We can solve $O(n)$ linear systems (over finite fields) with respect to $O(n)$ distinct primes, and integrate the $O(n)$ solutions with rational reconstruction via CRT to recover the exact rational solution. However, there is a bottleneck that prevents us from achieving the correct space complexity---storing the $O(n)$ solution vectors. A simple fix is to solve for the entire solution vector but only store one entry at a time. We repeat this procedure $n$ times to recover all $n$ entries of the final solution vector. This gives an algorithm using $\Otil(n^3 \cdot \nnz)$ bit operations and $O(n \log n)$ bits of space. The same time and space complexity could be achieved by using 
$p$-adic lifting from \cite{D1982exact} instead of CRT.

\paragraph{The second attempt: $p$-adic lifting and the determinant. }
We can see that the previous approach requires too much time due to computing the entire solution vector $n$ times. To avoid this, we make the following crucial observation and then apply Dixon's $p$-adic lifting from \cite{D1982exact}. 
By Cramer's rule, the entries of the solution can be turned into integers if we know the determinant. Specifically, since the denominator of each entry in the rational solution $\AA^{-1} \bb$ is a factor of $\det(\AA)$, we compute $\det(\AA)$ using \callalg{Determinant} (\Cref{thm:Determinant}) and multiply the system by $\det(\AA)$, 
\begin{align*}
    \AA \yy = \bb \det(\AA),
\end{align*}
so that $\yy \in \Z^n$ is an integer-valued vector.

We then apply the technique of $p$-adic lifting. The original technique proposed by \cite{D1982exact} employs $T = O(n)$ iterations to gradually extract the $p$-adic digits when the solution vector $\yy$ is expressed in $p$-adic form for some prime $p = \poly(n)$:
\begin{align*}
    \yy \equiv \sum_{i=0}^{T-1} \yy^{(i)} \cdot p^i \pmod{p^T}
\end{align*}
where $\yy^{(i)}$ consists only of entries in $[0, p)$. Then, $\bb^{(0)} = \bb \cdot \det(\AA)$ is initialized. The following would then be iteratively computed: 
\begin{align}
    & \yy^{(i)}  \equiv \AA^{-1}  \bb^{(i)} \pmod{p}, \\
    & \bb^{(i+1)}  = \frac{\bb^{(i)} - \AA \yy^{(i)}}{p}. \label{eq:padic-iteration}
\end{align}
All the vectors $\bb^{(i)}$ are computed over the integers, and the congruence relation ensures that the numerator is a multiple of $p$ so $\bb^{(i)}$ has integer entries. Dixon precomputes the inverse $(\AA^{-1} \bmod p)$ to optimize the running time, but here we can replace it with Wiedemann's solver from \cite{W1986solving}.

Now, we obtain $\yy^{(0)}, \yy^{(1)}, \dots, \yy^{(T-1)}$ sequentially, but we cannot afford to store them all since each vector takes $O(n)$ space.  
Observe that the $j$-th entry of $\AA^{-1} \bb$ is $\det(\AA)^{-1} \cdot \sum_{i=0}^{T-1} \yy^{(i)}_j p^i$  (WLOG we assume nonnegative entries), so we can keep track of a multiplicative approximation of
$
    \sum_{i=0}^{t} \yy^{(i)}_j p^i
$
with an $O(\log(n/\eps))$-bit floating point for iteration $t$. This fits in $O(n \log n)$ space, approximating $n$ entries simultaneously, which resolves the issue of having to compute the solution vector $n$ times.

However, there is still an issue -- storing even a single entry of $\bb^{(i)}$ requires $\Omega(n)$ bits. This approach would again require $\Omega(n^2)$ space.

\paragraph{Our approach: a variant of $p$-adic lifting.}
To reduce the space usage, we propose a variant of $p$-adic lifting that does not act directly on the vector $\bb^{(i)}$. We observe that we do not need the exact vector $\bb^{(i)}$ in the $i$-th iteration of $p$-adic lifting -- it suffices to obtain $(\bb^{(i)} \bmod p)$ to compute $\yy^{(i)} \equiv \AA^{-1}\bb^{(i)} \pmod{p}$. But it is still not clear how to obtain $(\bb^{(i+1)} \bmod p)$ from $(\bb^{(i)} \bmod p)$ since the information of higher bits is lost.

We therefore design a low space representation for $(\bb^{(i)} \bmod p)$. In particular, when we expand \Cref{eq:padic-iteration} we get 
\begin{align} \label{eq:expand}
    \bb^{(i)} = \bb \det(\AA) \cdot p^{-i} - \sum_{j=0}^{i-1} \AA \yy^{(j)} \cdot p^{j-i}.
\end{align}
Here we can see that the vector $\bb^{(i)}$ consists of two terms: one from $\bb \cdot \det(\AA)$ and the other from $\yy^{(0)}, \dots, \yy^{(i-1)}$.
Since we know all the entries of $\bb^{(i)}$ in \Cref{eq:expand} are integers, the fractional parts of the two terms cancel each other out. Therefore, $(\bb^{(i)} \bmod p)$ can be obtained as long as we know the integral part
\begin{align} \label{eq:inte}
    \left\lfloor \sum_{j=0}^{i-1} \AA \yy^{(i)} \cdot p^{j-i} \right\rfloor.
\end{align}
However, again due to our space constraints, we cannot store $\yy^{(0)}, \dots, \yy^{(i-1)}$ from the previous iterations to compute \Cref{eq:inte}. Instead, it is possible to compute \Cref{eq:inte} iteratively. It can be shown that the maximum entry of this vector is bounded by $\poly(n)$, so storing it uses $O(n \log n)$ bits and thus fits within our memory. Therefore, we can implement $p$-adic lifting with this modification in linear space.

\paragraph{Handling large $\bb$.} As an additional advantage, our algorithm also applies when each entry of $\bb$ is bounded by $U^n$. In \Cref{eq:expand}, we compute $\bb^{(i)}$ \textbf{entry-wise}: for the $j$-th entry, evaluate
\[
    \lfloor \bb_j \cdot \det(\AA) \cdot p^{-i} \rfloor - (\Cref{eq:inte}\text{ term}).
\]
Since $\bb_j$ and $\det(\AA)$ both have a linear number of bits, computing one entry at a time requires only linear space as well. This resembles the bounds in \cite{S2005shifted}, but their approach uses fast matrix multiplication and therefore, by nature, requires quadratic space.

\paragraph{Reducing the space to $O(n \log (nU))$ bits. } When $\eps$ is small, our algorithm works with $O(\log(nU/\eps))$-bit floating point numbers, requiring a total of $O(n \log(nU/\eps))$ bits of space. Surprisingly, we can get rid of the dependency on $\eps$ completely, no matter how small $\eps$ is.

To reduce the space usage, we carefully track the space usage and observe that only $\yy^{(i)}$ requires $O(\log(nU/\eps))$ bits per entry---the other vectors require only $O(\log(nU))$ bits per entry. We therefore partition the coordinates into $\Theta(\min\{n, \log(1/\eps)\})$ blocks. Storing and computing one block of coordinates fits in $O(n \log(nU))$ space. However, the trade-off is an $O(\log(1/\eps))$ blow-up in asymptotic running time.

\subsection{Applications to Numerical Problems Using Linear Space}

For input matrix $\AA$, we denote $U$ as a bound on the maximum magnitude of the entries, and $n$ as the maximum dimension. We apply our linear-space linear system solver to several numerical problems, including linear regression, linear programming, eigenvalues/eigenvectors, and singular value decomposition. Some of the applications are direct (e.g., linear regression and linear programming), whereas others require more careful work.

\paragraph{Linear regression.}
We give a linear-space algorithm for solving linear regression by leveraging our linear systems solver. 
Given a full-rank matrix \( \AA \in \Z^{n \times d} \) and a vector \( \bb \in \Z^n \) both with integer entries in $[-U, U]$, the least-squares solution to  $\min_{\xx \in \R^d} \|\AA \xx - \bb\|_2$ is given by the normal equation  $(\AA^\tp \AA)^{-1} \AA^\tp \bb.$ A direct approach would require explicitly forming \( \AA^\tp \AA \) which has \( O(d^2) \) entries. Instead, we avoid storing \( \AA^\tp \AA \) by providing access to its matrix vector product 
$\AA^\tp \AA \vv$  for some vector $\vv$. Passing this matrix vector product access as well as $\AA^\tp \AA$ and $\AA^\tp \bb$ into our linear system solver gives an entry-wise multiplicative error approximation while using only   
$O(d \log(nU))$ bits of space and bit complexity 
$\Otil(d^3 \log(1/\eps) \cdot \nnz)$. 
 The space is near-linear in the smaller dimension $d$ since $\log n$ can be asymptotically larger than $\log d$.

\paragraph{Linear programming.} 

We plug our linear-space linear system solver into the Interior Point Method (IPM).
Given $\AA \in \Z^{n \times d}, \bb \in \Z^d, \cc \in \Z^n$ with $n \ge d$, we consider the linear program $\min_{\AA^\tp \xx = \bb, \xx \ge 0} \cc^\tp \xx$.
For linear programs with condition number bounded by $\poly(n)$, we obtain an algorithm with near-linear space, where the extra logarithmic factors depend on the inner and outer radius, $r$ and $R$ respectively, of the linear program. Specifically, we pass the matrix-vector access for $\AA^\tp \WW \AA$ for some diagonal matrix $\WW$ into our solver. We obtain a running time of $\Otil( d^3 \cdot \nnz \cdot n^{0.5})$ bit operations, and space of $O\left(n \log(\frac{n U R}{\eps r}) \right)$ bits. 

\paragraph{Inverse power method.} 
We employ our linear-space linear system solver to give a slightly stronger version of inverse power method, \callalg{InvPower}, for real symmetric matrices without condition number bounds and spectral gap assumptions. Our algorithm can detect the existence of an eigenvalue very close to $0$ using the entry-wise multiplicative approximation guarantee of \callalg{LinSolve} and floating-point representations. Specifically, let $|\lambda_{\min}|$ denote the minimum magnitude of the eigenvalues. With high probability, we either certify that $|\lambda_{\min}| \le 1/\poly(n)$, or output $\lamtil$ such that
\begin{align*}
    \left| \lamtil -  | \lambda_{\min}| \right| \le \eps | \lambda_{\min}|.
\end{align*}
Even though the convergence analysis of inverse power method is standard, we put more effort to incorporate the bit complexity and error analysis into it.

For the running time, the algorithm runs in $\Otil(n^2\cdot \nnz \cdot \eps^{-1})$ in general, and runs in $\Otil(n^2 \cdot \nnz \cdot \log(1/\eps))$ if there is a spectral gap between the smallest and the second smallest eigenvalue. This algorithm requires $O(n \log (nU / \eps))$ bits of space. This space requirement is slightly higher than that of our linear system solver as we must also store the output of the solver as intermediate vectors. 

\paragraph{Computing eigenvalues, eigenvectors, and SVD.}

We emphasize that the eigenpair or SVD problem produces $\Theta(n)$ vectors, and therefore has output size at least $\Omega(n^2)$. Therefore, we require our algorithm to use less \emph{working space}. The algorithm needs to output the desired eigenpairs (or SVD matrices) in some order, and never reads its own output. It may access that output only by memorizing it, and any such memorization counts toward the space usage. In-place modifications to the input matrices are not allowed.

We first perturb the input matrix $\AA$ by a random diagonal matrix based on Minami bound such that the eigenvalues are perturbed by at most $\eps/2$, with the property that all eigenvalues are bounded away from each other by some $\gamma < \eps$ with $\gamma \ge \poly(1/n, \eps)$. We can then test the existence of an eigenvalue in an interval by applying the inverse power method to shifted matrices. We finally recursively extract the eigenvalues using a divide-and-conquer strategy. This gives an algorithm with running time of $\Otil(n^3 \cdot \nnz )$ bit operations and space of $O(n \log (nU / \eps))$ bits.

An associated eigenvector can be computed given a high-accuracy approximation of the eigenvalue. As a corollary, we can compute the singular value decomposition (SVD) for any real matrix $\AA$ since $\AA^\tp \AA$ is real symmetric. Because a matrix–vector product access to $\AA^\tp \AA$ takes only $O(\nnz)$ time, our SVD solver has exactly the same asymptotic time and space complexity as full eigenvalue decomposition.

\section{Preliminaries} \label{sec:prelims}
\subsection{Notations}
Unless the base is explicitly stated, we use base $e$ for $\log$. We use $\Z$ to denote the integers, $\Q$ to denote the rationals, $\R$ to denote the reals, and $\F_p$ to denote the finite field of the integers $\{0,1, \ldots, p-1\}$ for a prime $p$. For the definition of $\fl_L$, see \Cref{def:FP}. 

 \paragraph{Sets and sampling.} For any integer $n \ge 1$, we use $[n]$ to denote the set $\{1, 2, \dots, n\}$.
 For a finite set $S$, we use $x \sim S$ to denote that random variable $x$ is uniformly sampled from $S$. When considering multiple variables, we write $x_1, \dots, x_n \sim S$ to indicate that each $x_i$ is independently and uniformly sampled. 

\paragraph{Modular arithmetic, floor and ceiling.} For $n \in \Z$ and $p > 0 $, we use $(n \bmod p)$ to denote the remainder $r$, such that $r \in [0, p)$ and $n-r$ is a multiple of $p$. We use the notation $p \mid n$ for $n \bmod p = 0$.

For $\alpha \in \R$, the floor function $\lfloor \alpha \rfloor$ returns the largest integer at most $\alpha$, and the ceiling function $\lceil \alpha \rceil$ returns the smallest integer at least $\alpha$. We also extend the notations $(\cdot \bmod p)$ and $\lfloor \cdot \rfloor$ to vectors and matrices, meaning that the operation is applied element-wise to each entry.

\paragraph{Vectors and matrices.}
Generally, we use bold uppercase symbols (e.g., $\AA, \DD$) to denote matrices, bold lowercase symbols (e.g., $\xx, \yy$) to denote vectors, and regular symbols (e.g., $c, \Delta$) for scalars. We use $\OO$ to denote the zero matrix, and $\mathbf{0}$ to denote the zero vector.

\paragraph{Multiplicative approximations.} For $x, y \in \R$, we use $x \approxbar_\eps y$ to denote the $e^\eps$-multiplicative approximation relation
\begin{align*}
    \sgn(x) = \sgn(y), \quad  e^{-\eps} |y|\le |x| \le e^{\eps} |y| 
\end{align*}  
where $\sgn(\cdot) \in \{-1, 0, 1\}$ denotes the sign function. Note that when $\eps$ is small, we have
\begin{align*}
    |x-y| \le O(\eps) |x| .
\end{align*}

\begin{fact}
    If $\eps<0.7$ and $|x-y| \le \eps |x|$, then $x \approxbar_{2 \eps} y$.  
\end{fact}

We extend this notation to vectors and matrices for entrywise relations. Specifically, for vectors $\uu, \vv \in \R^n$, $\uu \approxbar_\eps \vv$ if $\uu_i \approxbar_\eps \vv_i$ for all $i \in [n]$. 

\begin{fact} \label{multiplicative-approx-linearity}
    For $\eps, \lambda_1, \lambda_2, x_1, x_2, y_1, y_2 \in \R_{\ge 0}$, if $x_1 \approxbar_\eps y_1$ and $x_2 \approxbar_\eps y_2$, we have
    \begin{align*}
        \lambda_1 x_1 + \lambda_2 x_2 \approxbar_\eps \lambda_1 y_1 + \lambda_2 y_2.
    \end{align*}
\end{fact}
\begin{fact} \label{multiplicative-approx-multiplications}
    For $\eps_1, \eps_2 \in \R_{\ge 0}$, and $x_1, x_2, y_1, y_2 \in \R$, if $x_1 \approxbar_{\eps_1} y_1$ and $x_2 \approxbar_{\eps_2} y_2$, we have
    \begin{align*}
        x_1 \cdot x_2 \approxbar_{\eps_1 + \eps_2} y_1 \cdot y_2.
    \end{align*}
\end{fact}
\begin{fact} \label{multiplicative-approx-reciprocal}
    For $\eps \in \R_{\ge 0}$, and $x, y \in \R$, if $x \approxbar_\eps y$, we have
    \begin{align*}
        x^{-1} \approxbar_{\eps} y^{-1}.
    \end{align*}
\end{fact}

\begin{fact} \label{multiplicative-triangle-inequality}
    For $\eps_1, \eps_2 \in \R_{\ge 0}$, and $x, y, z \in \R$, if $x \approxbar_{\eps_1} y$ and $y \approxbar_{\eps_2} z$, we have
    \begin{align*}
        x \approxbar_{\eps_1 + \eps_2} z.
    \end{align*}
\end{fact}

\paragraph{Finite fields and polynomial rings.} For a prime $p$, we use $\F_p$ to denote the finite field of the integers $\{0, 1, \dots, p-1\}$ with operations of addition and multiplication modulo $p$. For a field $F$, we write $F[X]$ to denote the polynomial ring with coefficients in $F$ and indeterminate $X$.

\subsection{Computational Models}
In this paper, we work with the RAM model with a read-only input. The time complexity is measured in the number of bit operations, and the space complexity is measured in the number of bits in the working space, not including the read-only input space.

\paragraph{Space complexity.} In the problems that we study, an $n \times n$ matrix with integer entries in $[-U, U]$ is given in the input. For our main result, we will show that our algorithm has space complexity of $O(n \log(nU))$ bits. That is, it uses an additional space of $O(n \log(nU))$ bits in addition to the input matrix, which is read-only and remains unmodified during the computation. 

\paragraph{Time complexity.}  We bound the number of bit operations used by our algorithms, without tracking the number of logarithmic factors in the input/output size. In particular, when the input has $O(n^2)$ integers in $[-U, U]$ and we aim for a target accuracy $\eps$, the input size is $O(n^2 \log U)$ and the output size is $O(n \log(1/\eps))$. Therefore, we ignore factors of $\poly \log( n \log (U/\eps) )$, and use $\Otil(T_{n,U,\eps})$ to denote
\begin{align*}
    O(T_{n,U,\eps} \cdot \log^{O(1)} ( n \log (U/\eps) )).
\end{align*}

\paragraph{Matrix-vector access.} In our algorithms, we cannot afford to store an $n \times n$ matrix explicitly. For a matrix $\AA$, we only access it in a black-box way, assuming that we can compute the matrix-vector product $\AA \vv$ for any query vector $\vv$. Over a finite field $F$, we use $\TimeMatVec{\AA}$ to denote the number of field operations required to compute the matrix-vector access. If $\AA \in F^{n \times n}$ has $\nnz$ nonzero entries, we have $\TimeMatVec{\AA} = O(\nnz)$, assuming that $\nnz \ge n$.

\subsection{Fixed Points and Floating Points}

In our model, real numbers cannot be stored without errors. We store \textit{representations} of real numbers, approximating them by rational numbers. Specifically, for $x \in \R$, we denote $\rep{x} \in \Q$ as the representation of $x$.

\paragraph{Integers.} We can accurately represent any integer $z \in [-2^L, 2^L]$ by $L$ bits together with the sign. Under integer arithmetic, there is no error and no overflow, i.e., we always use a sufficient number of bits to represent the integers.
\begin{fact}
    For $U \ge 2$ and all integers $a, b \in [-U, U]$, we can compute
    \begin{align*}
        a+b,\  a-b,\  a \cdot b,\  \lfloor a/b \rfloor,\  \lfloor \log a \rfloor
    \end{align*}
    (which we assume to be defined) in
    \begin{align*}
        O(\log U \log^{O(1)} \log U) = \Otil(\log U)
    \end{align*}
    bit operations, using $O(\log U)$ bits of space.
\end{fact}

\paragraph{Fixed points.} Fixed-point representation extends the integer representation to approximate real numbers by interpreting a $2L$-bit integer $x$ as the real number $x / 2^L$. We obtain additive error guarantee for fixed points.

\begin{fact}
    For any $L > 0$ and any real number $x \in [-2^L, 2^L]$, there exists an $L$-bit fixed-point representation $\rep{x}$ such that $|x - \rep{x}| \le 2^{-L}$.
\end{fact}

\paragraph{Floating points.} We employ floating points to achieve multiplicative approximation guarantees. We formalize the definition as follows.
\begin{definition}[$L$-bit floating point] \label{def:FP}
    For $L > 0$ and integers $a, b \in [-2^L, 2^L]$, an $L$-bit floating-point representation for $x \in \R$ is defined to be $\rep{x} := (a, b)$ such that $x = a \cdot 2^{b}$.

    For $L > 0$, we define
    \begin{align} \label{eq:fl-L}
        \fl_L := [-2^{2^L}, -2^{-2^L}] \cup \{0\} \cup [2^{-2^L}, 2^{2^L}]
    \end{align}
    to be the set of real numbers that can be well approximated by $L$-bit floating points.
\end{definition}

\begin{fact} \label{lemma:floating-point-representation}
    For any $L > 0$ and $x \in \fl_L$, there exists an $L$-bit floating-point representation $\rep{x}$ such that $x \approxbar_{2^{-L}} \rep{x}$.
\end{fact}

 We use $\merr{x}$ to denote an upper bound on the multiplicative error of $x$, such that
$
x \approxbar_{\merr{x}} \rep{x}.
$

When there is no cancellation, i.e., adding two numbers with different signs, the multiplicative error under floating-point arithmetic remains small by \Cref{multiplicative-approx-linearity}.

\begin{lemma} [Addition of same-sign floating points] \label{lemma:floating-point-addition}
    For any $L>0$ and $x, y \in \fl_L$ given in $L$-bit floating points, if $z := x + y \in \fl_L$ and $x \cdot y \ge 0$, then we can compute $z$ with $\merr{z} \le \max\{\merr{x}, \merr{y}\} + 2^{-L}$ in 
    $O(L \log^{O(1)} L)$
    bit operations, using $O(L)$ bits of space.
\end{lemma}

\begin{lemma} [Multiplication of floating points] \label{lemma:floating-point-multiplication}
    For any $L>0$ and $x, y \in \fl_L$ given in $L$-bit floating points, if $z := x \cdot y \in \fl_L$, then we can compute $z$ with $\merr{z} \le \merr{x} +  \merr{y} + 2^{-L}$ in 
    $O(L \log^{O(1)} L)$
    bit operations, using $O(L)$ bits of space.
\end{lemma}

\begin{lemma} [Reciprocal of floating points] \label{lemma:floating-point-reciprocal}
    For any $L>0$ and $x \in \fl_L$ given in $L$-bit floating points, if $z := x^{-1}$ and $\merr{x} <1/2$, then we can compute $z$ with $\merr{z} \le \merr{x} + 2^{-L}$ in 
    $O(L \log^{O(1)} L)$
    bit operations, using $O(L)$ bits of space.
\end{lemma}

\subsection{Chinese Remainder Theorem (CRT)}
\label{pre:crt}

We state the well-known Chinese Remainder Theorem here. In this paper, we sample primes and compute the determinant modulo the primes separately. By CRT, we can integrate the results from each independent system and recover the exact determinant.

\begin{lemma}[Chinese Remainder Theorem, Theorem 10.25 of \cite{GG2013modern}] \label{lemma:CRT}
    Given distinct primes $p_1, p_2, \dots, p_\ell$, and integers $r_1, \dots, r_\ell$ where $r_i \in [0, p_i-1]$ for $i \in [\ell]$, let $P$ denote the product of the primes. For the following system 
    \begin{align*}
        \begin{cases}
            x \equiv r_1 \pmod {p_1}  \\
            x \equiv r_2 \pmod {p_2}  \\
            \quad \vdots \\
            x \equiv r_\ell \pmod {p_\ell}
        \end{cases}
    \end{align*}
    there exists at least one solution $\bar{x} \in \Z$, and any two solutions are congruent modulo $P$. 
    The remainder $R := (\bar{x} \bmod P)$ can be computed in $\Otil(\ell \log U)$ bit operations and space of $O(\ell \log U)$ bits, where $U := \max_{i=1}^\ell p_i$.
    Here, $\Otil$ hides factors $\log^{O(1)}( \ell \log U)$.
\end{lemma}

\subsection{Prime Sampling} \label{prelim:PS}

We employ the following prime sampling subroutine in our algorithm. For completeness, we include the proof in \Cref{sec:prime-sampling}.

\begin{restatable}{theorem}{SamplePrime}
    \label{thm:SamplePrime}
    For any constant $c>0$, there exists an algorithm \callalg{SamplePrime}\labelalg{SamplePrime} that, given an integer $n \ge 16$ and an integer $k \in [1, n]$, satisfies the following properties:
    \begin{itemize}
        \item The algorithm outputs $k$ integers, with failure probability $\le 1/n^c$. 
        \item Conditioned on the success of the algorithm, the output distribution follows the uniform distribution of $k$ distinct primes in the range $[n, n^2]$.
        \item The algorithm runs in $O(k \log^{O(1)} n)$ bit operations, and space of $O(k \log n)$ bits.
    \end{itemize}
\end{restatable}

When reducing linear system solving to finite fields, it is possible that a matrix $\MM$ over $\Q$ is invertible, but $(\MM \bmod p)$ becomes singular over the field $\F_p$ for a prime $p$. The following lemma shows that this issue can be avoided with high probability.

\begin{fact} \label{fact:number-of-prime}
Given integer $n > 1$, we have 
    \begin{align*}
       \pi(n^2) - \pi(n) \ge \frac{n^2 - n}{2 \log_2 n}
    \end{align*}
    where $\pi(n)$ denotes the number of primes that are at most $n$.
\end{fact}
\begin{proof}
    One can verify the inequality for $n \le 55$. For $n>55$, it follows from the fact
    \begin{align*}
        \frac{n}{\log n + 2} < \pi(n) < \frac{n}{\log n - 4}.
    \end{align*}
\end{proof}

\begin{lemma}  \label{lemma:prime-factor-prob}
For integers $n, m > 1$, if $p$ is a prime uniformly sampled from $[n, n^2]$, the probability that $p$ is a factor of $m$
    \begin{align*}
        \Pr_{p} [ p \mid m] \le \frac{2 \log_2 m}{n}.
    \end{align*}
\end{lemma}

\begin{proof}
    It follows from \Cref{fact:number-of-prime} that we have,
    \begin{align} \label{eq:number-of-primes}
        \pi(n^2) - \pi(n) \ge \frac{n^2 - n}{2 \log_2 n} \ge n/2.
    \end{align}
    Then, the lemma follows from the fact that $m$ has at most $\log_2 m$ distinct prime factors. 
\end{proof}

Finally, we will use the following upper bound for the determinant.

\begin{fact}[Hadamard's inequality] \label{hadamard-inequality}
    For an $n \times n$ matrix $\MM$, if $|\MM_{ij}| \le B$ for all $i,j \in [n]$, we have
    \begin{align*}
        |\det(\MM)| \le B^n \cdot n^{n/2}.
    \end{align*}
\end{fact}

\section{Finite Field Linear Algebra in Linear Space} \label{sec:ffla}
Here for the use of the community and in our algorithms we present the results from \cite{W1986solving} for solving linear systems over finite fields. In-depth discussions and the proofs are deferred to \Cref{sec:wiedemann}. 
\paragraph{Linear System Solving.}
Our algorithm builds on the Wiedemann algorithm, a linear system solver over finite fields in $O(n)$ space, first introduced by \cite{W1986solving}. To solve $\AA \xx = \bb$ over the field $\F_p$ for some prime $p$, the algorithm computes the minimal linear recurrence of the scalar array
\begin{align}
    [\xx^\tp \yy, \xx^\tp \AA \yy, \xx^\tp \AA^2 \yy, \dots, \xx^\tp \AA^{2n} \yy]
\end{align}
for independent uniform random vectors $\xx, \yy \sim \F_p^n$. It can then be used to compute the minimal polynomial of $\AA$, and solve the linear system.

\begin{restatable}{theorem}{minpoly} \label{thm:wiedemann}
    For prime $p > 1$, let $F = \mathbb{Z}_p$. There exists a randomized algorithm \callalg{Wiedemann} that, given matrix-vector access for a matrix $\AA \in F^{n \times n}$, outputs the minimal polynomial of $\AA$ using $O(\TimeMatVec{\AA} \cdot n)$ field operations, $O(n)$ space of elements in $F$, and achieves success probability at least $\frac{1}{36 \lceil \log_p(n)\rceil}$. 
    
    In addition, \callalg{Wiedemann} always outputs a factor of the minimal polynomial. When $p > n$, the success probability is constant, and can be boosted to $1-\delta$ by running the algorithm $O(\log(1/\delta))$ times and  choosing the polynomial with the largest degree. 
\end{restatable}
Note that $\TimeMatVec{\AA}$ denotes the time to compute a matrix-vector product using the matrix-vector access of $\AA$. If $\AA$ has $\nnz \ge n$ nonzero entries, \callalg{Wiedemann} runs in $O(n \cdot \nnz)$ operations in $F$. 

We state the guarantees for linear system solving over finite fields via Wiedemann as follows. 

\begin{restatable}{theorem}{LinSolveZp}
    \label{thm:LinSolveZp}
    There exists a randomized algorithm \callalg{LinSolveZp} such that, given matrix-vector access for an invertible matrix $\AA \in \F_p^{n \times n}$, a vector $\bb \in \F_p^{n}$, $\delta \in (0,1/2)$ and a prime $p$, it outputs the solution $\xx \in \F_p^{n}$ to the system
    \begin{align*}
        \AA \xx \equiv \bb \pmod p
    \end{align*}
    with probability at least $1 - \delta$. Furthermore, the algorithm runs in $O( \TimeMatVec{\AA}\cdot n\log(1/\delta) )$ field operations over $\F_p$, and $O(n)$ space of field elements in $\F_p$.
\end{restatable}

\paragraph{Computing Determinants.}

Our algorithm also relies on computing the determinant of a matrix over finite fields. The techniques are originally discussed in \cite{W1986solving} and \cite{KS1991wiedemann}.
We reformulate Wiedemann's algorithm for this application and employ the improved preconditioner in \cite{CEKSTV2002efficient}. The results are reproduced and formalized with an aim to bound the space complexity.

\begin{restatable}{theorem}{DeterminantZp}
    \label{thm:DeterminantZp}
    For $\delta \in (0,1/2)$, there exists a randomized algorithm \callalg{DeterminantZp} that, given matrix-vector access for an invertible matrix $\AA \in F^{n \times n}$ where $F = \F_p$ for prime $p \ge 6n^2$, outputs $(\det(\AA) \bmod p)$ with probability at least $1-\delta$ using $O( \TimeMatVec{\AA} \cdot n\log(1/\delta) )$ field operations over $F$ and $O(n)$ space of field elements in $F$.
\end{restatable}

\section{Linear System Solving over $\Q$} \label{sec:rationals}

In this section, we prove our main results on solving linear systems over $\Q$ in linear space. The input matrix $\AA$ and vector $\bb$ have integer entries, but the output is over the rationals. We note that $\AA$ and $\bb$ can be given with entries over the rationals in fixed point precision, and it is without loss of generality that they can instead be represented with integral entries by scaling the entries.

\begin{restatable}{theorem}{LinSolve} \label{thm:LinSolve}
    For any constant $c>0$, there exists a randomized algorithm \callalg{LinSolve} that, given $\eps \in (0, 1)$, a matrix $\AA \in \Z^{n \times n}$ with integer entries in $[-U, U]$ and a vector $\bb \in \Z^n$ with integer entries in $[-U^n, U^n]$, outputs either \texttt{SINGULAR} or an entry-wise $e^\eps$-multiplicative approximation $\xxhat$, such that the following  holds:
    \begin{itemize}
        \item If $\AA$ is singular,  with probability at least $1 - n^{-c}$, it outputs \texttt{SINGULAR}.
        \item If $\AA$ is invertible, with probability at least $1 - n^{-c}$,  it outputs $\xxhat$ with each entry represented by an $O(\log(nU/\eps))$-bit floating point, such that $\xxhat \approxbar_\eps \AA^{-1} \bb$.
        \item The algorithm uses 
        $O(n \log (nU))$ bits of space and runs in $\Otil(n^2\log(1/\eps) \cdot  \nnz )$ bit operations, where $\nnz \ge n$ denotes the number of nonzero entries of $\AA$, and $\Otil$ hides logarithmic factors $\log^{O(1)}( nU \log (1/\eps))$.
    \end{itemize}
\end{restatable}

We note that our algorithm uses the input matrix as a black box. We present the guarantees achieved when the algorithm is given oracle access to the matrix-vector product in the following corollary. 

\begin{corollary}[\Cref{thm:LinSolve} as a black-box algorithm] \label{cor:LinSolve-black-box}
    For any constant $c>0$, there exists a randomized algorithm \callalg{LinSolve} that, given $\eps \in (0, 1)$, a vector $\bb \in \Z^n$ with integer entries in $[-U^n, U^n]$, and oracle access to the following function 
    \begin{align*}
        f_{\AA}(\vv) = \AA \cdot \vv
    \end{align*}
    for any vector $\vv \in \Z^n$ with entries bounded by $n^6U^2$, outputs either \texttt{SINGULAR} or an entry-wise $e^\eps$-multiplicative approximation $\xxhat$, such that the following properties hold:
    \begin{itemize}
        \item If $\AA$ is singular,  with probability at least $1 - n^{-c}$, it outputs \texttt{SINGULAR}.
        \item If $\AA$ is invertible, with probability at least $1 - n^{-c}$,  it outputs $\xxhat$ with each entry represented by an $O(\log(nU/\eps))$-bit floating point, such that $\xxhat \approxbar_\eps \AA^{-1} \bb$.
        \item The algorithm queries the oracle $O(n^2 \log(1/\eps))$ times, uses 
        $O(n \log (nU))$ bits of space, and runs in $\Otil(n^3 \log(1/\eps) )$ bit operations where $\Otil$ hides logarithmic factors $\log^{O(1)}( nU \log (1/\eps))$.
    \end{itemize}
\end{corollary}

\subsection{Computing the Determinant} \label{sub:det}
First, we show that the determinant of an integer matrix can be computed exactly in linear space. We present the algorithm \callalg{Determinant} based on CRT\footnote{See \Cref{pre:crt}.} and the finite field algorithm \callalg{DeterminantZp}. The space requirement is $O(n \log (nU))$ bits, corresponding to the bit length of the largest possible determinant of an $n \times n$ matrix with entries bounded by $U$.

\begin{restatable}{theorem}{Determinant} \label{thm:Determinant}
   For any constant $c>0$, there exists a randomized algorithm \callalg{Determinant} that, given a matrix $\AA \in \Z^{n \times n}$ with integer entries in $[-U, U]$ and $\nnz$ nonzero entries, outputs $\det(\AA)$, the determinant of $\AA$, with probability at least $1 - n^{-c}$, and runs in $\Otil(n^2 (\nnz+n))$ bit operations and space of $O(n \log (nU))$ bits. Here, $\Otil$ hides logarithmic factors $\log^{O(1)}(nU)$.
\end{restatable}
\begin{algorithm}[t]
    \caption{ \textsc{Determinant}}
    \labelalg{Determinant}
    
    \SetKwInOut{Input}{Input}
    \SetKwInOut{Output}{Output}
            
    \Input{A matrix $\AA \in \Z^{n \times n}$ with integer entries in $[-U, U]$ }
    \Output{The determinant $\det(\AA)$}
    
    $T \gets n$ \\
    Let $p_1, p_2, \dots, p_T \gets$ \callalg{SamplePrime}$(T, \max\{16, n^2U \})$. 
    
    \For{$i=1$ \KwTo $T$} {
            $r_{i} \gets \callalg{DeterminantZp}(\AA \bmod p_i)$ 
         \\
    }
    $(P, R) \gets \textsc{CRT}((p_1,r_1), \dots, (p_T, r_T))$ \quad \tcp{$P$ is the product of the primes.}
    \Return $R$ if $R < P/2$, otherwise return $R-P$\\ 
    
\end{algorithm}

\begin{proof}
    We first prove correctness. In particular, we prove that CRT recovers the determinant, $\det(\AA)$, exactly. By \Cref{hadamard-inequality}, since we have that the entries of $\AA$ are bounded by $U$, we have $|\det(\AA)| \le U^n \cdot n^{n/2}$
    and thus $P = \prod_{i=1}^{T} p_i > (n^2U)^n >  2 |\det(A)|$.
    
    By the correctness of CRT (\Cref{pre:crt}), if $r_i = \det(\AA) \bmod p_i$, we have $R = \det(\AA) \bmod P$.
    If $\det(\AA) < 0$, we return $R - P$ because $R > P/2$, and otherwise we return $R$. Therefore, the algorithm will return the exact $\det(\AA)$.

    To ensure the overall failure probability is at most $n^{-c}$, we bound the failure probability of \callalg{SamplePrime} (\Cref{thm:SamplePrime}) and each call of \callalg{DeterminantZp} (\Cref{thm:DeterminantZp}) by $n^{-c-2}$. Then, a union bound over $T+1 \le n^2$ calls gives the desired bound.

    Now we prove the space and running time bounds.
    \begin{itemize}
        \item By \Cref{thm:SamplePrime}, \callalg{SamplePrime} runs in $O(T \log^{O(1)} (n^2U)) = \Otil(n)$ bit operations, and using space of $T \log (n^2U) = O(n \log (nU))$ bits.
        
        \item By \Cref{thm:DeterminantZp}, each call of \callalg{DeterminantZp} with $\delta$ set to $n^{-c}$ runs in 
        \begin{align*}
            O(n (\TimeMatVec{\AA \bmod p} + n) \log n) \cdot O(\log p_i \cdot \log^{O(1)} \log p_i) = \Otil(n (\nnz + n))
        \end{align*}
        bit operations (since each field operation in $\F_{p_i}$ takes $O(\log p_i \cdot \log^{O(1)} \log p_i)$ bit operations), and requires space of $O(n \log p_i) = O(n \log (nU))$ bits.    

        \item By \Cref{lemma:CRT}, the CRT step runs in $\Otil(T \log(n^2U)) = \Otil(n \log n)$ bit operations, and requires space of $O(T \log(n^2U)) = O(n \log (nU))$ bits.
    \end{itemize}
    To conclude, for the running time, the dominant step is $T$ calls of \callalg{DeterminantZp}, giving us
    \begin{align*}
        T \cdot \Otil(n (\nnz + n)) = \Otil(n^2 (\nnz + n))
    \end{align*}
    bit operations.
    For the space complexity, the primary cost is storing $r_1, \dots, r_T$, which requires $O(n \log (nU))$ bits.
\end{proof}

\paragraph{Remarks. } 
We need the primes to be at least $\Omega(n^2)$ to satisfy the requirement of large fields in \callalg{DeterminantZp}.
We also note that \callalg{DeterminantZp} does not need the assumption that $\AA$ is invertible, so the algorithm works correctly even when $p_i$ is a factor of $\det(\AA)$.

\subsection{Linear System Solving}
In this section, we present our algorithm \callalg{LinSolve} and prove our main result, \Cref{thm:LinSolve}. Our algorithm first computes the determinant using finite field methods with the algorithm in \Cref{sub:det}. Then it recovers the integer solution vector $\det(\AA) \cdot \AA^{-1} \bb$ via $p$-adic lifting. 

\begin{algorithm}[t]
    \caption{ \textsc{LinSolve} }
    \labelalg{LinSolve}
    
    \SetKwInOut{Input}{Input}
    \SetKwInOut{Output}{Output}
            
    \Input{A matrix $\AA \in \Z^{n \times n}$ with entries in $[-U, U]$, a vector $\bb \in \Q^n$ with entries in $[-U^n, U^n]$, 
    the target multiplicative accuracy $\eps \in (0,1)$.} \label{line:inputs} 
    \Output{ Either \texttt{SINGULAR}, or $\xxhat \in \Q^n$ with each entry represented by an $O(\log(nU/\eps))$-bit floating point such that the entry-wise $e^\eps$-multiplicative approximation condition $\xxhat \approxbar_\eps  \xx$ holds. } 
    
    $T \gets n $  \\
     
    $\Delta \gets $ \callalg{Determinant}$(\AA)$  \label{line:determinant}  \\ 
    \If{$\Delta = 0$}{
        \Return \texttt{SINGULAR}
    }
    \label{line:sample-p} 
        $p \gets $ \callalg{SamplePrime} $(1, \max\{16, n^3U\})$ \quad \tcp{Sample one prime $\ge n^3U$.} 
    Initialize integer vector $\rrtil^{(0)} \gets \mathbf{0} \in \Z^n$ \\ 
    
    Initialize $\yy_+, \yy_- \gets \mathbf{0} \in \Q^n$, with each entry represented by an $L$-bit floating point, where $L := 2\log_2(nU/\eps)$. \\
    \For{$i = 0$ \KwTo $T-1$}   { \label{line:fp-p3-loop}
        entry-wise compute $\bbtil^{(i)} \gets \lfloor (\bb \cdot \Delta)/p^i \rfloor \bmod p $ \label{line:computeb}\\ 
         \label{line:solve-xxtil}
         $\yytil^{(i)} \gets \callalg{LinSolveZp}(\AA, (\bbtil^{(i)} - \rrtil^{(i)}) \bmod p) $ \quad \tcp{each entry is in $[0, p)$} 
        Under $L$-bit floating arithmetic: $\yy_+ \gets \yy_+ + \yytil^{(i)} \cdot p^i$    \label{line:update-yplus}\\
        Under $L$-bit floating arithmetic: $\yy_- \gets \yy_- + (p - 1 - \yytil^{(i)}) \cdot p^i$  \label{line:update-yminus} \\
        Under integer arithmetic: $\rrtil^{(i+1)} \gets \lfloor (\rrtil^{(i)} + \AA \yytil^{(i)}) / p \rfloor $ \label{line:update-rrtil}
    }
    For each $i \in [n]$, let $\yy[i] \gets \begin{cases} \yy_{+}[i] & \yy_+[i] > \yy_-[i] \\ -\yy_{-}[i]-1 & Otherwise\end{cases}$  \label{line:combine-y}
    \\
    \Return $\yy / \Delta$
    \label{line:output}

\end{algorithm}

\paragraph{Correctness of \callalg{LinSolve}.} 

 In the following we assume that all the subroutines do not fail, and we will address their failure probability at the end. The guarantee for the singular case is easy. By the correctness of \callalg{Determinant}$(\AA)$ (\Cref{thm:Determinant}), we have $\Delta = \det(\AA)$ exactly and the algorithm  outputs \texttt{SINGULAR} if $\Delta = 0$. Now we prove the case where $\AA$ is invertible.

Let $\xxbar := \AA^{-1}\bb$. By Cramer's rule, 
\begin{align*}
    \xxbar_i \cdot \det(\AA) = \frac{\det\big(\text{replace}(\AA, i, \bb)\big)}{\det(\AA)} \cdot \det(\AA) =  \det(\text{replace}(\AA, i, \bb))
\end{align*}
where $\text{replace}(\AA, i, \bb)$ is the matrix obtained by replacing the $i$-th column of $\AA$ by $\bb$. Hence, $\xxbar_i \cdot \det(\AA)$ is an integer for all $i \in [n]$. Moreover, by \Cref{hadamard-inequality}, $|\xxbar_i \cdot \det(\AA)|$ is bounded by $U^n \cdot n^{n/2}$. We define
$\yybar \in \Z^n$ as the following: 
\begin{align*}
    \yybar \equiv \xxbar \cdot \det(\AA) \pmod{p^T}
\end{align*}

Then, we can recover $\xxbar \cdot \det(\AA)$ from $\yybar$ exactly since
\begin{align} \label{eq:pT-larger-than-xxbar-det}
    p^T \ge (n^3U)^{n} \ge (nU)^n > 2 \max_i |(\xxbar \cdot \det(\AA))_i|.
\end{align}
We decompose $\yybar$ with its $p$-adic expansion,
\begin{align}
    \yybar \equiv \sum_{i=0}^{T-1} \yybar^{(i)} \cdot p^i \pmod{p^T}
\end{align}
where $\yybar^{(i)}$ has entries in $[0, p)$. For $i \in [0,T]$, define $\rr^{(i)} \in \Q^n$ such that
\begin{align*}
    \rr^{(i)} \triangleq \sum_{j=0}^{i-1} \frac{\AA \yybar^{(j)}}{p^{i-j}}.
\end{align*}
Note that $\rr^{0} = 0$.
We will show that $ \rrtil^{(i)} = \lfloor \rr^{(i)} \rfloor $ later.
We prove the following guarantees, given that every subroutine executes successfully.

\begin{lemma}  \label{lemma:p-is-not-factor-of-A}
    In \Cref{line:sample-p}, it holds that
    \begin{align*}
        \Pr_{p} [(\AA \bmod p)\text{ is invertible}] \ge 1 - \frac{1}{3n}.
    \end{align*}
\end{lemma}

\begin{proof}
    By \Cref{lemma:prime-factor-prob} and \Cref{hadamard-inequality}, 
    \begin{align*}
        \Pr_{p} [(\AA \bmod p)\text{ is singular}]
        = 
        \Pr_{p} [p \mid \det(\AA)]
        &\le 
        2 \log_2(\det(\AA)) / (n^3U)
        \\ & \le
        4 n \log(nU) / (n^3U)
        \le 
        1/3n. 
    \end{align*}
\end{proof}

\begin{lemma}\label{lemma:padic-mod-correct}
    If $\AA \bmod p$ is invertible, then $\rrtil^{(i)} = \lfloor \rr^{(i)} \rfloor$ and $ \yytil^{(i)} = \yybar^{(i)}$ for all $i \in [0, T-1]$, and therefore 
    \begin{align*}
        \yybar \equiv \sum_{i=0}^{T-1} \yytil^{(i)} p^i \pmod {p^T}.
    \end{align*}
\end{lemma}

\begin{proof}
    We prove by induction on $k \in [0, T]$ that
    \begin{itemize}
        \item For $i \in [0, k]$, it holds that $\rrtil^{(i)} = \lfloor \rr^{(i)} \rfloor$.
        \item For $i \in [0, k-1]$, it holds that $ \yytil^{(i)} = \yybar^{(i)}$.
    \end{itemize}
    For the base case $k = 0$, we trivially have that $\rr^{(0)} = \mathbf{0} = \lfloor \rr^{(0)} \rfloor = \rrtil^{(0)}$. 

    Now assume that the induction hypothesis holds for $k$, and we want to prove for $k+1$. It suffices to show $\yytil^{(k)} = \yybar^{(k)}$ and $\rrtil^{(k+1)} = \lfloor \rr^{(k+1)} \rfloor$. 
    
    We express $\yybar^{(k)}$ in terms of $\yybar^{(0)}, \dots, \yybar^{(k-1)}$,
    \begin{align*}
        \yybar^{(k)} \equiv \left( \frac{\yybar - \sum_{j=0}^{k-1} \yybar^{(j)} \cdot p^j }{ p^k} \right) \pmod p.
    \end{align*}
    where the term in the parenthesis must be an integer. Left multiplying by $\AA$ to both sides we have 
    \begin{align*}
        \AA \yybar^{(k)}
        & \equiv
        \AA \left( \frac{\yybar - \sum_{j=0}^{k-1} \yybar^{(j)} \cdot p^j }{ p^k} \right)  \pmod{p}
        \\ & \equiv 
        \left( \frac{\AA \yybar}{p^k} - \sum_{j=0}^{k-1} \frac{\AA \yybar^{(j)}}{p^{k-j}} \right)  \pmod{p}
        \\ & \equiv 
        \left( \frac{\det(\AA) \cdot \bb }{p^k} - \rr^{(k)} \right)  \pmod{p}. \tag{by definition of $\bar{\yy}$ and $\rr^{(k)}$}
    \end{align*}
   Therefore, we must have
    \begin{align*}
        \AA \yybar^{(k)}
        & \equiv
        \left\lfloor \frac{\det(\AA) \cdot \bb }{p^k} \right\rfloor - \lfloor \rr^{(k)} \rfloor \pmod{p}.
   \end{align*}
   since the term in the parenthesis is an integer.
   By the guarantee of \callalg{LinSolveZp} in \Cref{line:solve-xxtil}, we have
   \begin{align*}
       \AA \yytil^{(k)} \equiv \bbtil^{(k)} - \rrtil^{(k)} \equiv \left\lfloor \frac{\det(\AA) \cdot \bb }{p^k} \right\rfloor - \lfloor \rr^{(k)} \rfloor \pmod{p}
   \end{align*}
   by the definition of $\bbtil^{(k)}$ and the induction hypothesis, so we conclude $\yytil^{(k)} = \yybar^{(k)}$ since $(\AA \bmod p)$ is invertible and both have entries in range $[0, p-1]$.

   For the second part, by \Cref{line:update-rrtil},
   \begin{align*}
        \rrtil^{(k+1)} & = \lfloor (\rrtil^{(k)} + \AA \yytil^{(k)}) / p \rfloor
        \\ & =
        \lfloor ( \lfloor \rr^{(k)} \rfloor  + \AA \yybar^{(k)}) / p \rfloor
        \\ & =
         \left\lfloor 
            \frac{1}{p} \left( \left\lfloor
                \sum_{j=0}^{k-1} \frac{\AA \yybar^{(j)}}{p^{k-j}} \right\rfloor
                +
                \AA \yybar^{(k)}
            \right)
        \right\rfloor  
        \\ & =
         \left\lfloor 
            \frac{1}{p} \left\lfloor
                \sum_{j=0}^{k-1} \frac{\AA \yybar^{(j)}}{p^{k-j}} 
                +
                \AA \yybar^{(k)}
            \right\rfloor
        \right\rfloor
        \\ & =
         \left\lfloor 
            \frac{1}{p} \left(
                \sum_{j=0}^{k-1} \frac{\AA \yybar^{(j)}}{p^{k-j}} 
                +
                \AA \yybar^{(k)}
            \right)
        \right\rfloor
        =
        \lfloor \rr^{(k+1)} \rfloor,
    \end{align*}
    so we conclude the induction proof.
\end{proof}

The two above lemmas \Cref{lemma:p-is-not-factor-of-A} and \Cref{lemma:padic-mod-correct} show that, with high probability, we can completely recover $\yybar$. The solution $\AA^{-1} \bb$ can be obtained exactly from dividing $\yybar$ by $\Delta$. However, we are not given enough space to store the full precision. It remains to be shown that an $e^\eps$-multiplicative approximation of the solution is maintained by $\yy_+, \yy_-$ with floating-point numbers throughout the iteration, so that only $\yytil^{(i)}$ is stored in iteration $i$ and all the previous vectors are discarded.

\begin{lemma} \label{lemma:y-plus-minus-good}
    When \callalg{LinSolve} reaches the end of the loop, the following properties hold. 
    \begin{itemize}
        \item If $\xxbar_i \cdot \det(\AA) > 0$, then $\yy_{+}[i] \approxbar_\eps \xxbar_i \cdot \det(\AA)$.
        \item If $\xxbar_i \cdot \det(\AA) \le 0$, then $\yy_{-}[i] \approxbar_\eps -\xxbar_i \cdot \det(\AA)-1$.
    \end{itemize}
    Furthermore, in \Cref{line:combine-y}, we have $\yy \approxbar_\eps \xxbar \cdot \det(\AA)$.
\end{lemma}

\begin{proof}
    Fix an $i \in [n]$. Recall that by \eqref{eq:pT-larger-than-xxbar-det}, $p^T > 2 \xxbar_i \cdot \det(\AA)$. By the definition of $\yybar$, we have $\yybar_i \equiv \xxbar_i \cdot \det(\AA) \pmod{p^T}$.
    Therefore,
    \begin{align}
        \yybar_i =
        \begin{cases}
        \xxbar_i \cdot \det(\AA)  & \xxbar_i \cdot \det(\AA) >0,\\
        p^T +  \xxbar_i \cdot \det(\AA)  & \xxbar_i \cdot \det(\AA) \le 0.\\
        \end{cases}
    \end{align}
    It follows from \Cref{lemma:padic-mod-correct} that,
    \begin{align*}
        \sum_{j=0}^{T-1} \yytil^{(j)} p^j = \yybar.
    \end{align*}
    If $\det(\AA) \cdot \bar{\xx}_i > 0$, then we have
    \begin{align*}
        \yy_+[i] \approxbar_\eps \left(\sum_{j=0}^{T-1} \yytil^{(j)} p^j\right)_i
        =
        \yybar_i
        =
        \det(\AA) \cdot \xx_i,
    \end{align*}
    and otherwise
    \begin{align*}
        \yy_-[i] \approxbar_\eps
        \left(\sum_{j=0}^{T-1} (p - 1 - \yytil^{(j)}) p^j\right)_i 
        =
        p^T-1-\yybar_i
        =
        -\det(\AA) \xx_i-1,
    \end{align*}
    where the guarantees of $e^\eps$-multiplicative approximation for $\yy_+$ and $\yy_-$ follow from \Cref{lemma:floating-point-addition} and \Cref{lemma:floating-point-multiplication}, because we use $L$-bit floating-point arithmetic. Recall that $L := 2 \log_2(nU/ \eps)$. 
    
    Finally, it follows from \Cref{line:combine-y} that $\yy \approxbar_\eps \det(\AA) \xx_i$ directly.
\end{proof}

So far, the correctness is almost done -- we have proven that the algorithm outputs $\yy / \Delta$, $e^\eps$-multiplicatively approximating $\xxbar = \AA^{-1} \bb$ by the above lemma \Cref{lemma:y-plus-minus-good}. It remains to bound the success probability of the subroutines.

\paragraph{Success probability of \callalg{LinSolve}. }
We note that it suffices to lower bound the success probability by $1 - 1/n$. All the success probabilities for the subroutines below and the lemmas above can be boosted to $1-n^{-c}$ for any constant $c > 0$ while only increasing by a constant factor in the time and space. 

The algorithm calls the following subroutines:
\begin{itemize}
    \item \callalg{Determinant}: Fails with probability at most $n^{-2}$ (with choice $c = 2$ in \Cref{thm:Determinant}).
    \item \callalg{SamplePrime}: Fails with probability at most $n^{-2}$ (with choice $c = 2$ in \Cref{thm:SamplePrime}).
    \item \callalg{LinSolveZp}: Runs for $T$ times, each with failure probability at most $n^{-3}$ (with choice of $\delta = n^{-3}$ in \Cref{thm:LinSolveZp}).
\end{itemize}
A union bound gives failure probability at most $1/n$. We conclude the correctness of \callalg{LinSolve}, finishing the proof of the first two properties in \Cref{thm:LinSolve}.

\paragraph{Time and space complexity of \callalg{LinSolve}. } 
Below we first show that $\callalg{LinSolve}$ has space complexity of $O(n \log (nU/\eps))$ bits. Then, a small modification to the algorithm reduces the space to $O(n \log (nU))$ bits by running $O(\log(1/\eps))$ instances and computing a smaller set of entries of the solution vector.  

We measure time in the number of bit operations, and space in the number of bits of storage. Let $P := \max\{16, n^3U\}^2$ be an upper bound of $p$. The algorithm involves the following variables:
\begin{itemize}
    \item $T, p, i:$ variables requiring an insignificant amount of space, i.e., $o(n)$ bits.
    \item $\Delta:$ an integer for the determinant, requiring $O(n \log (nU))$ bits by \Cref{hadamard-inequality}.
    \item $\yy_+, \yy_-$: nonnegative $n$-dimensional floating-point vectors, with each entry represented by $L$ bits. They require a total of $O(n \log (nU/\eps))$ bits.
    \item $\bbtil^{(i)}, \rrtil^{(i)}:$ $n$-dimensional integer vectors. The copies of them in previous iterations are discarded. We will bound the entries by $U \cdot P$, so they require a total of $O(n \log (nU))$ bits. 
\end{itemize}
Note that by definition, the entries of $\bbtil^{(i)}$ are bounded by $p \le P$ for all $i$. It remains to bound $\rrtil^{(i)}$.  
\begin{lemma}
    It holds that $\|\rrtil^{(i)}\|_\infty \le U \cdot P$ for all $i \in [0, T]$.
\end{lemma}
\begin{proof}
    For all $k \in [n]$, by \Cref{lemma:padic-mod-correct}, 
    \begin{align*}
        \rrtil^{(i)}_k  = \lfloor \rr^{(i)}_k \rfloor \le \sum_{j=0}^{i-1} \frac{(\AA \yybar^{(j)})_k}{p^{i-j}} \le
        \sum_{j=0}^{i-1} \frac{nUp}{p^{i-j}} \le
        nU \cdot \sum_{j \ge 0} p^{-j} \le
        2nU.
    \end{align*}
\end{proof}

The algorithm involves the following subroutines, where the last one is dominating:
\begin{itemize}
    \item \callalg{Determinant}: By \Cref{thm:Determinant}, the time is $\Otil(n^2(\nnz+n))$ bit operations, and the space is $O(n \log (nU))$ bits.
    \item \callalg{SamplePrime}: By \Cref{thm:SamplePrime}, the time is $O(\log^{O(1)} n)$ bit operations, and the space is $O(\log (nU))$ bits. 
    \item \callalg{LinSolveZp}: By \Cref{thm:LinSolveZp}, setting $\delta = n^{-3}$, each call runs in $O(n (\TimeMatVec{\AA \bmod p}+n) \log{n} )$ field operations, and $O(n)$ space of field elements in $\F_p$, where each field element is $O(\log P) = O(\log (nU))$ bits.
    Since $\TimeMatVec{\AA \bmod p} = O(\nnz + n)$, the total time (over $T$ runs) is thus
    \begin{align} \label{eq:LinSolve-running-time}
        \Otil(T \cdot n (\nnz + n) \log P) =  \Otil(n^2 (\nnz + n))
    \end{align}
    bit operations. The space is $O(n\log(nU))$ bits. 
\end{itemize}
For the other steps in \callalg{LinSolve}, it is easy to verify they can all be done efficiently in time and space. Specifically, the dominating step is \Cref{line:computeb}, where computing an entry of $\bb \cdot \Delta$ uses $O(n \log (nU))$ bits of space with time complexity $\Otil(n^2)$. Over $T$ iterations, this step runs in $\Otil(n^3 \log(1/\eps))$ bit operations. 

\paragraph{Reducing the space complexity to $O(n \log (nU))$.} Observe that in \callalg{LinSolve}, only the vectors $\yy_+, \yy_-$ require $O(n \log (nU/\eps))$ bits to store. The other vectors in the loop, $\yytil, \bbtil, \rrtil$, are computed without $\yy_+, \yy_-$ as inputs, and therefore require only $O(n \log(nU))$ bits.

To reduce the space usage, we use the following two facts: 
\begin{enumerate}
    \item $\log(1/\eps)$ need not exceed $O(n\log(nU))$, since any higher precision would effectively give exact values of $\yy_+$ and $\yy_-$ and therefore give the exact solution.
    \item \Cref{line:update-yplus}, \Cref{line:update-yminus}, and \Cref{line:combine-y} only involve entry-wise computation, so the coordinates of $\yy_+, \yy_-$, and $\yy$ are updated independently of each other.
\end{enumerate}
We run \callalg{LinSolve}
\[
K := \min\{n, \max\{1, \Theta(\log(1/\eps)) \} \}
\]
times, each time storing only $\lceil n/K\rceil $ entries of $\yy_+, \yy_-$, and computing the corresponding entries of $\yy$ and the solution vector $\yy / \Delta$. Therefore, the modified algorithm uses space
\begin{align*}
    O\left(n \log(nU) + \lceil n / K \rceil \log(nU/\eps) \right)
    & =
    O\left(n \log(nU) + \lceil n / K \rceil \log(nU) + \lceil n / K \rceil\log(1/\eps) \right)
    \\ & =
    O\left(n \log (nU) + n \log(nU) + \max\{n, n \log(nU)\} \right).
    \\ & =
    O(n \log (nU)).
\end{align*}

Now we analyze the time and space complexity of the modified algorithm. The loop in \Cref{line:fp-p3-loop} is repeated $K$ times.
Thus, the dominant subroutine $\callalg{LinSolveZp}$ and the dominant step \Cref{line:computeb} are also repeated $K$ times. 
For computing $\yy_+, \yy_-$, each entry requires $\Otil(T \cdot \log(1/\eps)) = \Otil(n \log(1/\eps))$ bit operations, so it requires a total of  $\Otil(n^2 \log(1/\eps))$ bit operations, dominated by the other steps.
 We conclude the final bit complexity
\begin{align*}
    \Otil(n^2 (\nnz + n) + n^3) \cdot K   = \Otil(n^2\log(1/\eps) \cdot \nnz),
\end{align*}
 and the space complexity of $O(n \log (nU))$ bits, completing the proof of the third property of \Cref{thm:LinSolve}.

\paragraph{Proof of \Cref{cor:LinSolve-black-box}, black-box access for $\AA$.} $\AA$ is used as a black box in \callalg{Determinant} and \callalg{LinSolveZp}. Every time we access $\AA$ we query the matrix-vector product $\AA \cdot \vv$.
In \Cref{line:solve-xxtil}, the matrix-vector product is under the finite field $\Z_p$, so the entries of right-hand vector $\vv$ are bounded by $p$. In \Cref{line:update-rrtil}, the entries of $\yytil^{(i)}$ are also bounded by $p$. Therefore, it suffices to query the function $\ff_\AA(\vv) = \AA \cdot \vv$ for $\vv$ with entries bounded by $p \le n^6 U^2$.

There are $O(n^2)$ queries in \callalg{Determinant}, and $O(n)$ queries in each \callalg{LinSolveZp} with a total of $T \log(1/\eps)$ calls. We conclude that in total there are $O(n^2 \log(1/\eps))$ queries, and the time and space complexity of the remaining part of our algorithm follow from the analysis above. Note that the computation of matrix-vector product does not count toward our running time here. The time complexity of the remaining part is dominated by \Cref{line:computeb}, requiring $\Otil(n^3)$ bit operations.

\paragraph{Remarks.} Assuming that $\log(1/\eps) \le O(n \log(nU))$, our linear system solver \callalg{LinSolve} has a near-linear time-space trade-off between the space bounds $O(n \log(nU))$ and $O(n \log(nU/\eps))$. Specifically, for any working-space budget $S \in [O(n \log(nU), O(n \log (nU/\eps))]$, the running time scales as $T = \Otil(n^3 \log(1/\eps) \cdot \nnz / S)$, where $\log^{O(1)}(nU\log(1/\eps))$ factors are suppressed.

\section{Applications} \label{sec:app}

\subsection{Linear Regression}
As a direct application to our linear system solver, we show that we can solve linear regression 
\begin{align*}
    \min_{\xx \in \R^d}  \|\AA \xx - \bb\|_2
\end{align*}
for $\AA \in \Z^{n \times d}$ and $\bb \in \Z^{n}$ in near-linear space in $d$. We plug in our solver to compute an entry-wise multiplicative approximation of 
\begin{align*}
    \arg \min_{\xx \in \R^d}  \|\AA \xx - \bb\|_2 = (\AA^\tp \AA)^{-1} \AA^\tp \bb. \tag{the normal equation}
\end{align*}
We first compute and store $\vv := \AA^{\tp}\bb$ in $\Otil(d)$ space. Then, we call our linear-space linear system solver $\callalg{LinSolve}(\AA^\tp \AA, \vv)$ without computing the matrix multiplication $\AA^\tp \AA$. Instead, we employ matrix-vector access for $\AA^\tp \AA$ -- given any $\xx \in \Z^d$, we can compute $\AA^\tp \AA \xx$ in space $\Otil(d)$ by enumerating $i \in [d], j \in [n], k \in [d]$ and summing over $(\AA^\tp)_{ij} \AA_{jk} \xx_k$.

\begin{restatable}{corollary}{linreg} \label{cor:linear-regression}
     For any constant $c>0$ and $n \ge d$, there exists an algorithm that, given $\eps>0$, a full-rank matrix $\AA \in \Z^{n \times d}$ 
    and a vector $\bb \in \Z^{d}$ both with integer entries in $[-U, U]$, with probability at least $1 - d^{-c}$, computes $\xxtil$ given in $O(\log(nU / \eps))$-bit floating points such that 
    \begin{align*}
        \xxtil \approxbar_{\eps} \arg \min_{\xx \in \R^d}  \|\AA \xx - \bb\|_2,
    \end{align*}
    and runs in $\Otil(d^3 \log(1/\eps) \cdot \nnz)$ bit operations using $O(d \log(nU))$ bits of space, where $\Otil$ hides $\log^{O(1)} (nU \log(1/\eps))$.
\end{restatable}

\begin{proof}
    The desired algorithm simply invokes $\callalg{LinSolve}(\AA^\tp \AA, \AA^\tp \bb, \eps)$. The matrix $\AA^\tp \AA$ and the vector $\AA^\tp \bb$ have maximal dimension $d$ and entries bounded by $n \cdot U^2$.
    
    The correctness follows from \Cref{cor:LinSolve-black-box} since $\AA^\tp \AA$ is invertible and thus the returned solution is an entry-wise $e^\eps$-multiplicative approximation of
    \begin{align*}
            (\AA^\tp \AA)^{-1} (\AA^\tp \bb)
            =
            \arg \min_{\xx \in \R^d}  \|\AA \xx - \bb\|_2.
    \end{align*}
    By \Cref{cor:LinSolve-black-box}, the algorithm uses $O(d^2 \log(1/\eps))$ calls of the oracle 
    \begin{align*}
        f_{\AA^\tp \AA}(\vv, p) = (\AA^\tp \AA \cdot \vv) \bmod p
    \end{align*}
    for any prime $p \le \poly(d, n, U)$ and any vector $\vv \in \F_p^d$. The oracle can be implemented by evaluating the $i$-th entry for $i \in [d]$, 
    \begin{align*}
        (f_{\AA^\tp \AA}(\vv, p))_i \equiv \sum_{j=1}^{n} \sum_{k=1}^{d} 
        (\AA^\tp)_{ij} \AA_{jk} \vv_k  \pmod{p}.
    \end{align*}
    It suffices to enumerate $(j,k)$ over $\{(j,k) : \AA_{jk} \neq 0\}$, so each entry can be computed in $\Otil( \nnz )$ bit operations, and the entire vector can be computed in $\Otil(\nnz \cdot d)$ bit operations. It requires $O(d \log (nU))$ bits of space to store the $d$-dimensional vector and the iterators. 

    We conclude the corollary by plugging in the time and space complexity in \Cref{cor:LinSolve-black-box}, i.e., 
    \begin{align*}
        \Otil( d^3 \log(1/\eps)) + O(d^2 \log(1/\eps)) \cdot \Otil(\nnz \cdot d) = \Otil(d^3 \log(1/\eps) \cdot \nnz)
    \end{align*}
    bit operations and $O(d \log (nU))$ bits of space.
\end{proof}

\paragraph{Remarks.} An alternative way to compute the matrix-vector product $\AA^\tp \AA \xx$ is to first compute $\yy := \AA \xx$ and then $\AA^\tp \yy$. This reduces the time complexity for one matrix-vector product from $\Otil(\nnz \cdot d)$ to $\Otil(\nnz)$. However, this method needs to store $\AA \xx$, which would require $\Omega(n)$ space. For linear regression, we typically have $d << n$, so in \Cref{cor:linear-regression} we present the version that uses $\Otil(d)$ space.

\subsection{Linear Programming}

In this section, we apply our linear-space linear system solver to linear programming. Ghadiri, Peng and Vempala analyzed the bit complexity of the interior point method (IPM) in \cite{GPV2023bit}. We replace the linear system solver in the inverse free path following IPM with our linear-space solver.

\begin{definition}[Definition 2 in \cite{GPV2023bit}] \label{def:inner-radius-outer-radius}
    Let $\AA \in \R^{n \times d}$, $\bb \in \R^d$, $\cc \in \R^n$ with $n \ge d$. For a linear program of the form $\min_{\AA^\tp \xx = \bb, \xx \ge 0} \cc^\tp \xx$, we define the following quantities:
    \begin{itemize}
        \item Inner radius $r$: There exists an $\xx$ such that $\AA^\tp \xx = \bb$ and $x_i \ge r \ge 0$ for all $i \in [n]$.
        \item Outer radius $R$: For all $\xx \ge 0$ with $\AA^\tp \xx = \bb$, it holds that $\| \xx \|_2 \le R$.
    \end{itemize}
\end{definition}

\begin{theorem}[Variant of Theorem 1.2 and Theorem 1.3 in \cite{GPV2023bit}, plugged in \callalg{LinSolve}]  \label{thm:IPM}
    Given $\AA \in \Z^{n \times d}$ with full column-rank, $\bb \in \Z^d, \cc \in \Z^n$ all with entries in $[-U, U]$, and an error parameter $0 < \eps < 1$, suppose $\AA$ has $\nnz$ nonzero entries and $\AA^\tp \AA$ has condition number bounded by $\kappa$, and the inner radius and outer radius of the linear program $\min_{\AA^\tp \xx = \bb, \xx \ge 0} \cc^\tp \xx$ are $\rr$ and $\RR$, respectively. Then there is an algorithm (adapted from Algorithm 6, \cite{GPV2023bit}) that finds $\xxhat \in \R^n$ such that
    \begin{align*}
        \cc^\tp \xxhat \le \min_{\AA^\tp \xx = \bb, \xx \ge 0} \cc^\tp \xx + \eps, \quad \text{ and } \quad
        \| \AA^\tp \xxhat - \bb \|_2 \le \eps,
    \end{align*}
    in $O( d^3 \nnz \cdot n^{0.5} \cdot \log^{O(1)} ( \frac{n\kappa UR}{ \eps \cdot r}))$ bit operations
     and in space of $O(n \log(\frac{n \kappa U R}{\eps r}))$ bits.
\end{theorem}

Theorem 1.3 in \cite{GPV2023bit} plugs in the exact solver from \cite{S2005shifted}, so they do not assume any condition number bound. Each entry is stored with $\Omega(n)$ bits, which requires at least $\Omega(n^2)$ space in total. For our purposes, we study the robust IPM setting with the assumption that the condition number is bounded by $\kappa$, so we can achieve near-linear space when $\kappa \le \poly(n)$. This is also the setting of Theorem 1.2 in \cite{GPV2023bit}, but they used a projection maintenance data structure to optimize the time, which can be replaced by our solver.

We justify our theorem by the following facts. Let $L :=  O(\log(\frac{n \kappa U R}{\eps r}))$ denote the bit length.
\begin{itemize}
    \item Since the matrix has condition number bound $\kappa$, to maintain the invariant of IPM,
    \begin{align*}
        \| \xx \circ \ss - \mathbf{1} \|_2 < 0.1,
    \end{align*}
    it suffices to round everything to $L$ bits. Therefore, the algorithm stores $O(1)$ vectors with each entry represented by $L$-bit fixed points.
    \item The dominant step of IPM is the following:
    \begin{align*}
        \delta_s \gets \AA ( \AA^\tp \overline{\XX} ( \overline{\SS})^{-1} \AA )^{-1} \AA^\tp \overline{\SS} \delta_\mu
    \end{align*}
    where $\overline{\XX} = \diag(\xxbar), \overline{\SS} = \diag(\ssbar)$ for some vector $\xxbar$ and $\ssbar$. Similarly to \Cref{cor:linear-regression}, we implement matrix-vector access to the matrix $\AA^\tp \overline{\XX} ( \overline{\SS})^{-1} \AA$. Since the diagonal matrix only reweights the products of the entries, the analysis of time and space complexity of the matrix-vector access is identical to the proof of \Cref{cor:linear-regression}. We obtain $\Otil( \nnz \cdot d)$ time and $O(d L)$ space for each matrix-vector access.
    \item IPM runs in $\Otil(n^{0.5} \cdot \log(\frac{R}{ \eps \cdot r}))$ iterations. We conclude the desired time and space bounds.
\end{itemize}

\section{Eigenvalues, Eigenvectors, and SVD in Linear Space} \label{sec:eigen}
In this section, we study how to apply our linear-space linear system solver to compute eigenvalues and eigenvectors. All the algorithms in this section use $O(n \log(nU/\eps))$ bits of space. We focus on the parameter regime $\eps > 1/\poly(n)$; accordingly, $\log^{O(1)}(1/\eps)$ factors are suppressed in the stated running-time bounds.

We first prove a slightly stronger version of the Inverse Power Method in \Cref{subsec:InvPower} using our linear-space linear system solver. Then, in \Cref{subsec:PerturbSpectrum}, we show that if we perturb by a random diagonal matrix, the eigenvalues are still good approximations, with the property that all eigenvalues are bounded away from each other.  In \Cref{subsec:ComputeSpectrum}, by applying the inverse power method to shifted matrices, we can recursively extract the eigenvalues using a divide-and-conquer strategy. As a corollary, we can compute eigenvectors and SVDs as discussed in \Cref{subsec:SVD}.

\subsection{Inverse Power Method}  \label{subsec:InvPower}

In this section, we state the guarantees of the inverse power method applied to real symmetric matrices. It is well-known that the power method has convergence rate depending linearly on the spectral gap, but if we only want the eigenvalues for real symmetric matrices, we can achieve better convergence independent of the spectral gap. 

We give a slightly stronger inverse power method without assuming any condition number bound. Our algorithm can detect the existence of an eigenvalue very close to $0$, due to the floating-point representation of \callalg{LinSolve}.

In inverse power method, in each iteration we apply $\AA^{-1}$ to some vector $\vv$.
When the matrix has a small eigenvalue, the inverse can have a large eigenvalue that can be exponential in $n$. We cannot afford to store $\AA^{-1} \vv$ in fixed-point representation. Instead, we store the vector using floating points, which is returned by \callalg{LinSolve}. 

\begin{restatable}[Inverse Power Method]{theorem}{InvPower}  \label{thm:InvPower}
For a real symmetric matrix $\AA$, 
let $\lam^A_1, \lam^A_2, \dots, \lam^A_n$ be the eigenvalues and $\ee_1, \ee_2, \dots, \ee_n$ be a corresponding orthonormal eigenbasis of $\AA$ with $|\lam^A_1| \le | \lam^A_2| \le \dots \le | \lam^A_n|$. For any constant $c>0$, there exists a randomized algorithm \callalg{InvPower} that, given $\eps \in (0,1)$, $\delta \ge n^{-c}$, and a real symmetric matrix $\AA$ with $\nnz$ nonzero integer entries in $[-U, U]$, 
with success probability $\ge 1 - n^{-c}$, computes $\lamtil \ge 0$ such that
    \begin{align*}
        \lamtil \approxbar_{\eps} \max\{\delta, | \lam^A_1| \},
    \end{align*}
    and runs in $\Otil( n^2(\nnz + n)/\eps)$ bit operations using space of $O(n \log (nU/\eps))$ bits, where $\Otil$ hides logarithmic factors $\log^{O(1)} (nU/\eps)$. 
\end{restatable}

Although folklore regards the inverse power method as numerically stable, we still provide a detailed analysis in \Cref{sec:proof-InvPower} that accounts for rounding errors and details the time and space complexity of our variant, \callalg{InvPower}.

\begin{algorithm}[t] \labelalg{InvPower}
    \caption{\textsc{InvPower}}
    \SetKwInOut{Input}{Input}
    \SetKwInOut{Output}{Output}
    \Input{A symmetric matrix $\AA \in \mathbb{Z}^{n \times n}$ with entries in $[-U,U]$,\\
    the approximation parameters $\eps, \delta$}
    \Output{$\lamtil \ge 0$ such that $\lamtil \approxbar_{\eps} \max\{ \delta, \min_i|\lambda_i(\AA)| \}$. }
        \If{$\callalg{LinSolve}(\AA,\mathbf{0}) = \texttt{SINGULAR}$}{
            \Return $0$.
        }
        $T\gets \lceil\frac{28\log (4n / \eps_S)}{\eps}\rceil.$ \\
        Sample $\uu^{(0)} \sim \mathcal{N}(0, \II)$ with each entry represented by $L$-bit fixed-point where $L := O(\log(nU / \eps))$.
        \label{line:sample-vv-0}
        \\
        \For{$i = 1$ \KwTo $T$}{
            $\vv^{(i)} \gets \uu^{(i-1)} / \| \uu^{(i-1)} \|_2$ in floating-point arithmetic, and converts to $L$-bit fixed points. \label{line:compute-vi}
            \\
            $\uu^{(i)}\leftarrow \callalg{LinSolve}(\AA, \vv^{(i)}, \eps_L)$ with each entry represented by $L$-bit floating-point, and $\eps_L \ge \poly(1/n, \delta)$ to be chosen later.  \label{line:compute-ui}\\
            \If{ $\| \uu^{(i)} \|_2^2 \ge 2 / \delta^2$ } {
                \Return $\delta$   \label{line:return-0-if-eigval-small}
            }
        }
        \Return $\| \vv^{(T)} \|_2 / \| \uu^{(T)} \|_2$.
\end{algorithm}

\paragraph{Inverse power method with a spectral gap.}

The next corollary states the guarantees when there is a constant spectral gap between the smallest and the second smallest eigenvalue. In this variant, the time complexity does not depend on $1/\eps$. We also output an $\eps$-approximation of the corresponding eigenvector. The proof of the next corollary is in \Cref{sec:proof-InvPowerGap}.

\begin{corollary}[Inverse Power Method with a spectral gap] \label{cor:InvPowerGap}
    For any constant $c>0$, there exists a randomized algorithm \labelalg{InvPowerGap}\callalg{InvPowerGap} that, 
    given inputs $\eps , \delta , \AA$ as in \Cref{thm:InvPower} with an additional property of the spectral gap: $\delta \le |\lam^A_1|$ and $1.1|\lam^A_1| \le | \lam^A_2|$, 
    with success probability $\ge 1 - n^{-c}$ computes $\lamtil \ge 0$ such that
    $
    \lamtil \approxbar_{\eps} | \lam^A_1|.
    $
    The algorithm runs in $\Otil( n^2(\nnz + n))$ bit operations and uses space of $O(n \log (nU/\eps))$ bits, where $\Otil$ hides logarithmic factors $\log^{O(1)} (nU/\eps)$.  Furthermore, the vector in the last iteration $\vv^{(T)}$ is an approximation of $\ee_1$, such that
    \begin{align*}
        |\langle \vv^{(T)},  \ee_1 \rangle| \in [1 - \eps, 1 + \eps], \quad \langle \vv^{(T)}, \vv^{(T)} \rangle \in [1-\eps, 1+\eps].
    \end{align*}
\end{corollary}

\subsection{Perturbing Spectrum}   \label{subsec:PerturbSpectrum}

Our methods precondition the matrix based on Minami bound in \cite{M1996local}, such that the eigenvalues are bounded away from each other by $\gamma \ge (\eps / nU)^{O(1)}$. We use the following version from \cite{APSSS2017matrix}.

\begin{lemma}[Equation 1.11 in \cite{APSSS2017matrix}]  \label{lemma:Minami-bound}
    For any real symmetric matrix $\AA \in \R^{n \times n}$, any probability distribution $\rho$ on $\R$ with bounded maximum density $\|\rho \|_\infty$, any interval $I$ with length $|I|$, if $\VV^{\diag}$ is a diagonal matrix with each diagonal entry independently sampled from $\rho$, it holds that
    \begin{align*}
        \Pr_{\VV^{\diag}} [ (\AA + \VV^{\diag}) \text{ has }\ge 2\text{ eigenvalues in }I ] \le O( \| p \|_\infty |I| n)^2.
    \end{align*}
\end{lemma}

By this lemma, we can add a random diagonal matrix to perturb the eigenvalues. We will show that the entries in the diagonal matrix can be sampled from some uniform distribution over $O(\log(nU/\eps))$-bit fixed points.

\begin{lemma}[Spectrum Perturbation] \label{lemma:PerturbSpectrum}
    There exists a randomized algorithm \callalg{PerturbSpectrum}\labelalg{PerturbSpectrum} that, given $\eps \in (0,1)$ and a matrix $\AA \in \Z^{n \times n}$ with entries in $[-U, U]$, with probability $\ge 1 - O(n^{-2})$, returns a matrix $\BB \in \Q^{n \times n}$, such that the following properties hold. We define $\gamma := \eps^2 / (n^4U)$.
    \begin{itemize}
        \item $\BB = \AA + \DD$ for some diagonal matrix $\DD$ whose entries are $O(\log(1/\gamma))$-bit fixed points bounded by $\eps/2$.
        \item For all $i \in [n]$, we have $|\lambda_i(\AA) - \lambda_i(\BB)| \le \eps$.
        \item The eigenvalues of $\BB$ are bounded away by $\gamma$ from each other. That is, for all $i \in [n-1]$, we have $\lambda_i(\BB) + \gamma < \lambda_{i+1}(\BB)$.
    \end{itemize}
    The algorithm runs in $\Otil(n \log(U/\eps))$ bit operations, using space of $O(n \log (nU / \eps))$ bits.
\end{lemma}

\begin{proof}
    Let $\rho$ be the uniform distribution on $[0, \eps/2]$, we have density $\| \rho \|_\infty = 2 / \eps$. Let $\VV^{\diag}$ be the diagonal matrix with each diagonal entry independently sampled from $\rho$. The eigenvalues of $\AA + \VV^{\diag}$ are in the range $[-U, U + \eps/2] \subseteq [-2U, 2U]$.
    
    Let $K>0$ be a parameter to be chosen later. For any $i \in [0, K-2]$, let $I_i$ be the interval $[-2U + i\cdot 4U/K, -2U + (i+2)\cdot 4U/K]$. This family of intervals has the property that, for any $x, y \in [-2U, 2U]$ such that $|x - y| < 4U/K$, there must exist an interval that contains both $x$ and $y$. By \Cref{lemma:Minami-bound}, for all $i \in [0, K-2]$, 
    \begin{align*}
        \Pr_{\VV^{\diag}} [ (\AA + \VV^{\diag}) \text{ has }\ge 2\text{ eigenvalues in }I_i ] \le O( \| p \|_\infty |I| n)^2.
    \end{align*}
    By a union bound over the events that each interval does not contain two eigenvalues, we have the probability that 
    $(\AA + \VV^{\diag}) \text{ has eigenvalues bounded away from each other by }4U/K$, is bounded by
    \begin{align*}
        K \cdot O( \| p \|_\infty |I| n)^2 \le O(K (2/\eps)^2 (8U/K)^2 n^2) \le O(\eps^{-2}U^2n^2/K).
    \end{align*}
    We set $K = n^4U^2 \eps^{-2}$, so the gap between the eigenvalues is at least $4U/K = 4\eps^2 / (n^4U) \ge 4\gamma$.

    To construct \callalg{PerturbSpectrum}, we sample the rounded version of $\VV^{\diag}$ up to additive error $\gamma$. This can be done by sampling uniform distributions on fixed points with $O(\log(1/\gamma))$ bits.
    Then, $\BB = \AA + \VV^{\diag} + \Delta$ is a perturbed matrix for some diagonal matrix $\Delta$ with entries in $[-\gamma, \gamma]$. Since the matrices are real symmetric (and thus Hermitian), and the spectral radius of $\Delta$ is at most $\gamma$, by Weyl's inequality,
    \begin{align*}
        |\lambda_i(\BB) - \lambda_i(   \AA + \VV^{\diag}) | \le \gamma, 
    \end{align*}
    and therefore the eigenvalues of $\BB$ are bounded away from each other by $2 \gamma$. Again by Weyl's inequality,
    \begin{align*}
        |\lambda_i(\BB) - \lambda_i(\AA ) | \le \| \VV^{\diag} + \Delta \|_2 \le \eps/2 + \gamma < \eps, 
    \end{align*}
    concluding the lemma.
\end{proof}

\subsection{Computing Spectrum}    \label{subsec:ComputeSpectrum}
We present the algorithm \callalg{Spectrum} that computes the entire spectrum up to additive error via \callalg{ShiftInvert}. We use the divide-and-conquer strategy to recursively narrow the search ranges and pinpoint the eigenvalues.

\begin{restatable}{theorem}{Spectrum} \label{thm:Spectrum}
    For any constant $c>0$, there exists a randomized algorithm \callalg{Spectrum} that, given $\eps > 0$, a symmetric matrix $\AA \in \Z^{n \times n}$ with integer entries in $[-U, U]$ and $\nnz$ nonzero entries, with probability at least $1 - n^{-c}$, outputs $\lamtil_1, \dots, \lamtil_n$ such that
    \begin{align*}
        | \lam_i(\AA) - \lamtil_i| \le \eps.
    \end{align*}
    The algorithm runs in $\Otil(n^3 (\nnz+n) )$ bit operations and space of $O(n \log (nU / \eps))$ bits. Here, $\Otil$ hides logarithmic factors $\log^{O(1)}( n U \eps^{-1})$.
\end{restatable}

\begin{algorithm}[ht]
    \caption{ \textsc{Spectrum} }
    \labelalg{Spectrum}
    
    \SetKwInOut{Input}{Input}
    \SetKwInOut{Output}{Output}
    
    \Input{A symmetric matrix $\AA \in \Z^{n \times n}$ with entries in $[-U, U]$, \\
    the target accuracy $\eps \in (0,1)$.} 
    \Output{ $\lamtil_1, \dots, \lamtil_n \in \Q^n$ each represented by an $O(\log(U/\eps))$-bit fixed point. For all $i \in [n]$, we have $|\lambda_i - \lamtil_i| \le \eps$.} 
    \algorithmfootnote{$\ell, r$ are stored by $O(\log(nU/\eps))$-bit fixed points, so no error is incurred when divided by $2$}
    $\gamma \gets \eps^2 / (n^4U)$
    \\
    $\BB \gets $ \callalg{PerturbSpectrum}$(\AA, \eps/2)$
    \\
    $\Lambda \gets \texttt{ComputeSpectrum}(\BB, -2U, 2U)$  \label{line:Lambda}
    \\
    \Return Sorted list of $\{ \lambda \in \Lambda : \forall \lambda' \in \Lambda, \ \lambda' \notin (\lambda - \gamma/2,  \lambda)\}$
    \label{line:return-merge-Lambda}
    \\
    \ 
    \\
    \SetKwFunction{FMain}{ComputeSpectrum}
    \SetKwProg{Fn}{Function}{:}{}
    \Fn{\FMain{$\BB$, $\ell$, $r$}}{ 
        \If{ \upshape\texttt{ShiftInvert}$(\BB, \ell, r)$ = \texttt{YES} }{
            \If { $r-\ell < \gamma/8$ }{
                \Return $\{ \ell \}$
            } \Else {
                \Return $\texttt{ComputeSpectrum}(\BB, \ell, (\ell+r)/2) \cup \texttt{ComputeSpectrum}(\BB, (\ell+r)/2, r)$ \label{line:internal-node}
            }
        }
        \Else{
            \Return $\emptyset$
        }
    }
    \ 
    \\
    \SetKwFunction{FShift}{ShiftInvert}
    \Fn{\FShift{$\BB$, $\ell$, $r$}}{ \labelalg{ShiftInvert}
        $m \gets (\ell + r)/2$
        \\
        $\lambda \gets \callalg{InvPower}( \BB - m \II, 0.1, (r - \ell)/2)$  \label{line:compute-lambda}
        \\
        \Return \texttt{YES} if $\lambda < 1.2(r-\ell)/2$, otherwise return \texttt{NO}
    }
\end{algorithm}

\paragraph{Proof of \Cref{thm:Spectrum}.}

Let $\calI$ denote the set of intervals $[\ell, r]$ where $\texttt{ComputeSpectrum}(\ell, r)$ is invoked.
The intervals form a tree structure, such that each internal node $u = (\ell, r)$ has two children $(\ell, (\ell+r)/2)$ and $((\ell + r)/2, r)$.
A node $u = (\ell, r)$ is a leaf if and only if either of the two following cases holds, so the leaf intervals can be categorized into two types:
\begin{itemize}
    \item \textbf{Type 1:} $\callalg{ShiftInvert}(\BB, \ell, r)$ returns \texttt{NO}. In this case, the node returns $\emptyset$.
    \item \textbf{Type 2:} $\callalg{ShiftInvert}(\BB, \ell, r)$ returns \texttt{YES}, and $r - \ell < \gamma/8$. In this case, the node returns $\{ \ell \}$.
\end{itemize}
Let $\calL$ denote the set of leaf intervals in the second case. Then, by \Cref{line:Lambda}, $\Lambda$ is the set of left endpoints of the intervals in $\calL$, since each internal node returns the union set returned by its two children.

\paragraph{Correctness of \Cref{thm:Spectrum}.}
We assume that all calls of \callalg{PerturbSpectrum} and \callalg{InvPower} are executed successfully, such that the guarantees in \Cref{lemma:PerturbSpectrum} and \Cref{thm:InvPower} are satisfied, which holds with high probability since we will show there are a total of $O(n \log(nU/\eps))$ calls in \eqref{eq:size-calI}. 

By \Cref{lemma:PerturbSpectrum}, $\gamma= \eps^2 / (n^4U)$ is the gap of the spectrum of $\BB$, such that every eigenvalue is bounded away by $\gamma$ from each other.

We denote the eigenvalues of $\AA$ and $\BB$ as
\begin{align*}
    \lambda_1 \le \lambda_2 \le \dots \le \lambda_n \quad \text{and} \quad    \kappa_1 \le \kappa_2 \le \dots \le \kappa_n,
\end{align*}
respectively. We show the guarantees of \texttt{ShiftInvert} in the next lemma. Since \texttt{ShiftInvert} always returns when $r - \ell < \gamma/8$, so we always have $r - \ell \ge \gamma/16$.
\begin{lemma} \label{lemma:ShiftInvert}
    In {\upshape\texttt{ShiftInvert}}, if $r - \ell \ge \gamma/16$, either one of the following holds.
    \begin{itemize}
        \item It returns \texttt{NO} and $B$ has no eigenvalues in the range $[\ell, r]$.
        \item It returns \texttt{YES} and $B$ has at least one eigenvalue in the range $[\ell-(r-\ell)/4, r + (r-\ell)/4]$.
    \end{itemize}
\end{lemma}
\begin{proof}
    By \Cref{thm:InvPower}, $\lambda$ in \Cref{line:compute-lambda} satisfies
    \begin{align*}
        \lambda \approxbar_{0.1} \max\{ (r - \ell)/2, \min_i |\lambda_i(\BB - m\II)| \}
        =
        \max\{ (r - \ell)/2, \min_i |\lambda_i(\BB) - m| \}.
    \end{align*}
    If the function returns \texttt{YES}, we have
    \begin{align*}
        \min_i |\lambda_i(\BB) - m| \le 1.1 \lambda \le 1.32 \cdot (r - \ell)/2,
    \end{align*}
    and thus there exists an eigenvalue in $\frac{\ell + r}{2} \pm 1.32 \frac{r - \ell}{2} \in [\ell-(r-\ell)/4, r + (r-\ell)/4]$.
    On the other hand, if the function returns \texttt{NO}, we have
    \begin{align*}
        \max\{ (r - \ell)/2, \min_i |\lambda_i(\BB) - m| \} \ge 0.9\lambda \ge 0.9 \cdot 1.2(r - \ell)/2 > (r - \ell)/2,
    \end{align*}
    and thus $\min_i |\lambda_i(\BB) - m| > (r - \ell)/2$, so we conclude that no eigenvalue is in $[\ell, r]$.
\end{proof}

We prove the following claims.
\begin{claim}
    For all $i \in [n]$, there exists an interval $[\ell, r] \in \calL$ such that $\kappa_i \in [\ell, r]$. 
\end{claim}
\begin{proof}
    Note that the leaf nodes of $\calI$ partition the interval on the root $[-2U, 2U]$, and we have $\kappa_i \in [-2U, 2U]$. Therefore, there exists an interval $[\ell, r]$ that contains the point $\kappa_i$. It remains to show that the leaf is of the second type.

    By \Cref{lemma:ShiftInvert}, $\callalg{ShiftInvert}(\BB, \ell, r)$ must return \texttt{YES} because $\kappa_i$ is in the range $[\ell, r]$.
\end{proof}
\begin{claim}
    For all $[\ell, r] \in \calL$, there exists $i \in [n]$ such that $|\ell - \kappa_i| < \gamma/5$. 
\end{claim}
\begin{proof}
    Since $\callalg{ShiftInvert}(\BB, \ell, r)$ returns \texttt{YES} by the definition of $\calL$, we must have
    \begin{align*}
        \kappa_i \in [\ell-(r-\ell)/4, r + (r-\ell)/4].
    \end{align*}
    We have $r - \ell < \gamma/8$ by the property of the leaf interval, and thus $|\ell - \kappa_i| \le \gamma/8 + \gamma/8/4 < \gamma/5$.
\end{proof}
By \Cref{line:return-merge-Lambda} and the two above claims, $\Lambda$ can be partitioned into $n$ subsets with the $i$-th group in the range $[\kappa_i - \gamma/5, \kappa_i + \gamma/5]$. By \Cref{lemma:PerturbSpectrum}, the eigenvalues of $\BB$ are bounded away from each other by $\gamma$, and $| \lambda_i - \gamma_i| \le \eps/2$ for all $i \in [n]$. Therefore, the sorted list in \Cref{line:return-merge-Lambda} contains $n$ elements $\lamtil_1 \le \dots \le \lamtil_n$, such that
\begin{align*}
    |\lamtil_i - \lam_i| \le |\lamtil_i - \kappa_i| + |\kappa_i - \lam_i| \le \gamma/5 + \eps/2 < \eps.
\end{align*}

\paragraph{Performance of \Cref{thm:Spectrum}.}
Since the algorithm runs the subroutine \callalg{ShiftInvert} for $|\calI|$ times. We first bound the number of intervals in $\calI$.
We partition $\calI$ into levels, such that level $d \ge 0$, denoted as $\calI_d$, contains the set of intervals with length $4U / 2^d$. Note that level $0$ contains the only interval $[-2U, 2U]$ at the root.
\begin{claim} \label{claim:level-bound}
    For all $d > \log_2 (32U / \gamma)$, it holds that $\calI_d = \emptyset$.
\end{claim}
\begin{proof}
    An internal node associated with an interval $(\ell, r)$ goes to the branch \Cref{line:internal-node}, so the length $r - \ell$ is at least $\gamma/8$. When $d > \log_2 (32U / \gamma)$, any interval at level $d-1$ has length $< 4U / 2^d = \gamma/8$, so it can only be a leaf. Therefore, there are no intervals at level $d$.
\end{proof}
Let $\calI_{\mathrm{int}}$ be the set of intervals associating to the internal nodes, we have $2|\calI_{int}|+1 = |\calI|$, which is a property that holds for any binary tree. To bound the size of $\calI_{\mathrm{int}} \cap \calI_d$, we use double counting
\begin{align}
    \sum_{(\ell, r) \in \calI_{\mathrm{int}} \cap \calI_d} 1
    & \le   \notag 
    \sum_{(\ell, r) \in \calI_{\mathrm{int}} \cap \calI_d} \ \sum_{i \in [n]} \mathbbm{1} [\kappa_i \in [\ell-(r-\ell)/4, r + (r-\ell)/4]]
    \\ & =   \notag
    \sum_{i \in [n]} \ \sum_{(\ell, r) \in \calI_{\mathrm{int}} \cap  \calI_d}  \mathbbm{1} [\kappa_i \in [\ell-(r-\ell)/4, r + (r-\ell)/4]]
    \\ & \le    \label{eq:level-size-bound}
    \sum_{i \in [n]} 2 = 2n,
\end{align}
where the first inequality follows from the property of an internal node, and the second inequality follows from the fact that the intervals at level $d$ have the same length and do not overlap. Together with \Cref{claim:level-bound}, we have
\begin{align} \label{eq:size-calI}
    |\calI| = 2|\calI_{\mathrm{int}}| + 1 \le 2 \cdot 2n(1+ \log_2(32U/\gamma))+1 = O(n \log(nU/\eps)).
\end{align}

Now we bound the time and space complexity. Note that $\ell, r$ in \texttt{ComputeSpectrum} is stored exactly in $O(\log(nU/\eps))$-bit fixed points, due to the bound of the recursion tree depth in \Cref{claim:level-bound}, and the fact that $\ell, r$ multiplied by $2^d$ is an integer at level $d$. Therefore, when we call \callalg{InvPower} in \Cref{line:compute-lambda}, all its inputs are $O(\log(nU/\eps))$-bit fixed points, so the entries are bounded by $U' := U \cdot 2^{O(\log(nU/\eps))}  \le \poly(n, U, \eps^{-1})$ in \Cref{thm:InvPower} when we scale up the matrix to become an integer matrix. 

Our algorithm calls \callalg{InvPower} for $| \calI |$ times, with each call in $\Otil(n^2(\nnz + n))$ time and $O(n \log (nU))$ space. We conclude the running time
\begin{align*}
    O(n \log (nU/\eps)) \cdot \Otil(n^2(\nnz + n)) \le \Otil(n^3(\nnz + n)).
\end{align*}
For the space, by \Cref{lemma:PerturbSpectrum}, the representation of $\BB$ has $O(n \log(nU / \eps))$ bits. We conclude the space complexity $O(n \log(nU/\eps))$ bits.

\subsection{Eigenvector and Singular Value Decomposition}   \label{subsec:SVD}

In this section, we compute the approximation of the eigenvectors for a real symmetric matrix using \callalg{Spectrum} and \callalg{InvPower}. We first perturb the matrix to make the eigenvalues bounded away from each other by $\gamma = \Theta(\eps / \poly(n))$, and then compute the approximations of the eigenvalues up to additive error $\gamma/10$. By shifting the matrix with respect to the approximation of the eigenvalue, the spectral gap is large enough (relative to the shifted eigenvalue) to apply the inverse power method which admits a good approximation of the corresponding eigenvector.

\begin{theorem} \label{thm:Eigendecompose}
    For any constant $c>0$, there exists a randomized algorithm \labelalg{Eigendecompose}\callalg{Eigendecompose} that, given $\eps > 0$, a symmetric matrix $\AA \in \Z^{n \times n}$ with integer entries in $[-U, U]$ and $\nnz$ nonzero entries, with probability at least $1 - n^{-c}$, outputs $\lamtil_1, \dots, \lamtil_n$ and vectors $\vv_1, \dots, \vv_n$ with fixed point representations such that the following holds:
    \begin{itemize}
        \item For all $i \in [n]$, let $\ee_i$ denote the eigenvector associated with $\lam_i(\AA)$ of unit length,   
        \begin{align*}
            | \lam_i(\AA) - \lamtil_i| \le \eps, \quad 
            \| \vv_i \|_2^2 \in [1-\eps, 1+\eps], \quad
            \|\AA \vv_i - \lamtil_i \vv_i \|_2 \le \eps.
        \end{align*}
        \item For all $i, j \in [n]$ and $i \neq j$, it holds that $|\la \vv_i, \vv_j \ra| \le \eps$.
        \item The algorithm runs in $\Otil(n^3 (\nnz+n) )$ bit operations and space of $O(n \log (nU / \eps))$ bits. Here, $\Otil$ hides logarithmic factors $\log^{O(1)}( n U \eps^{-1})$.
    \end{itemize}
\end{theorem}

\begin{proof}
    We invoke $\callalg{PerturbSpectrum}(\AA, \eps/2)$ to perturb the spectrum. By \Cref{lemma:PerturbSpectrum}, we obtain an access to matrix $\BB$ such that 
    \begin{align*}
        | \lambda_i(\AA) - \lambda_i(\BB) | \le \eps/2, \quad \forall i \in[n]
    \end{align*}
    and
    \begin{align*}
        | \lambda_i(\BB) - \lambda_{i+1}(\BB) | \ge \gamma, \quad \forall i \in[n-1]
    \end{align*}
    where $\gamma:= \eps^2/(4n^4U)$. We also have $\| \AA - \BB \|_2 \le \eps/2$. Then, we obtain approximations of eigenvalues of $\BB$ with $\gamma/10$ accuracy by invoking
    \begin{align*}
        \lamtil^B_1, \lamtil^B_2, \dots, \lamtil^B_n \gets \callalg{Spectrum}(\BB, \gamma/10).
    \end{align*}
    By \Cref{thm:Spectrum}, we have $|\lamtil^B_i - \lam_i(\BB)| \le \gamma/10$ for all $i \in [n]$. 

    To approximate the eigenvector associated with $\lam_i(\BB)$, we invoke $\callalg{InvPowerGap}$ with the matrix $\BB - (\lamtil^B_i + \gamma/5) \II$ and the parameters $\delta := \gamma/10$ and a sufficiently small $\eps_0$ to be specified later. For this matrix, we have
    \begin{align*}
        \min_j \left| \lambda_j(\BB - (\lamtil^B_i + \gamma/5) \II) \right|
        =
        \left| \lambda_i(\BB) - (\lamtil^B_i + \gamma/5) \right| \in [0.1 \gamma, 0.3 \gamma].
    \end{align*}
    For any $j \neq i$, the eigenvalue satisfies
    \begin{align*}
        \left| \lambda_j(\BB - (\lamtil^B_i + \gamma/5) \II) \right| 
        & =
        \left| \lambda_j(\BB) - (\lamtil^B_i + \gamma/5) \right| 
        \ge
        | \lambda_j(\BB) - \lamtil^B_i| - | \gamma/5|
        \\ & \ge 
        | \lambda_j(\BB) - \lambda_i(\BB) | - |\lambda_i(\BB)- \lamtil^B_i| - | \gamma/5|
        \ge
        \gamma - \gamma/10 - \gamma/5 = 0.7 \gamma.
    \end{align*}
    Therefore, the spectral gap conditions in \Cref{cor:InvPowerGap} are satisfied. We obtained a vector $\vv_i$ such that
    \begin{align}
        | \la \vv_i, \ee_i \ra | \in [1-\eps_0, 1+\eps_0], \quad 
         \la \vv_i, \vv_i \ra  \in [1-\eps_0, 1+\eps_0]
         \label{eq:eigenvectors-prop-1}
    \end{align}
    where $\ee_i$ is a unit eigenvector of $\BB$ corresponding to $\lambda_i(\BB)$.
    We decompose $\vv_i := \uu_i + \ww_i$ such that $\uu_i = \la \vv_i, \ee_i \ra \ee_i$ and $\la \ww_i, \ee_i \ra = 0$, then
    \begin{align*}
        \| \ww_i\|_2^2 = \| \vv_i \|_2^2 - \| \uu_i \|_2^2 \le (1 + \eps_0)^2 - (1 - \eps_0)^2 = 4 \eps_0
    \end{align*}
    and thus $\| \ww_i \|_2 \le 2 \sqrt{\eps_0}$.
    We can bound by triangle inequality,
    \begin{align*}
        \| \AA \vv_i - \lamtil^B_i \vv_i \|_2
        & \le
        \| \AA \vv_i - \BB \vv_i \|_2 
        +
        \| \BB \vv_i - \BB \uu_i \|_2
        +
        \| \BB \uu_i - \lamtil^B_i \uu_i \|_2
        +
        \| \lamtil^B_i \uu_i - \lamtil^B_i \vv_i \|_2
        \\ & \le
        \| \AA - \BB \|_2 \| \vv_i \|_2
        +
        \| \BB \|_2 \| \ww_i \|_2
        +
        \| \lambda_i(\BB) \uu_i - \lamtil^B_i \uu_i \|_2
        +
        \lamtil^B_i \| \ww_i \|_2
        \\ & \le
        (1 + \eps_0) \eps/2
        +
        2 nU \cdot 2 \sqrt{\eps_0}
        +
        (1 + 5 \eps_0) \gamma/10
        +
        (2nU  + \gamma/10) \cdot 2 \sqrt{\eps_0}
        \\ & \le
        0.6\eps + 0.1 \eps + 0.1 \eps + 0.1 \eps < \eps
    \end{align*}
    by setting $\eps_0 = (\eps/(60nU))^2$. For the orthogonality condition, for any $i\neq j$, we have
    \begin{align}
        |\la \vv_i, \vv_j \ra | 
        & =
        | \la \uu_i + \ww_i, \uu_j + \ww_j \ra |
        \le
        | \la \uu_i , \uu_j \ra | +
        | \la \uu_i , \ww_j \ra | +
        | \la \ww_i , \uu_j \ra | +
        | \la \ww_i , \ww_j \ra |
        \notag
        \\ & \le
        0 + \| \ww_j\|_2 + \| \ww_i \|_2 + \| \ww_i \|_2 \|\ww_j \|_2 \le 8 \sqrt{\eps_0} \le \eps.
        \label{eq:eigenvectors-prop-2}
    \end{align}

    For the time complexity, the dominant step is $\callalg{Spectrum}(\BB, \gamma/10)$ and $n$ calls of $\callalg{InvPowerGap}$, requiring a total time of
    \begin{align*}
        \Otil(n^3 (\nnz + n)) + n \cdot \Otil(n^2(\nnz + n) ) = \Otil(n^3 (\nnz + n)).
    \end{align*}
\end{proof}

\begin{corollary} \label{cor:SVD}
    For any constant $c>0$, there exists a randomized algorithm \labelalg{SVD}\callalg{SVD} that, given $\eps>0$, $n \ge m$, a matrix $\AA \in \Z^{n \times m}$ with integer entries in $[-U, U]$ and $\nnz$ nonzero entries, with probability at least $n^{-c}$, outputs matrices $\UU \in \R^{n \times n},  \VV \in \R^{m \times m}$ and a rectangular diagonal matrix $\SSigma \in \R^{n \times m}$ such that 
    \begin{align*}
        \| \UU^\tp \UU - \II \|_2 \le \eps, \quad
        \| \VV^\tp \VV - \II \|_2 \le \eps, \quad
        \| \AA \VV - \UU \Sigma \|_2 \le \eps, \quad
        \| \AA^\tp \UU - \VV \Sigma^\tp \|_2 \le \eps, 
    \end{align*}
    in $\Otil(n^3(\nnz + n))$ bit operations and working space of $O(n \log(nU/\eps))$ bits. Here, $\Otil$ hides the logarithmic factors $\log^{O(1)}(nU \eps^{-1})$.
\end{corollary}

\begin{proof}
    Matrix-vector access to $\AA \AA^\tp + \diag(\cdot)$ can be implemented in $O(\nnz+n)$ time for the query of a right multiplication of a vector. Let $\eps_0$ be specified later. We invoke
    \[ 
        \callalg{Eigendecompose}(\AA \AA^\tp + \eps_0 \II, \eps_0/10)
    \]
    and let $\UU$ be the matrix with columns $\uu_1, \dots, \uu_n$ set to the approximate eigenvectors of $\AA \AA^\tp + \eps_0 \II$. Let $\SSigma$ be the rectangular diagonal matrix with diagonal entries $\sigma_1, \dots, \sigma_m$ set to the square roots of the corresponding approximate eigenvalues. Let $\VV$ be the $m$ by $m$ matrix with columns $\vv_1, \dots, \vv_m$ such that
    \begin{align*}
        \vv_i = \sigma_i^{-1} \AA^\tp \uu_i.
    \end{align*}

    By the guarantees of $\callalg{Eigendecompose}$, each $\sigma_i$ is nonzero since the eigenvalues $\ge \eps_0$ and the error is $\eps_0/10$, so $\vv_i$ is well-defined. All entries of $\UU^\tp \UU - \II$ are bounded by $\eps_0$ and thus $\| \UU^\tp \UU - \II \|_2 \le n \eps_0$. For distinct $i,j \in [m]$, we have
    \begin{align*}
        \vv_i^\tp \vv_j = \sigma_i^{-2} \uu_i^\tp \AA \AA^\tp \uu_j = \uu_i^\tp (\uu_j + \Delta) \le \uu^\tp_i \uu_j + \| \uu_i \|_2 \| \Delta \|_2
    \end{align*}
    for some vector $\Delta$ with $\| \Delta \|_2 \le \eps_0$, so $\| \VV^\tp \VV - \II\|_2 \le 3m \eps_0$. The last two properties follow from $\AA \vv_i = \sigma_i^{-1} \AA \AA^\tp \uu_i \approx \sigma_i \uu_i$, and $\AA^\tp \uu_i \approx \sigma_i \vv_i$. By choosing $\eps_0 = \eps / \poly(n)$ sufficiently small, all the inequalities have error bounded by $\eps$ as desired.
    
    This algorithm has the same time and space complexity as \Cref{thm:Eigendecompose}. Specifically, the working space is low since $\vv_i$ depends only on $\uu_i$ and $\sigma_i$.
\end{proof}

\section{Conclusion and Discussion}

In this paper, we presented a linear-space solver for general linear systems that runs in $\Otil(n^2 \cdot \nnz)$ time. We then show many other numerical linear algebra primitives can be solved in near-linear space and polynomial time.

Our algorithm approximates entries up to multiplicative factors, implying that the support of the solution $\AA^{-1} \bb$ is obtained from the approximation, since $0$ is the only multiplicative approximation of itself. However, one could compute the support of $\AA^{-1} \bb$ in $\Otil(n \cdot \nnz)$ time by solving the system modulo a large random prime. This uses the fact that zero modulo any prime is still zero, and a nonzero integer remains nonzero modulo a random prime with high probability.

For the space usage, our algorithm stores $O(n \log (nU))$ bits, which is linear with the assumption $U = \poly(n)$. However, for $U = O(1)$, each column of the matrix can be stored in $O(n)$ bits, but we use $O(n \log n)$ bits to compute $\AA^{-1} \bb$.

It is an open question whether there exists a polynomial-time algorithm using \emph{truly linear space}, i.e., $O(n \log U)$ bits or even $O(n)$ bits. It seems challenging to achieve truly linear space, since storing a determinant of a $01$-matrix would still require $\Omega(n \log n)$ bits. More fundamentally, finite field methods appear incapable of achieving truly linear space---one cannot even perform rational reconstruction for one entry of the solution vector. $O(n \log(nU))$ bits of space is the minimal requirement for rational reconstruction.

Another interesting direction is to optimize the $O(n^2 \cdot \nnz)$ running time for a linear-space linear system solver. It may be easier to deal with the polynomially-conditioned matrices and norm-wise approximations.  

\section*{Acknowledgments}
We thank Richard Peng for many useful discussions and useful reference pointers, Yang P. Liu for reviewing our paper and giving helpful comments, and Sepehr Assadi for helpful discussions. 

\bibliography{general}

\appendix
\section{Prime Sampling} 
\label{sec:prime-sampling}
Here we present the missing proofs from \Cref{prelim:PS}.
To sample a prime from an interval uniformly, we can use rejection sampling since the density of primes is asymptotically $1 / \ln(n)$. First, we show that the primality of a given number can be tested with high probability using the Miller-Rabin test \cite{R1980probabilistic}.

\begin{lemma}[\cite{C2011miller}] \label{lemma:prime-test}
    There exists a randomized algorithm \labelalg{TestPrime}\callalg{TestPrime} that, given a integer $0 < x \le U$, outputs either \texttt{PRIME} or \texttt{COMPOSITE}, and runs in $O(\log^2 U \log^{O(1)} \log U )$ bit operations and $O(\log U)$ space of bits, such that the following properties hold:
    \begin{itemize}
        \item If $x$ is prime, it outputs \texttt{PRIME} with probability $1$.
        \item If $x$ is composite, it outputs \texttt{COMPOSITE} with probability at least $3/4$. 
    \end{itemize}
\end{lemma}

We restate and prove \callthm{thm:SamplePrime} on the guarantees of uniformly sampling $n$ distinct primes. 

\SamplePrime*
\begin{proof}
    We sample the primes sequentially. For the $i$-th prime, we apply rejection sampling: we repeatedly draw a uniform number from $[n, n^2]$ until it is a prime and distinct from the previously sampled primes. 
    
    By \Cref{fact:number-of-prime}, there are at least $(n^2 - n) / (2 \log_2 n)$ primes in the range. A uniform random integer in the range has probability at least
    \begin{align*}
        \frac{(n^2 - n) / (2 \log_2 n) - (n-1)}{n^2} 
        \ge
        \frac{(n^2 - n) / (2 \log_2 n) - (n-1) \cdot n /(4 \log_2 n)}{n^2} 
        \ge
        \frac{1}{8 \log_2 n}
    \end{align*}
    to be a new prime, where we used the fact $n \ge 4 \log_2 n$ for $n \ge 16$. Therefore, for each prime we want to draw, we can sample $O(\log^2 n)$ times so that the failure probability (the probability of never picking a new prime) is at most $1 / n^{c+2}$. 

    As shown above, when drawing the $i$-th prime, we are sampling $O(\log^2 n)$ times. For each of the $O(\log^2 n)$ samples, we must determine whether the drawn integer is a new prime. We store the previously sampled primes in a BBST (balanced binary search tree), such that each insertion and query takes $O(\log^2 n)$ bit operations. By \Cref{lemma:prime-test}, we can run \callalg{TestPrime} $O(\log n)$ times to boost the success probability of one primality test to $1 - 1/n^{c+2}$. A union bound gives that the total failure probability is at most $1/n^c$.
    
    Therefore, by \Cref{lemma:prime-test}, the algorithm runs in 
    \begin{align*}
        O(k \log^2 n) \cdot O(\log n) \cdot O(\log^2 n \log^{O(1)} \log n) = O(k \log^{5} n \log^{O(1)} \log n  )
    \end{align*}
    bit operations, and in space of $O(k \log n)$ bits. We remark that this is only a crude analysis -- we do not optimize the number of logarithm factors.    
\end{proof}

\section{Wiedemann's Algorithm: Linear Systems and Determinants over Finite Fields} \label{sec:wiedemann}

In this section, we present Wiedemann's algorithm \cite{W1986solving} and its application. We reformulate the algorithm so that we can adapt it for our application of computing determinants. We track the space complexity carefully, and we measure it in bits or in number of finite field elements. 
We discuss the following problems.

\begin{problem}[Computing Minimal Polynomials, \callalg{MinPoly}\labelalg{MinPoly}]  \label{problem:min-poly}
    For prime $p$, given a matrix $\AA \in \F_p^{n \times n}$, find the minimal polynomial $\mu(\AA)$.
\end{problem}

\begin{problem}[Linear System Solving, \callalg{LinSolve1}\labelalg{LinSolve1}] \label{problem:linear-system}
    For prime $p$, given an invertible matrix $\AA \in \F_p^{n \times n}$ and a vector $\bb \in \F_p^{n}$, find $\xx \in \F_p^{n}$ such that
    \begin{align*}
        \AA\xx \equiv \bb \pmod p.
    \end{align*}
\end{problem}

\begin{problem}[Finding Nontrivial Kernel Vectors, \callalg{LinSolve0}\labelalg{LinSolve0}] \label{problem:kernel}
    For prime $p$, given a singular matrix $\AA \in \F_p^{n \times n}$, find a vector $\vv$ such that $\AA\vv = 0$ and $\vv$ is not the zero vector.
\end{problem}

\begin{problem}[Computing Determinants, \callalg{Det}\labelalg{Det}]  \label{problem:det}
    For prime $p$, given a matrix $\AA \in \F_p^{n \times n}$, compute $\det(\AA) \bmod p$.
\end{problem}

The original presentation in Wiedemann's paper \cite{W1986solving} addresses only \callalg{LinSolve1}, while the discussion for \callalg{Det} is insufficient for our purposes. The solution to \callalg{MinPoly} appears in a subroutine and is not clearly stated.
We will reformulate Wiedemann's algorithm for \callalg{MinPoly}, as all the other problems can be reduced to this one. 
The space and time complexity for each algorithm/reduction will be carefully presented. 

First note that we can reduce \callalg{LinSolve1} to \callalg{LinSolve0}. 
\begin{fact}  \label{lemma:linear-system-reduction-to-kernel}
    For an invertible matrix $\AA$ and a vector $\bb$, let $\yyhat$ be a nontrivial kernel vector such that 
    \begin{align*}
        \Mbegin
        \AA & -\bb \\
        \OO & 0
        \Mend
        \yyhat
        =
        \mathbf{0}.
    \end{align*}
    Let $\hat{\yy} = [\yy\ v]^\tp$, then $\yy / v$ is a solution to $\AA \xx = \bb$.
\end{fact}
\begin{proof}
    Note that $\AA \yy - v \bb = 0$. If $v = 0$, we must have $\yy = 0$ since $\AA$ invertible, but $\yyhat$ is nonzero, a contradiction. Therefore, $v \neq 0$, and $\AA^{-1} \bb = \yy/v$.
\end{proof}

We will discuss in detail how to solve \callalg{LinSolve0} and \callalg{Det} using \callalg{MinPoly} in the subsequent sections. 

\subsection{Finding Minimal Linear Recurrences}

The core of Wiedemann's algorithm is finding minimal linear recurrences. We interpret linear recurrences in the language of polynomials.

\begin{definition}[Linear recurrence of an array] For an array $[a_0, \dots, a_m]$ with elements over a field $F$, for $d \in [0, m]$, a degree-$d$ polynomial $g(X) = \sum_{i=0}^{d} c_i X^i$ is a linear recurrence of the array if for all $j \in [0, m-d]$, we have
\begin{align*}
    \sum_{i=0}^{d} c_i a_{i+j} = 0.
\end{align*}
Furthermore, $g$ is called a minimal linear recurrence if it is a linear recurrence with the least degree.
\end{definition}

We also refer to the minimal linear recurrence as the \textit{minimal polynomial}. We have the following fact.
\begin{fact}
    Up to scaling, an array has a unique minimal polynomial.
\end{fact}
It is well-known that the Berlekamp-Massey algorithm \cite{B1968nonbinary, M1969shift, B2015algebraic} finds the minimal linear recurrence in $O(m^2)$ time. 

\begin{theorem}[Berlekamp-Massey] \label{thm:BM}
    There exists an algorithm \callalg{LinearRecurrence}\labelalg{LinearRecurrence} such that, given an array $[a_0, a_1, \dots, a_{2m}]$ over $F = \F_p$ for prime $p > 1$ with a minimal polynomial $g(X) \in F[X]$ of degree at most $m$, the following holds:
    \begin{itemize}
        \item The algorithm outputs a scalar multiple of $g(X)$ deterministically.
        \item The algorithm runs in $O(m^2)$ operations in $F$ and has a space requirement of $O(m)$ field elements, i.e., $O(m \log p)$ bits.
    \end{itemize}
\end{theorem}

For the purposes of this paper, this original version suffices since we use it with Wiedemann's algorithm and the other components already require $O(m^2)$ time. 
Nevertheless, for the reader's convenience, we give the guarantees for the optimized version. The Berlekamp-Massey algorithm can be optimized to $O(m \log^2 (mp))$ time using Half-GCD and polyonomial multiplications based on Fast Fourier Transforms.

\begin{theorem}[Optimized Berlekamp-Massey \cite{L2005half, GG2013modern}] \label{thm:BM-opt}
    There exists an algorithm such that, given an array $[a_0, a_1, \dots, a_{2m}]$ over $F = \F_p$ for prime $p$ with a minimal polynomial $g(X) \in F[X]$ of degree at most $m$, the following properties hold:
    \begin{itemize}
        \item The algorithm outputs a scalar multiple of $g(X)$ deterministically.
        \item The algorithm runs in $\Otil(m \log p )$ bit operations and has a space requirement of $O(m)$ field elements, i.e., $O(m \log (mp))$ bits. Here, $\Otil$ hides $\log^{O(1)} (m \log p)$ factors.
    \end{itemize}
\end{theorem}

\subsection{Wiedemann's Algorithm for Minimal Polynomials} \label{sec:min-poly}

In this section, we present Wiedemann's algorithm for \callalg{MinPoly}. Recall that we defined the minimal polynomial for an array of scalars in the previous section. We first extend this notion to an array of vectors and an array of matrices.

\begin{definition}[Minimal polynomial of an array of vectors] 
Let $[\vv_0, \dots, \vv_m]$ be an array of vectors over a field $F$, meaning that for each $i \in [0, m]$ we have $\vv_i \in F^n$. The minimal polynomial of the array is a polynomial $\mu(X) = \sum_{i=0}^{d} c_i \cdot X^i$ with the least degree such that for all $j \in [0, m-d]$, it holds that 
\begin{align*}
    \sum_{i=0}^{d} c_i \vv_{i+j} = \mathbf{0}.
\end{align*}
\end{definition}

\begin{definition}[Minimal polynomial of an array of matrices] 

Let $[\MM_0, \dots, \MM_m]$ be an array of matrices over a field $F$, meaning that for each $i \in [m]$ we have $\MM_i \in F^{n \times n}$.
 The minimal polynomial of the array is a polynomial $\mu(X) = \sum_{i=0}^{d} c_i \cdot X^i$ with the least degree such that for all $j \in [0, m-d]$, it holds that 
\begin{align*}
    \sum_{i=0}^{d} c_i \MM_{i+j} = \mathbf{O}.
\end{align*}
\end{definition}

Note that a minimal polynomial of a matrix $\AA$ by definition, denoted as $\mu_\AA(X)$, is the minimal polynomial of the following array
\begin{align*}
    [\II, \AA, \AA^2, \dots, \AA^{m}]
\end{align*}
for any $m \ge n$. Wiedemann's algorithm can be summarized as follows: with decent probability, the minimal polynomial $\mu_\AA(X)$ is the same as the minimal polynomial of the scalar array 
\begin{align*}
    [\xx^\tp \yy, \xx^\tp \AA \yy, \xx^\tp \AA^2 \yy, \dots, \xx^\tp \AA^m \yy]
\end{align*}
where $\xx, \yy \sim F^n$ are vectors that are independent and drawn uniformly at random. Therefore, it is possible to use $\AA$ as a black-box with accesses to the matrix-vector product to find the minimal polynomial in $O(n)$ space.  We formally present this in \callalg{Wiedemann}.

\minpoly*

\begin{algorithm}[t] \labelalg{Wiedemann}
\caption{\textsc{Wiedemann}}
\SetKwInOut{Input}{Input}
\SetKwInOut{Output}{Output}
        
\Input{Matrix-vector access for a matrix $\AA \in F^{n \times n}$, where $F = \F_p$ for prime $p > n$}
\Output{A polynomial $g(X) \in F[X]$, such that $g \mid \mu_\AA(X)$ } 
    $\xx, \yy \gets$ independent and uniformly random vector from $F^n$ 
     \\
    \tcp{Compute $b_i = \xx^\tp (\AA^i \yy)$ for $i = 0 \dots 2n$ }
    
    \For{$i=0$ to $2n$}{
        \label{line:wiedemann-loop}
        $b_i \gets \xx^\tp\yy$\\
        $\yy \gets \AA\yy$ \\
    }

\Return $g(X) \gets \callalg{LinearRecurrence}(b_0, \dots, b_{2n})$ 
\end{algorithm}

\paragraph{Correctness of \callalg{Wiedemann}.} We present the correctness proof for \callalg{Wiedemann}. Most of the following proofs are adapted from \cite{W1986solving}. 

The returned polynomial of \callalg{Wiedemann} is the minimal polynomial for the array
\begin{align*}
   \{b_i\}_{i=0}^{2n} =  \{\xx^\tp \AA^i \yy \}_{i=0}^{2n},
\end{align*}
and we show that it coincides with the minimal polynomial of $\AA$ with decent probability for random $\xx$ and $\yy$. Let $\mu_{\xx,\yy}(X)$ denote the minimal polynomial for $\{b_i\}_{i=0}^{2n}$. For the array of column vectors
\begin{align*}
    \{\AA^i \yy\}_{i=0}^{2n},
\end{align*}
we let $\mu_\yy(X)$ denote its minimal polynomial. For the matrix $\AA$, its minimal polynomial is denoted as $\mu_\AA(X)$.

Before we go into the proofs, we first establish a bijection between the Krylov subspace $K \triangleq \Kr_\AA(\yy) = \mathrm{span}(\yy, \AA \yy, \AA^2\yy, \dots )$ and a quotient polynomial ring $R \triangleq F[X] / (\mu_\yy(X))$. Note that $R$ is a vector space, since it is closed under addition and scalar multiplication. For a polynomial $f \in F[X]$, we use $(f \bmod \mu_\yy(X))$ to denote the residue polynomial in $R$.

Let $d$ denote $\dim(K)$, then $K = \mathrm{span}(\yy, \AA \yy, \dots, \AA^{d-1} \yy)$ and $\deg(\mu_\yy(X)) = d$. Define $\psi: K\to R$ as a linear operation satisfying
\begin{align*}
    \psi(\AA^i \yy) = X^i,\quad \forall i\in[0,d-1].
\end{align*}

\begin{fact}
    $\psi$ is a bijection between $K$ and $R$.
\end{fact}

\begin{proof}
Note that $\{\AA^i\yy\}_{i=0}^{d-1}$ is a basis for $K$ and $\{X^i\}_{i=0}^{d-1}$ is a basis for $R$. So $\psi$ is a bijection.
\end{proof}

Note that $\psi(\AA \vv) = X\psi(\vv)$, so we can generalize for all $i\ge 0$, 
\begin{align*}
    \psi(\AA^i\yy) = X^i.
\end{align*}
Therefore, for any polynomial $p \in F[X]$, we have $\psi(p(\AA)\yy) = p$.

Now we introduce a lemma on polynomial sampling in $R$, which is useful later.

\begin{lemma}\label{lemma:prob-of-perp-in-R}
    The probability that a uniformly random polynomial in $R$ shares no common factors with $\mu_\yy$ is bounded by
    \begin{align*}
        \Pr_{g \sim R} [\gcd(g, \mu_\yy)=1] \ge \frac{1}{6 \lceil \log_{|F|}(d) \rceil}.
    \end{align*}
\end{lemma}

\begin{proof}
    Let $q$ denote $|F|$, the size of the field $F$. Let $h_1,\dots,h_k$ be the pairwise distinct irreducible factors of $\mu_\yy$. Let $d_i$ be the degree of $h_i$. By CRT we know that
    \begin{align*}
        \Pr_{g \sim R} [\gcd(g, \mu_\yy)=1] = \Pr_{g \sim R} \left[ \bigvee_{i=1}^{k} \gcd(g, h_i)=1 \right] = \prod_{i=1}^k \left(1 - \frac{1}{q^{d_i}}\right).
    \end{align*}
    
    It is known that every irreducible polynomial with degree $j$ divide $\left(X^{q^{j}} - X\right)$, which means the sum of their degrees is at most $q^j$. Let $s_j = j\cdot (\text{number of $h_i$'s with degree $j$})$. Then $0\le s_j\le q^j$, and
    \begin{align*}
        \sum_{i\ge 1} s_i \le d \le \sum_{i=1}^{\ell} q^i.
    \end{align*}
    where $\ell = \lceil \log_{|F|}(d) \rceil$.
    From this, it follows that
    \begin{align*}
        \sum_{i\ge 1} \frac{s_i}{q^i i}\le \sum_{i= 1}^{\ell}\frac{1}{i} \le \log 3\ell.
    \end{align*}
    Since the coefficients of $s_i$ on the left-hand side are decreasing, by taking $s_i=q^i$ for $i\le \ell$ and $s_i=0$ for $i>\ell$, we can achieve the maximum value of the sum. Therefore,
    \begin{align*}
        \log \Pr_{g \sim R} [\gcd(g, \mu_\yy)=1] &= \sum_{i\ge 1} \log\left(1 - \frac{1}{q^{i}}\right) \cdot \frac{s_i}{i}\\
        &\ge -\sum_{i\ge 1} \left(\frac{1}{q^i} + \frac{1}{q^{2i}}\right) \frac{s_i}{i}\\
        &\ge -\sum_{i\ge 1} \frac{s_i}{q^ii} - \sum_{i\ge 1}\frac{1}{q^{i}i}\\
        &\ge -\log 3\ell - \log 2 \ge -\log{6\ell}.
    \end{align*}
    Taking exponential on both sides completes the proof.
\end{proof}

\begin{lemma}\label{lemma:wiedemann-first-part}
    For any fixed $\yy$, it holds that $\mu_{\xx,\yy}(X)$ divides $\mu_\yy(X)$, and
    \begin{align*}
        \Pr_{\xx}[\mu_{\xx,\yy}(X) = \mu_\yy(X)] \ge 1/\lceil 6\log_{|F|} n \rceil.
    \end{align*}
\end{lemma}
\begin{proof}
    Define the inner product operator $\xi_\uu : K\to F$ specified by a vector $\uu\in F^n$ as
    \begin{align*}
        \xi_\uu(\vv) = \uu^{\tp}\vv.
    \end{align*}
    Since $\xi_\uu$ is a linear functional on $K$, $\xi_\uu\circ \psi^{-1}$ is a linear functional on $R$, which is defined as 
    \begin{align*}
        (\xi_\uu\circ \psi^{-1})(f) = \xi_\uu(\psi^{-1}(f)).
    \end{align*} 
    For $g\in R$, define $\eta_g: R\to F$ as
    \begin{align*}
        \eta_g(f) = [X^{d-1}]\big(g\cdot f\bmod \mu_\yy\big).
    \end{align*}
    It's easy to see $\eta_g$ is linear and thus $\eta_g\in R^*$, where $R^* = \{\mu: R\to F \text{ is a linear functional}\}$ is the dual space for $R$. Consider any linear functional $\mu\in R^*$, pick
    \begin{align*}
        g = \sum_{i=0}^{d-1} \mu(X^i) X^{d-1-i}\in R.
    \end{align*}
    It's easy to verify that for all $0\le i<d$, $\eta_g(X^i) = \mu(X^i)$, therefore $\eta_g = \mu$. Thus the mapping $\eta:R\to R^*$ defined by  $\eta(g(X)) = \eta_g$ is an injection, so it's also a bijection. There uniquely exists $g_\uu\in R$ such that $\eta_{{g_\uu}} = \xi_\uu\circ \psi^{-1}$. Therefore,
    \begin{align*}
        \{\xx^{\tp}\AA^i\yy\}_{i=0}^{2n}
        &= \left\{\left(\xi_\xx\circ \psi^{-1}\right)(X^i)\right\}_{i=0}^{2n}\\
        &= \left\{\eta_{g_\xx}(X^i)\right\}_{i=0}^{2n}\\
        &= \left\{[X^{d-1}]\Big(g_\xx(X)\cdot X^i\bmod \mu_\yy(X)\Big)\right\}_{i=0}^{2n}.
    \end{align*}

    Recall that $\mu_{\xx,\yy}$ is the minimal polynomial of $\{\xx^{\tp}\AA^i\yy\}_{i=0}^{2n}$. Let $\ell$ be its degree, then for every $j\in[0,2n-\ell]$,
    \begin{align*}
        0 &= \sum_{i=0}^{\ell} [X^i]\mu_{\xx,\yy} \cdot \xx^{\tp}\AA^{j+i}\yy\\
        &= \sum_{i=0}^{\ell} [X^i]\mu_{\xx,\yy} \cdot  [X^{d-1}]\Big(g_\xx\cdot X^{j+i}\bmod \mu_\yy\Big)\\
        &= [X^{d-1}]\left(\left(\sum_{i=0}^{l}  [X^i]\mu_{\xx,\yy}X^i \right) \cdot g_\xx X^j \bmod \mu_\yy\right)\\
        &= [X^{d-1}]\left(\mu_{\xx,\yy} g_\xx X^j \bmod \mu_\yy\right).
    \end{align*}
    This implies that
    \begin{align*}
         \mu_{\xx,\yy} g_\xx \bmod \mu_\yy = 0.
    \end{align*}
    The polynomial satisfying this with minimal degree is given by
    \begin{align*}
        \mu_{\xx,\yy} = \frac{\mu_\yy}{\gcd\big(\mu_\yy, g_{\xx}\big)}.
    \end{align*}
    From this, we can see that $\mu_{\xx,\yy}$ divides $\mu_\yy$.

    It's easy to verify that the map $\gamma : F^n\to R$ defined by
    \begin{align*}
        \gamma(\uu) = \eta^{-1}(\xi_\uu\cdot \psi^{-1})
    \end{align*}
    is a linear injection. Thus, picking $\xx$ uniformly at random from $F^n$ is equivalent to picking $g$ uniformly at random from $R$. Therefore,
    \begin{align*}
        \Pr[\mu_{\xx,\yy} = \mu_\yy] &= \Pr\left[\gcd(g_\xx,\mu_\yy)=1\right]\ge \frac{1}{\lceil 6\log_{|F|} n \rceil}
    \end{align*}
    by \Cref{lemma:prob-of-perp-in-R}.
\end{proof}

\begin{lemma}  \label{lemma:wiedemann-correctness}
    It holds that $\mu_{\yy}(X)$ divides $\mu_\AA(X)$, and 
    \begin{align*}
        \Pr_{\yy}[\mu_\yy(X) = \mu_\AA(X)] \ge 1/\lceil 6\log_{|F|} n \rceil.
    \end{align*}
\end{lemma}

\begin{proof}
    Let $f^{\tp}_A$, $f^{\tp}_{\yy}$ and $f^{\tp}_{\xx,\yy}$ be the corresponding $\mu_{\AA}$,$\mu_{\yy}$ and $\mu_{\xx,\yy}$ specified by $\AA^\tp$. Formally, $f^{\tp}_A$, $f^{\tp}_{\yy}$ and $f^{\tp}_{\xx,\yy}$ are the minimal polynomials of $\{(\AA^\tp)^i\}_{i=0}^{2n}$, $\{(\AA^\tp)^i \yy\}_{i=0}^{2n}$ and $\{\xx^\tp(\AA^\tp)^i \yy\}_{i=0}^{2n}$, respectively.

    Using the rational canonical form of $\AA$, we could find $\vv\in F^n$ such that $f^{\tp}_\AA = f^{\tp}_\vv$. For $\yy$ uniform randomly sample in $F^n$,
    \begin{align*}
        \Pr[f_{\vv,\yy} = \mu_{\AA}] &= \Pr[f^\tp_{\yy,\vv} = f^{\tp}_{\AA}]
        = \Pr[f^\tp_{\yy,\vv} = f^\tp_{\vv}]
        \ge \frac{1}{\lceil 6\log_{|F|} n \rceil}.
    \end{align*}
    where the last step is followed from \Cref{lemma:wiedemann-first-part}.

    Also by \Cref{lemma:wiedemann-first-part}, $f_{\vv,\yy}$ divides $\mu_{\yy}$. Moreover, $\mu_{\yy}$ divides $\mu_{\AA}$ since $\mu_{\AA}$ is the least common multiple of all $\mu_\yy$'s. Thus $f_{\vv,\yy} = \mu_{\AA}$ implies $\mu_{\yy} = \mu_{\AA}$. From this, $\Pr[\mu_{\yy} = \mu_{\AA}]\ge \Pr[f_{\vv,\yy} = \mu_{\AA}] \ge 1/\lceil 6\log_{|F|} n \rceil$.
    
\end{proof}

\paragraph{Performance of \callalg{Wiedemann}. } We first focus on the time complexity measured in terms of the number of field operations in $\F_p$. Sampling uniform random $\yy$ and $\xx$ takes $O(n)$ time. The loop in \Cref{line:wiedemann-loop} computes the inner product $\xx^\tp \yy$ and calls the matrix-vector access $\AA \cdot \yy$ for $n$ times. This part takes $O(n \cdot \TimeMatVec{\AA})$ time. In the end it returns the minimal linear recurrence by calling \callalg{LinearRecurrence}, which takes $\Otil(n^2)$ bit operations due to \Cref{thm:BM}. 

For space complexity, the algorithm only stores the variables $\yy, \xx, \{b_i\}_{i=0}^{2n}$. This requires storing $O(n)$ field elements, that is, $O(n \log p)$ bits.

\paragraph{Proof for \Cref{thm:wiedemann}. } By \Cref{lemma:wiedemann-first-part} and \Cref{lemma:wiedemann-correctness}, $\mu_{\xx,\yy}(X)$ is always a factor of $\mu_\AA(X)$, and the algorithm correctly outputs the minimal polynomial with probability at least
\begin{align*}
     \Pr_{\yy} [\mu_\yy(X) = \mu_\AA(X)] \cdot \Pr_{\xx} [\mu_{\xx,\yy}(X) = \mu_\yy(X) \mid y] \ge \frac{1}{36 \lceil \log_{p} (n)\rceil^2}
\end{align*}
since $\xx$ and $\yy$ are independent. Combining with the performance analysis above, we conclude the proof.

\subsection{Computing Determinant} \label{sec:DeterminantZp}

In this section, we demonstrate a linear-space algorithm for \callalg{Det}, computing the determinant of a matrix $\AA \in F^{n \times n}$ over a large finite field $\F_p$ for $p > \poly(n)$, based on preconditioning the Wiedemann's algorithm for minimal polynomials.

In \cite{W1986solving} and \cite{KS1991wiedemann}, they showed that, with high probability, the characteristic polynomial and the minimal polynomial of $\UU \AA \VV \DD$ coincide up to scaling, where $\UU, \VV$ are random Toeplitz matrices and $\DD$ is a random diagonal matrix.
\cite{CEKSTV2002efficient} provided a new analysis and simplified the preconditioner to a single diagonal matrix.

\begin{lemma}[Lemma 4.2 of \cite{CEKSTV2002efficient}] \label{lemma:precondition-D}
    For a finite field $F$, let $S \subseteq F$, $\AA \in F^{n \times n}$, and $\DD = \diag(d_1, \dots, d_n)$ be a diagonal matrix where $d_1, \dots, d_n$ are independently and uniformly random sampled from $S$. Then, with probability at least $1 - n(n-1)/(2|S|)$, there exists $t \ge 0, c \in F$ such that $cX^t \cdot \mu(\DD \AA) = \chi(\DD \AA)$, where $\mu(\cdot)$ and $\chi(\cdot)$ denote the minimal polynomial and the characteristic polynomial (in indeterminate $X$) respectively.
\end{lemma}
We remark that the original lemma in \cite{CEKSTV2002efficient} states that $\DD \AA$ is cyclic up to nilpotent blocks, which means that the invariant factors $f_1 \mid f_2 \mid \dots \mid f_s$ have the property that $f_1, \dots, f_{s-1}$ are monomials. This is equivalent to the condition $cX^t \cdot \mu(\DD \AA) = \chi(\DD \AA)$ for $t \ge 0$ and $c \in F$. 

We present \callalg{Determinant} by combining the diagonal preconditioner with \callalg{Wiedemann}. We restate and prove \callthm{thm:DeterminantZp}.

\DeterminantZp*

\begin{algorithm}[t]  \labelalg{DeterminantZp}
    \caption{\textsc{DeterminantZp}}
    \SetKwInOut{Input}{Input}
    \SetKwInOut{Output}{Output}
            
    \Input{A matrix-vector access for a matrix $\AA \in F^{n \times n}$ with $F = \F_p$ for prime $p >n^2/\delta$.}
    \Output{The determinant $(\det(\AA) \bmod p)$, with success probability $\ge 2/3$.}
    
    $\DD \gets $ a random diagonal matrix $\diag(d_1, \dots, d_n)$ where $d_1, \dots, d_n \sim (F \setminus \{0\})$
    \label{DeterminantZp:Step-1} \\
    
    $f(X) \gets \callalg{Wiedemann}(\DD \AA)$ be monic, with success probability $\ge 5/6$ 
    \label{DeterminantZp:Step-2} \\
    
    \Return $f(0) / \prod_{i=1}^{n} d_i$ if $\deg(f) = n$, otherwise return $0$.
    \label{DeterminantZp:Step-3}
\end{algorithm}

\begin{proof}
    \textbf{Correctness. } Note that $\det(\DD \AA) = \det(\AA) \prod_{i=1}^{n} d_i$ and $d_i$'s have inverse in F, so we have
    \begin{align} \label{eq:det-eq-1}
        \det(\AA) = \frac{\det(\DD \AA)}{\prod_{i=1}^{n} d_i} 
        = \frac{\chi(\DD \AA)(0)}{\prod_{i=1}^{n} d_i}.
    \end{align}
    By \Cref{lemma:precondition-D}, with probability $1-n(n-1)/(2(|F|-1)) \ge 1- 1/6$, we have $c X^t \cdot \mu(\DD \AA) = \chi(\DD \AA)$. We assume that $\mu(\cdot)$ and $\chi(\cdot)$ are both monic, and since $\deg(\chi(\DD \AA)) = n$, the condition becomes
    \begin{align} \label{eq:det-eq-2}
        X^{n - \deg(\mu(\DD \AA))} \cdot \mu(\DD \AA) = \chi(\DD \AA).
    \end{align}
    By \Cref{thm:wiedemann}, with probability $1 - \delta/2$, $f(X) = \mu(\DD \AA)$. Both of the above hold simultaneously with probability $2/3$ by a union bound, and by \eqref{eq:det-eq-1} and \eqref{eq:det-eq-2},
    \begin{align*}
        \chi(\DD \AA) = \begin{cases}
            \frac{f(0)}{\prod_{i=1}^{n} d_i}  & \text{if } \deg(f) = n, \\
            0 & \text{otherwise,}
        \end{cases}
    \end{align*}
    which concludes the correctness. The success probability can be boosted to $1 - \delta$ by running the algorithm for $\log(1/\delta)$ times and taking the majority vote.
    
    \textbf{Performance. }
    Note that $\DD$ is implicitly stored by $d_1, \dots, d_n$, and we can implement the matrix-vector access for $\DD \AA$ so that each call takes $O(\TimeMatVec{\AA})$ field operations. One can verify that Step \ref{DeterminantZp:Step-1} and Step \ref{DeterminantZp:Step-3} can be done in $O(n)$ field operations and incur $O(1)$ extra space. 

    For Step \ref{DeterminantZp:Step-2}, by \Cref{thm:wiedemann}, since $p \ge n$, \callalg{Wiedemann} can achieve success probabiilty $\ge 5/6$ by repeating for $O(1)$ times. This step makes $O(n)$ calls to $\DD\AA$, and thus runs in
    \begin{align*}
        O(n \cdot \TimeMatVec{\AA} )
    \end{align*}
    field operations. When the success probability is boosted to $1-\delta$, the algorithm is repeated for $\log(1/\delta)$ times, and it follows the desired time complexity.
    
    The space requirement is $O(n)$ field elements, since $\DD$ and $f(X)$ can be stored with $O(n)$ field elements. 
\end{proof}

\subsection{Linear System Solving over Finite Fields} \label{sec:LinSolveZp}

In this section, we use \callalg{Wiedemann} for \callalg{MinPoly} to solve \callalg{LinSolve0}, finding a nontrivial kernel vector of a singular matrix, which would give a solution to \callalg{LinSolve1} by the reduction in \Cref{lemma:linear-system-reduction-to-kernel}. We present the algorithm \callalg{FindKernel}.

\begin{algorithm}[t] \labelalg{FindKernel}
\caption{\textsc{FindKernel}}
\SetKwInOut{Input}{Input}
\SetKwInOut{Output}{Output}
\SetKwFor{Loop}{Repeat}{}{}
        
\Input{ Matrix-vector access to a singular matrix $\AA \in \F_p^{n \times n}$ for prime $p$ }
\Output{A vector in the kernel of $\AA$}

\tcp{Computing the minimal polynomial}
$\mu(X)  \gets \callalg{Wiedemann}(\AA)$ with success probability $\ge 1 - \delta/2$\\
$\bar{\mu}(X) \gets \mu(X) / X^c$ such that $\bar{\mu}(0) \neq 0$ \\
\tcp{Finding a nontrivial kernel vector}
$\zz \gets$ uniformly random vector from $F^{n}$ \\
\Loop {$\log_p(2/\delta)$ times}{
    $\yy \gets \bar{\mu}(\AA) \zz$ \label{line:ls-yy} \\
    Find $t\ge 0$ to be the smallest integer that $\AA^t \yy = 0$  \label{line:ls-t} \\
    \If{$t>0$}{
    \Return $\AA^{t-1} \yy$
    }
}
\end{algorithm}

\paragraph{Analysis of \callalg{FindKernel}.} Note that $\AA$ is a singular matrix over the finite field $F = \F_p$ for a prime $p$. Assume the minimal polynomial
\begin{align} \label{eq:mu}
     \mu_\AA(X) = \sum_{i=0}^{d} a_i  X^i
\end{align}
is given, we aim to find a kernel vector $\xx$ such that $\AA \xx = 0$ and $\xx \neq 0$. We have $\bar{\mu}_\AA(X) = \mu_\AA(X) / X^c$ for some unique integer $c$ such that the constant term is nonzero, $\bar{\mu}_\AA(0) \neq 0$.

Let $\zz$ be a uniformly random vector from $F^n$ and $\yy = \bar{\mu}(\AA) \zz$. The integer $t \ge 0$ is the smallest integer such that $\AA^t \yy = \mathbf{0}$.
We know $t \le c \le n$ since
\begin{align}
    \AA^{c} \yy = \AA^{c} \bar{\mu}(\AA) \zz = \mu(\AA) \zz = \mathbf{0}
\end{align}
If $t>0$, we can find the nontrivial kernel vector $\AA^{t-1} \yy$. We show this holds with high probability so that the algorithm terminates fast.
\begin{lemma} \label{lemma:kernel-prob}
    It holds that 
    \begin{align}
        \Pr[t=0] = \Pr[\yy = \mathbf{0}] \le \frac{1}{|F|}
    \end{align}
    where the randomness is over $\zz \sim F^n$.
\end{lemma}
\begin{proof}
    Let $V = F^n$ denote the vector space. Let $K$ denote the kernel space of $\AA$, which is a subspace of $V$. Let $K^\perp$ denote the orthogonal complement of $K$. We can decompose $V$ into the direct sum $V = K \oplus K^\perp$, and now every vector $\vv \in V$ can be uniquely written as $\vv = \uu + \uu^\perp$ where $\uu \in K$ and $\uu^\perp \in K^\perp$.

    We can write $\yy$ in the form
    \begin{align*}
        \yy = \bar{\mu}(\AA) \zz = \sum_{i=0}^{\deg(\bar{\mu})} \bar{a}_i \AA^i \zz = \bar{a}_0 \zz + \BB (\AA \zz)
    \end{align*}
    for matrix $\BB := \sum_{i=1}^{\deg(\bar{\mu})} \bar{a}_i \AA^{i-1}$. Note that by definition $\bar{a}_0 = \bar{\mu}(0) \neq 0$, so we have 
    \begin{align*}
        \Pr_{\zz \sim F^n}[\yy = \mathbf{0}]
        & =
        \Pr_{\uu \sim K, \uu^\perp \sim K^\perp}[ \yy = 0 \mid \zz = \uu + \uu^\perp]
        \\ & =
        \Pr_{\uu \sim K, \uu^\perp \sim K^\perp}[\bar{a}_0 (\uu + \uu^\perp) + \BB (\AA(\uu + \uu^\perp)) = \mathbf{0}]
        \\ & =
        \Pr_{\uu \sim K, \uu^\perp \sim K^\perp}[\bar{a}_0 \uu + \bar{a}_0 \uu^\perp + \BB \AA \uu^\perp = \mathbf{0}]
        \\ & =
        \Pr_{\uu^\perp \sim K^\perp} \left[ \Pr_{\uu \sim K}[\bar{a}_0 \uu = - \bar{a}_0 \uu^\perp - \BB \AA \uu^\perp] \right] 
        \\ & =
        \Pr_{\uu^\perp \sim K^\perp} \left[ \frac{1}{|K|} \cdot \mathbf{1}[ - \bar{a}_0 \uu^\perp - \BB \AA \uu^\perp \in K]  \right] 
        \\ & \le
        \frac{1}{|K|} \le \frac{1}{|F|}
    \end{align*} 
    where the last inequality holds since $\AA$ is singular and has nullity $>0$.
\end{proof}

The guarantees for linear system solving are based on \callalg{FindKernel} and the reduction \Cref{lemma:linear-system-reduction-to-kernel}. We restate and prove \callthm{thm:LinSolveZp}.

\LinSolveZp*
\labelalg{LinSolveZp}

\begin{proof}  
    We call \callalg{FindKernel} to find a nontrivial kernel vector, 
    \begin{align*}
        \Mbegin
        \AA & \bb \\
        \OO & 0
        \Mend
        \Mbegin
        \yy \\
        v
        \Mend
        =
        \mathbf{0},
    \end{align*}
    then $-\yy/v$ is the solution $\AA^{-1} \bb$, since $\AA$ is invertible and thus $v$ must be nonzero.

    By \Cref{lemma:kernel-prob}, assuming the minimal polynomial is given, one inner loop can correctly compute a nontrivial kernel vector with probability $\ge 1 - 1/p$. By repeating $\lceil \log_p(2/\delta) \rceil$ times, the success probability is boosted to $\ge 1 - \delta/2$. By \Cref{thm:wiedemann}, \callalg{Wiedemann} correctly computes the minimal polynomial with probability $\ge 1 - \delta / 2$, so a union bound concludes that the failure probability is at most $\delta$.
    
    For the performance analysis, the algorithm requires one call to \textsc{Wiedemann} and $\Otil(n)$ calls to the matrix-vector access, so the time complexity follows from \Cref{thm:wiedemann}. The space requirement is $O(n)$ field elements, since all the steps can be computed with $O(1)$ many vectors. All the polynomial evaluations of a matrix, for example the $\bar{\mu}(\AA)$ in \Cref{line:ls-yy}, should be computed by Horner's method so that the space is low.
\end{proof}

\section{Proofs of the Inverse Power Method}

\subsection{Proof of \Cref{thm:InvPower}} \label{sec:proof-InvPower}

\paragraph{Correctness of \callalg{InvPower}.} It suffices to prove the algorithm succeeds with probability $\ge 1 - O(1/n)$, since running it for constant number of times can boost the success probability to $1 - n^{-c}$. 
We assume all calls of \callalg{LinSolve} output successfully, satisfying the guarantees in \Cref{thm:LinSolve}, since this holds with high probability (for example $\ge 1 - n^{-c-1}$).

When $\AA$ is singular, by \Cref{thm:LinSolve}, \callalg{LinSolve} can detect the singularity and we return $0$.
Now we assume that $\AA$ is non-singular.

Let $\lambda_1, \dots, \lambda_n$ be the eigenvalues of $\AA^{-1}$ such that $|\lambda_1| \ge |\lambda_2| \ge \dots \ge |\lambda_n|$, so we have $\lam_i = (\lam^A_i)^{-1}$ for all $i\in[n]$.
Note that $\ee_1,\dots,\ee_n$ form an orthonormal eigenbasis of $\AA$. They are also eigenvectors of $\AA^{-1}$ with corresponding eigenvalues $\lambda_1, \lambda_2, \dots, \lambda_n$.

Let $S := \{i:|\lambda_i|\geq (1-\eps_S)|\lambda_1|\}$ where $\eps_S := \eps/4$.
Note that $\max_i | \lambda_i(\AA)| \le nU$, so we have $|\lambda_1| \ge 1/(nU)$.
Define potential function
\begin{align*}
    \Phi(\xx) := \frac{\sum_{k\notin S} \langle\xx,\ee_k\rangle^2}{ \langle\xx,\ee_1\rangle^2}
\end{align*}
for all $\xx \in \R^n$. Intuitively, if there were no error, the potential function would simplify to 
\begin{align*}
    \Phi(\vv^{(i+1)}) = \Phi(\AA^{-1} \vv^{(i)} / \| \AA^{-1} \vv^{(i)} \|_2 )
    =
    \Phi(\AA^{-1} \vv^{(i)} )
    =
    \frac{
        \sum_{k \in S}  \lambda_k^2 \la \vv^{(i)}, \ee_k \ra^2
    }{
        \lambda_1^2 \la \vv^{(i)}, \ee_1 \ra^2
    }
    \le
    (1 - \eps_S)^2\Phi(\vv^{(i)}),
\end{align*}
so the potential would be small enough after $O(\frac{\log (n/\eps)}{\eps})$ iterations and we are done. However, we need more effort to incorporate the analysis of the rounding errors. Specifically, for any $\xx \in \R^n$, we define another potential function
\begin{align*}
    \Psi(\xx) := 
    \frac{\sum_{k \in [n]} \langle\xx,\ee_k\rangle^2}{ \langle\xx,\ee_1\rangle^2} = \frac{\| \xx \|_2^2}{ \la \xx, \ee_1 \ra^2}.
\end{align*}
We want $\Psi(\vv^{(i)})$ to be polynomially bounded, since otherwise the denominator of $\Phi(\vv^{(i)})$ would be too small after normalization and would incur large errors.

We first prove the following two lemmas concerning the initial conditions.

\begin{lemma} \label{lemma:Phi-0}
    With probability at least $1 - 2/n$, it holds that $\| \uu^{(0)} \| \le n$, $| \la \uu^{(0)}, \ee_1 \ra | \ge n^{-2}/2$ and $\Phi(\uu^{(0)}) \le \Psi(\uu^{(0)}) \le 4n^6$.
\end{lemma}
\begin{proof}
    In \Cref{line:sample-vv-0} we sample $\uu^*\sim \mathcal{N}(0,\II)$ and take its fixed-point approximation $\uu^{(0)}$. For all $i \in [n]$, denote $c_i := \langle\uu^*,\ee_i\rangle$, and note that $c_i\sim \mathcal{N}(0,1)$ independently, we have
    \begin{align*}
        \Pr\left[|c_i|< n^{-2}\right] & = \sqrt{2/\pi}\int_{0}^{1/n^2} e^{-x^2/2}\mathrm{d}x\le 1/n^2, \\
        \Pr\left[|c_i| > 2 \log n \right]
        & = \sqrt{2/\pi}\int_{2 \log n}^{+\infty} e^{-x^2/2}\mathrm{d}x
        \le \int_{2 \log n}^{+\infty} e^{-x}\mathrm{d}x
        \le 1/n^2.
    \end{align*}
    Note that 
    $
    \la \uu^{(0)}, \ee_i \ra \in \la \uu^*, \ee_i \ra \pm \| \uu^{(0)} - \vv^* \|_2 
    $,
    and by the guarantee of $L$-bit fixed-point representation that $\| \uu^{(0)} - \uu^* \|_2 \le 2^{-L} \cdot n$, we have
    \begin{align*}
        \Pr_{\uu^{(0)}} \left[ |\langle \uu^{(0)}, \ee_i \rangle| \in [n^{-2}/2, 4\log n] \right] \ge 1- 2/n^2.
    \end{align*}
    By a union bound, we have $\| \uu^{(0)} \|_2 \le O(\sqrt{n \log^2 n}) \le O(n)$ and $\Psi( \uu^{(0)} ) \le \| \uu^{(0)} \|_2^2 / \la \uu^{(0)}, \ee_1 \ra^2 \le 4n^6$. We conclude the lemma by noting that $\Phi( \xx) \le \Psi(\xx)$ for any $\xx$.
\end{proof}

\begin{lemma} \label{lemma:InvPower-eigenvalue-very-small}
    With probability $\ge 1 - 2/n$, it holds that:
    \begin{itemize}
        \item If the algorithm reaches \Cref{line:return-0-if-eigval-small}, then $|\lambda_1| \ge 1/\delta$ and the output satisfies the desired property.
        \item If $|\lambda_1| > 16 n^3 / \delta$, then the algorithm reaches \Cref{line:return-0-if-eigval-small}.
    \end{itemize}
\end{lemma}
\begin{proof}
    By $L$-bit fixed-point and floating-point arithmetic, we have
    $\left\| \vv^{(1)} - \uu^{(0)}  / \| \uu^{(0)} \|_2 \right\|_2 \le \eps/n^4$ and thus $\| \vv^{(1)} \|_2 \le 1 + \eps/n^4$.
    Note that 
    \begin{align*}
        \| \uu^{(1)} \|_2^2
        \le
        (1+\eps_L)^2 \| \AA^{-1} \vv^{(1)} \|_2^2 
        \le
        (1+\eps_L)^2 \lambda_1^2 \|\vv^{(1)}\|_2^2
        \le
        2 \lambda_1^2. 
    \end{align*}
    When the algorithm goes to \Cref{line:return-0-if-eigval-small}, we must have
    \begin{align*}
        2\lambda_1^2 \ge \| \uu^{(0)} \|_2^2 \ge 2 / \delta^2 \Rightarrow |\lambda_1| \ge 1/\delta.
    \end{align*}
    Since the algorithm returns $\delta$ and $\max\{\delta, |1/\lambda_1|\} = \delta$, the output satisfies the desired property.
    
    For the second claim, if the algorithm does not go to \Cref{line:return-0-if-eigval-small}, it holds that
    $
    \| \uu^{(1)} \|_2^2 \le 4 / \delta^2
    $. We have
    \begin{align*}
        \| \uu^{(1)} \|_2^2
        \ge
        (1-\eps_L)^2 \| \AA^{-1} \vv^{(1)} \|_2^2 
        \ge
        (1-\eps_L)^2 \langle \AA^{-1} \vv^{(1)}, \ee_1 \rangle^2
        =
        (1-\eps_L)^2 \lambda_1^2 \langle \vv^{(1)}, \ee_1 \rangle^2.
    \end{align*}
    By \Cref{lemma:Phi-0}, with probability $\ge 1 - 2/n$, we have $|\langle \uu^{(0)}, \ee_1 \rangle| \ge n^{-2}/2$. Then,
    \begin{align*}
        |\langle \vv^{(1)}, \ee_1 \rangle |
        \ge |\langle \uu^{(0)} / \| \uu^{(0)} \|_2, \ee_1 \rangle| -
        \left\| \vv^{(1)} -  \uu^{(0)} / \| \uu^{(0)} \|_2 \right\|_2 \cdot \| \ee_1\|_2
        \ge
        n^{-3}/2 - \eps/n^4 \ge n^{-3}/4.
    \end{align*}
    where we used the fact
    \begin{align*}
        \langle \uu^{(0)} / \| \uu^{(0)} \|_2, \ee_1 \rangle^2
        =
        \Psi(\uu^{(0)})^{-1} \ge n^{-6}/4.
    \end{align*}
    Therefore,
    \begin{align*}
        4 / \delta^2
        \ge
        \| \uu^{(1)} \|_2^2
        \ge
         (1-\eps_L)^2 \lambda_1^2 \langle \vv^{(1)}, \ee_1 \rangle_2^2
         \ge
         \lambda_1^2 n^{-6}/32
        \Rightarrow
        |\lambda_1| \le 16 n^3 / \delta.
    \end{align*}
\end{proof}

By \Cref{lemma:InvPower-eigenvalue-very-small}, the case $|\lambda_1| > 16n^3/\delta$ is handled by \Cref{line:return-0-if-eigval-small}, so we can assume $|\lambda_1| \le 16n^3/\delta$. By \Cref{lemma:Phi-0}, we have $\|\uu_0 \|_2 \le n$. Now, we know that all $\uu^{(i)}, \vv^{(i)}$ have polynomially bounded entries. Specifically, since $\vv^{(i)}$ is a normalized vector of $\uu^{(i-1)}$, we have
\begin{align}
    \| \vv^{(i)} \|_2  \in [1 - \eps/3, 1+\eps/3]. \label{eq:vv-i-norm}
\end{align}
Since $|\lambda_i(\AA^{-1})| \in [1/(nU), 16n^3/\delta]$, we have
\begin{align*}
    \| \uu^{(i)} \|_2 \in [1/\sqrt{M}, M]
\end{align*}
for any $M \ge n^2U^2 + 16n^3/\delta$. We will pick a sufficiently large $M = \poly(n, U, \eps^{-1})$ later, and set $\eps_0 := \eps / (100M^2)$ such that the following approximation guarantees hold. The guarantees follow from $L$-bit fixed-point arithmetic (note that we used floating-point arithmetic but the guarantees are only stronger).
\begin{itemize}
    \item In \Cref{line:compute-vi}, it holds that
    \begin{align} \label{eq:invpow-approx-2}
        \left\| \vv^{(i)} - \uu^{(i-1)}  / \| \uu^{(i-1)} \|_2 \right\|_2  \le 2^{-L} \cdot \poly(n,U)  \le \eps_0.
    \end{align}
    \item By \Cref{thm:LinSolve}, in \Cref{line:compute-vi}, it holds that
    \begin{align} \label{eq:invpow-approx-1}
        \| \uu^{(i)}  - \AA^{-1} \vv^{(i)}\|_2 \le \eps_L \cdot \poly(n, U) \le \eps_0
    \end{align}
    by choosing a small enough $\eps_L$.
\end{itemize}

Next, we show that if a vector $\xx$ approximates $\yy$ under small additive error with bounded $\ell_2$ norm and potential $\Psi$, then $\Phi(\xx)$ and $\Psi(\xx)$ approximate $\Phi(\yy)$ and $\Psi(\yy)$, respectively. 

\begin{claim} \label{claim:phi-additive}
    For $\zeta \in [0, 0.1], M>1$, any vector $\xx$ and $\yy$ with $\| \xx \|_2 \in [1/\sqrt{M},M]$ and $\Psi(\xx) \le M$, if $\|\xx - \yy\|_2 \le \zeta/M^2$, it holds that
    \begin{align*}
        \Phi(\xx) \approxbar_{8 \zeta}  \Phi(\yy), \quad
        \Psi(\xx) \approxbar_{8 \zeta}  \Psi(\yy).
    \end{align*}
\end{claim}
\begin{proof}
    We have
    $ \la \xx, \ee_1 \ra^2 \ge \| \xx \|_2^2 / \Psi(\xx) \ge 1/M^2$ and thus
    $|\la \xx, \ee_1 \ra | \ge 1/M$.
    Define $\Delta := \yy - \xx$. For any $k \in [n]$, we have
    \begin{align*}
        |\la \xx, \ee_k \ra - \la \yy, \ee_k \ra|
        \le
        |\la \Delta, \ee_k \ra|
        \le
        \| \Delta\|_2 \cdot \| \ee_k \|_2
        =
        \| \Delta \|_2
        \le
        \zeta/M^2 \le \zeta |\la \xx, \ee_1 \ra|.
    \end{align*}
    We also have
    \begin{align*}
        |\la \xx, \ee_k \ra| & \le \| \xx \|_2 \cdot \| \ee_k \|_2 \le  M, \\
        |\la \yy, \ee_k \ra| & \le |\la \xx, \ee_k \ra| + |\la \yy-\xx, \ee_k \ra|
         \le M + \zeta/M^2 \le 1.1M,
    \end{align*}
    Therefore,  
    \begin{align*}
        |\la \xx, \ee_k \ra^2 - \la \yy, \ee_k \ra^2|
        & \le
        |\la \xx, \ee_k \ra^2 - \la \xx, \ee_k \ra \la \yy, \ee_k \ra|
        +  |\la \xx, \ee_k \ra \la \yy, \ee_k \ra - \la \yy, \ee_k \ra^2| 
        \\ & \le
        |\la \xx, \ee_k \ra| \cdot |\la \xx, \ee_k \ra - \la \yy, \ee_k \ra| +  
        |\la \xx, \ee_k \ra - \la \yy, \ee_k \ra| \cdot |\la \yy, \ee_k \ra|
        \\ & \le
        2 (\zeta/M^2) \cdot 1.1M
        \le 2.2 \zeta / M \le 2.2 \zeta | \la \xx, \ee_1 \ra |.
    \end{align*}
    Then,
    \begin{align*}
        \Phi(\yy) 
        =
        \frac{
            \sum_{k \notin S} \la \yy, \ee_k \ra^2
        } {
            \la \yy, \ee_1 \ra^2
        }
        =
        \frac{
            \sum_{k \notin S} \la \yy, \ee_k \ra^2 / \la \xx, \ee_1 \ra^2
        } {
            \la \yy, \ee_1 \ra^2 / \la \xx, \ee_1 \ra^2
        }
        =
        \Phi(\xx) \cdot \frac{1 \pm 2.2 \zeta}{1 \pm 2.2 \zeta}
        \in
        \Phi(\xx) \cdot (1 \pm 4 \zeta).
    \end{align*}
    Therefore, we have $\Phi(\xx) \approxbar_{8 \zeta} \Phi(\yy)$. Similarly, we have 
    \begin{align*}
        \Psi(\yy) 
        =
        \frac{
            \sum_{k \in [n]} \la \yy, \ee_k \ra^2
        } {
            \la \yy, \ee_1 \ra^2
        }
        =
        \frac{
            \sum_{k \in [n]} \la \yy, \ee_k \ra^2 / \la \xx, \ee_1 \ra^2
        } {
            \la \yy, \ee_1 \ra^2 / \la \xx, \ee_1 \ra^2
        }
        =
        \Psi(\xx) \cdot \frac{1 \pm 2.2 \zeta}{1 \pm 2.2 \zeta}
        \in
        \Psi(\xx) \cdot (1 \pm 4 \zeta).
    \end{align*}
    and thus $\Psi(\xx) \approxbar_{8 \zeta} \Psi(\yy)$.
\end{proof}

The next lemma shows that the potential $\Phi$ is decreasing over iterations, and the potential $\Psi$ is not growing too fast. 

\begin{lemma} \label{lemma:Phi-iterative}
    For any $i \in [1, T]$ and $M>0$, if $\Psi(\uu^{(i)}) \le M$ and $\eps_0 \le \eps/(100 M^2)$, we have
    \begin{itemize}
        \item $\Phi(\vv^{(i)}) \le \exp(\eps/8)\Phi(\uu^{(i-1)})$ and $\Phi(\uu^{(i)}) \le \exp( - \eps_S)\Phi(\uu^{(i-1)})$.
        \item $\Psi(\vv^{(i)}) \le \exp(\eps/8)\Psi(\uu^{(i-1)})$ and $\Psi(\uu^{(i)}) \le \exp(\eps_S) \Psi(\uu^{(i-1)})$.
    \end{itemize}
\end{lemma}

\begin{proof}
    Define $\uuhat^{(i-1)} := \uu^{(i-1)} / \| \uu^{(i-1)} \|_2$.  By \eqref{eq:invpow-approx-2} and \Cref{claim:phi-additive}, we have 
    \begin{align}  \label{eq:phi-3}
        \Phi( \vv^{(i)} )
        & \approxbar_{\eps/8}
        \Phi(\uuhat^{(i-1)})
        =
        \Phi(\uu^{(i-1)}), \quad
        \Psi( \vv^{(i)} )
        \approxbar_{\eps/8}
        \Psi(\uuhat^{(i-1)})
        =
        \Psi(\uu^{(i-1)}).
    \end{align}
    Since $\langle \AA^{-1} \xx, \ee_k \rangle^2 = \lambda_k^2 \langle \xx, \ee_k \rangle^2$ for any vector $\xx$ and $k \in [n]$, we have
    \begin{align} \label{eq:phi-1}
        \Phi( \AA^{-1} \vv^{(i)} )
        =
         \frac{\sum_{k\notin S} \langle \AA^{-1}\vv^{(i)} ,\ee_k\rangle^2}{ \langle \AA^{-1}\vv^{(i)} ,\ee_1\rangle^2}
        =
        \frac{\sum_{k\notin S} \lambda_k^2 \langle \vv^{(i)} ,\ee_k\rangle^2}{ \lambda_1^2 \langle \vv^{(i)} ,\ee_1\rangle^2}
        \le
        (1-\eps_S)^2 \Phi(\vv^{(i)})
    \end{align}
    by the definition of $S$. Similarly, we have $\Psi( \AA^{-1} \vv^{(i)} ) \le \Psi(\vv^{(i)})$ without using the property of $S$.
    Now $\Psi(\AA^{-1} \vv^{(i)})$ is bounded by $\exp(\eps/8)M$, so it follows from \eqref{eq:invpow-approx-1} and \Cref{claim:phi-additive} that
    \begin{align}  \label{eq:phi-2}
        \Phi( \uu^{(i)} )
        & \approxbar_{\eps/8}
        \Phi(\AA^{-1} \vv^{(i)}), \quad
        \Psi( \uu^{(i)} )
        \approxbar_{\eps/8}
        \Psi(\AA^{-1} \vv^{(i)}).
    \end{align}
    Combining all the above inequalities, we have
    \begin{align*}
        \Phi(\uu^{(i)}) \le \exp(\eps/8)^2 (1-\eps_S)^2 \Phi( \uu^{(i-1)})
         & \le
        \exp( \eps/4 - 2\eps_S) \Phi( \uu^{(i-1)})
        \le
        \exp( - \eps_S) \Phi( \uu^{(i-1)})
    \end{align*}
    where we used the fact $\eps_S = \eps/4$.
    Similarly, 
    $
    \Psi(\uu^{(i)}) \le \exp(\eps/8)^2 \Psi( \uu^{(i-1)})
        \le
        \exp( \eps_S) \Psi( \uu^{(i-1)})
    $.
\end{proof}

Now we prove the correctness by combining all the above lemmas. We first set the parameter $M$ by bounding the largest potential $\Psi$.
Recall that $T = \lceil\frac{28\log (4n / \eps_S)}{\eps}\rceil = \lceil\frac{7\log (n / \eps_S)}{\eps_S}\rceil$.
By \Cref{lemma:Phi-0} and \Cref{lemma:Phi-iterative} assuming $\eps_0$ is small enough to satisfy the conditions, we have
\begin{align*}
    \max_{i} \Psi(\uu^{(i)}) \le \Psi(\uu^{(T)}) \le \exp(\eps_S)^T \Psi(\uu^{(0)})
    \le
    (n/\eps_S)^7 \cdot 4n^6.
\end{align*}
Therefore, it suffices to set
\begin{align*}
    M = n^2U^2 + 16n^3/\delta + (n/\eps_S)^7 \cdot 4n^6 \quad \text{and} \quad
    \eps_0 = \eps / (100M^2)
\end{align*}
to satisfy all the conditions of \Cref{lemma:Phi-iterative} and the desired property for the approximation guarantees. Then, we bound the potential
\begin{align*}
    \Phi(\uu^{(T)}) \le \exp(-\eps_S)^T \Phi(\uu^{(0)})
    \le
    n^{-7}\eps_S \cdot 4n^6
    \le \eps_S/2
\end{align*}
by \Cref{lemma:Phi-0} and \Cref{lemma:Phi-iterative}. Thus we have $\Phi(\vv^{(T)}) \le \exp(\eps/8) \eps_S/2 \le \eps_S$.

Note that $\min_i|\lambda_i(\AA)| = 1 / \lambda_1$.
We want to show $\lamtil \approxbar_{\eps} 1 / \lambda_1$. It holds that
\begin{align*}
    \left(\frac{\left\|\AA^{-1}\vv^{(T)}\right\|_2}{\|\vv^{(T)}\|_2}\right)^2
    &= \frac{\sum_{i=1}^n \lambda_i^2\langle \vv^{(T)},\ee_i\rangle^2}{\sum_{i=1}^n \langle \vv^{(T)},\ee_i\rangle^2}
    \le \lambda_1^2,
\end{align*}
and
\begin{align*}
    \left(\frac{\left\|\AA^{-1}\vv^{(T)}\right\|_2}{\|\vv^{(T)}\|_2}\right)^2
    \ge \frac{\sum_{i\in S} \lambda_i^2 \langle \vv^{(T)},\ee_i\rangle^2}{\langle \vv^{(T)},\ee_1\rangle^2}
    \ge (1-\eps_S)^2\lambda_1^2 (1 - \Phi(\vv^{(T)})) 
    \ge
    \exp( - 6\eps_S) \lambda_1^2.
\end{align*}
Thus we have
$
\left\|\AA^{-1}\vv^{(T)}\right\|_2\approxbar_{3\eps_S} \lambda_1 \|\vv^{(T)}\|_2.
$
Since $\| \uu^{(T)} - \AA^{-1}\vv^{(T)} \| \le \eps_0$,
we have
\begin{align*}
 \| \uu^{(T)}\|_2  / \| \vv^{(T)} \|_2 \approxbar_{3\eps_S + 2 \eps_0} \lambda_1   
\end{align*}
and thus 
\begin{align*}
    \lamtil \approxbar_{\eps_0} \|\vv^{(T)}\|_2 / \| \uu^{(T)} \|_2 \approxbar_{3\eps_S + 2\eps_0}  1/\lambda_1
\end{align*}
We conclude that
$
\lamtil \approxbar_{\eps}  1/\lambda_1
$
since $3\eps_0 + 3\eps_S \le \eps$.

\paragraph{Time and space complexity of \Cref{thm:InvPower}.}  We have $T = O(\log(n/\eps)/\eps)$ and $\log(1/\eps_0) = O(\log(nU/\eps))$. The dominant step of the algorithm the subroutine \callalg{LinSolve}. Note that we scale up the entries of $\AA$ and $\vv^{(i)}$ by $2^L$ to make the input of \callalg{LinSolve} become integers.  
Since the entries are still bounded by $U \cdot 2^L = U \cdot 2^{O(\log(nU/\eps))} \le \poly(n, U, \eps^{-1})$, and the algorithm invokes the subroutine for $T$ times, the time complexity is
\begin{align*}
    O(T \cdot n^2(\nnz + n + \log(1/\eps_L))) = \Otil(n^2(\nnz + n)/\eps).
\end{align*}
It has the same space guarantees as \callalg{LinSolve}, concluding the proof of \Cref{thm:InvPower}.

\subsection{Proof of \Cref{cor:InvPowerGap}}  \label{sec:proof-InvPowerGap}
        The corollary assumes the property of spectral gap: $\delta \le | \lambda^A_1|$ and $1.1 | \lambda^A_1| \le | \lambda^A_2|$. Thus we have
    \begin{align*}
        |\lam_1 | \ge 1/\delta, \quad |\lam_1| \ge 1.1 |\lam_2|.
    \end{align*}
    The first property assures that an eigenvector $\vv^{(T)}$ will be produced. For the second property, it allows us to use a larger $\eps_S := 0.1$, and the set $S$ by definition
    \begin{align*}
        S = \{i : |\lam_i| \ge (1 - \eps_S) |\lam_1| \} = [n] \setminus \{1\}.
    \end{align*}
    We still have $\Phi(\uu^{(i+1)}) \le \exp(-\eps_S) \Phi(\uu^{(i)})$, so we can use a smaller $T$ for \callalg{InvPowerGap} to achieve $\eps$ error, specifically,
    \begin{align*}
        T := O( \log(n/\eps)).
    \end{align*}
    This gives a shorter running time than \callalg{InvPower} without the dependency $1/\eps$.
    
    We prove the ``Furthermore'' part of \Cref{cor:InvPowerGap}. By \eqref{eq:vv-i-norm}, we have $\|\vv^{(T)}\|_2\in [1 - \eps/3, 1 + \eps/3]$ and thus
    \begin{align*}
        \langle \vv^{(T)},\vv^{(T)}\rangle = \|\vv^{(T)}\|_2^2 \in [1-\eps,1+\eps].
    \end{align*}
    Since $|\lam^A_2|>(1+\eps)|\lam^A_1|$, it holds that
    \begin{align*}
        |\lambda_2|< (1 - \eps/2)|\lambda_1| < (1-\eps_S)|\lambda_1|,
    \end{align*}
    which means that $S = [n] \setminus \{1\}$. We have already concluded that $\Phi(\vv^{(T)}) \le \eps_S$, so
    \begin{align*}
        \| \vv^{(T)}\|_2^2
        &= \sum_{i=1}^n \langle \vv^{(T)}, \ee_i\rangle^2
        = (\Phi(\vv^{(T)})+1) \langle \vv^{(T)}, \ee_1\rangle^2
        \le (\eps_S+1) \langle \vv^{(T)}, \ee_1\rangle^2.
    \end{align*}
    Thus,
    \begin{align*}
        \left|\langle \vv^{(T)}, \ee_1\rangle\right| \ge \frac{1}{\sqrt{\eps_S + 1}} \| \vv^{(T)}\|_2 \ge \frac{1 - \eps/3}{\sqrt{1 + \eps / 3}} \ge 1 - \eps.
    \end{align*}
    On the other hand,
    $
        \left|\langle \vv^{(T)}, \ee_1\rangle\right| \le \| \vv^{(T)}\|_2 \le 1+\eps.
    $
    Now we conclude that
    $
    \left|\langle \vv^{(T)}, \ee_1\rangle\right| \in [1-\eps,1+\eps].
    $

\end{document}